\pdfoutput=1
\documentclass[12pt,a4paper]{article}
\usepackage[colorinlistoftodos]{todonotes}

\usepackage{ifthen} % for conditional statements
\newboolean{pdflatex}
\setboolean{pdflatex}{true} % False for eps figures

\newboolean{articletitles}
\setboolean{articletitles}{true} % False removes titles in references

\newboolean{uprightparticles}
\setboolean{uprightparticles}{false} %True for upright particle symbols

\newboolean{inbibliography}
\setboolean{inbibliography}{false} %True once you enter the bibliography

\def\paperauthors{LHCb collaboration} % Leave as is for PAPER and CONF
\def\paperasciititle{Search for CP violation through an amplitude analysis of D0->KKpipi decays} % Set ASCII title here
\def\papertitle{Search for \CP violation through an amplitude analysis of \DztoKKpipi decays} % Latex formatted title
\def\paperabstract{A search for \CP violation in the Cabibbo-suppressed \DztoKKpipi decay mode is performed using an amplitude analysis. The measurement uses a sample of $pp$ collisions recorded by the \lhcb experiment during 2011
and 2012, corresponding to an integrated luminosity of 3.0\invfb .
The \Dz mesons are reconstructed from semileptonic $\mbox{$b$-hadron}$ decays into $\Dz\mun X$ final states.
The selected sample contains more than 160\,000 signal decays, allowing the most precise amplitude modelling of this \Dz decay to date.
The obtained amplitude model is used to perform the search for \CP violation. The result is compatible with \CP symmetry, with a sensitivity ranging from 1\% to 15\% depending on the amplitude considered.}

\def\paperkeywords{{High Energy Physics}, {LHCb}, {CP violation}, {charm}, {Amplitude Analysis}} % Comma separated list
\def\papercopyright{\the\year\ CERN for the benefit of the LHCb collaboration} % new since 9/Apr/2018
\def\paperlicence{CC-BY-4.0 licence}
\def\paperlicenceurl{https://creativecommons.org/licenses/by/4.0/}

\usepackage[top=1in, bottom=1.25in, left=1in, right=1in]{geometry}

\usepackage{microtype}
\usepackage{lineno}  % for line numbering during review
\usepackage{xspace} % To avoid problems with missing or double spaces after
\usepackage{caption} %these three command get the figure and table captions automatically small

\usepackage{graphicx}  % to include figures (can also use other packages)
\usepackage{color}
\usepackage{colortbl}
\graphicspath{{./figs/}} % Make Latex search fig subdir for figures
\DeclareGraphicsExtensions{.pdf,.PDF,png,.PNG}

\usepackage{amsmath} % Adds a large collection of math symbols
\usepackage{amssymb}
\usepackage{amsfonts}
\usepackage{upgreek} % Adds in support for greek letters in roman typeset

\newcommand*\patchAmsMathEnvironmentForLineno[1]{%
\expandafter\let\csname old#1\expandafter\endcsname\csname #1\endcsname
\expandafter\let\csname oldend#1\expandafter\endcsname\csname
end#1\endcsname
 \renewenvironment{#1}%
   {\linenomath\csname old#1\endcsname}%
   {\csname oldend#1\endcsname\endlinenomath}%
}
\newcommand*\patchBothAmsMathEnvironmentsForLineno[1]{%
  \patchAmsMathEnvironmentForLineno{#1}%
  \patchAmsMathEnvironmentForLineno{#1*}%
}
\AtBeginDocument{%
\patchBothAmsMathEnvironmentsForLineno{equation}%
\patchBothAmsMathEnvironmentsForLineno{align}%
\patchBothAmsMathEnvironmentsForLineno{flalign}%
\patchBothAmsMathEnvironmentsForLineno{alignat}%
\patchBothAmsMathEnvironmentsForLineno{gather}%
\patchBothAmsMathEnvironmentsForLineno{multline}%
\patchBothAmsMathEnvironmentsForLineno{eqnarray}%
}

\usepackage{hyperxmp}

\usepackage[pdftex,
            pdfauthor={\paperauthors},
            pdftitle={\paperasciititle},
            pdfkeywords={\paperkeywords},
            pdfcopyright={Copyright (C) \papercopyright},
            pdflicenseurl={\paperlicenceurl}]{hyperref}

\usepackage[all]{hypcap} % Internal hyperlinks to floats.

\usepackage{xspace} 
\usepackage{upgreek}

\newcommand{\offsetoverline}[2][0.1em]{\kern #1\overline{\kern -#1 #2}}%

\def\lhcb   {\mbox{LHCb}\xspace}

\def\babar  {\mbox{BaBar}\xspace}

\def\MagUp {\mbox{\em Mag\kern -0.05em Up}\xspace}

\ifthenelse{\boolean{uprightparticles}}%
{

 \def\Pmu         {\ensuremath{\upmu}\xspace}

 \def\Ppi         {\ensuremath{\uppi}\xspace}

 \def\PDelta      {\ensuremath{\Delta}\xspace}                 
 \def\PXi         {\ensuremath{\Xi}\xspace}                 
 \def\PLambda     {\ensuremath{\Lambda}\xspace}                 
 \def\PSigma      {\ensuremath{\Sigma}\xspace}                 
 \def\POmega      {\ensuremath{\Omega}\xspace}                 
 \def\PUpsilon    {\ensuremath{\Upsilon}\xspace}

 \def\PB      {\ensuremath{\mathrm{B}}\xspace}                 
                  
 \def\PD      {\ensuremath{\mathrm{D}}\xspace}

 \def\PK      {\ensuremath{\mathrm{K}}\xspace}

 \def\Pb      {\ensuremath{\mathrm{b}}\xspace}                 
 \def\Pc      {\ensuremath{\mathrm{c}}\xspace}

 \def\Pi      {\ensuremath{\mathrm{i}}\xspace}

 \def\Pp      {\ensuremath{\mathrm{p}}\xspace}                 
 \def\Pq      {\ensuremath{\mathrm{q}}\xspace}

}
{

 \def\Pmu         {\ensuremath{\mu}\xspace}

 \def\Ppi         {\ensuremath{\pi}\xspace}

 \mathchardef\PDelta="7101
 \mathchardef\PXi="7104
 \mathchardef\PLambda="7103
 \mathchardef\PSigma="7106
 \mathchardef\POmega="710A
 \mathchardef\PUpsilon="7107
                  
 \def\PB      {\ensuremath{B}\xspace}                 
                  
 \def\PD      {\ensuremath{D}\xspace}

 \def\PK      {\ensuremath{K}\xspace}

 \def\Pb      {\ensuremath{b}\xspace}                 
 \def\Pc      {\ensuremath{c}\xspace}

 \def\Pi      {\ensuremath{i}\xspace}

 \def\Pp      {\ensuremath{p}\xspace}                 
 \def\Pq      {\ensuremath{q}\xspace}

}

\makeatletter
\ifcase \@ptsize \relax% 10pt
  \newcommand{\miniscule}{\@setfontsize\miniscule{4}{5}}% \tiny: 5/6
\or% 11pt
  \newcommand{\miniscule}{\@setfontsize\miniscule{5}{6}}% \tiny: 6/7
\or% 12pt
  \newcommand{\miniscule}{\@setfontsize\miniscule{5}{6}}% \tiny: 6/7
\fi
\makeatother

\DeclareRobustCommand{\optbar}[1]{\shortstack{{\miniscule (\rule[.5ex]{1.25em}{.18mm})}
  \\ [-.7ex] $#1$}}

\def\mup        {{\ensuremath{\Pmu^+}}\xspace}
\def\mun        {{\ensuremath{\Pmu^-}}\xspace} % muon negative (\mum is taken)

\def\quark     {{\ensuremath{\Pq}}\xspace}

\def\cquark    {{\ensuremath{\Pc}}\xspace}

\def\bquark    {{\ensuremath{\Pb}}\xspace}

\def\pion   {{\ensuremath{\Ppi}}\xspace}

\def\pip    {{\ensuremath{\pion^+}}\xspace}
\def\pim    {{\ensuremath{\pion^-}}\xspace}

\def\kaon    {{\ensuremath{\PK}}\xspace}
  \def\Kbar    {{\kern 0.2em\overline{\kern -0.2em \PK}{}}\xspace}

\def\KorKbar {\kern 0.18em\optbar{\kern -0.18em K}{}\xspace}

\def\Kp      {{\ensuremath{\kaon^+}}\xspace}
\def\Km      {{\ensuremath{\kaon^-}}\xspace}

\def\KS      {{\ensuremath{\kaon^0_{\mathrm{S}}}}\xspace}

\def\Kstar   {{\ensuremath{\kaon^*}}\xspace}
\def\Kstarb  {{\ensuremath{\Kbar{}^*}}\xspace}

  \def\Dbar    {{\kern 0.2em\overline{\kern -0.2em \PD}{}}\xspace}
\def\D       {{\ensuremath{\PD}}\xspace}

\def\DorDbar {\kern 0.18em\optbar{\kern -0.18em D}{}\xspace}
\def\Dz      {{\ensuremath{\D^0}}\xspace}
\def\Dzb     {{\ensuremath{\Dbar{}^0}}\xspace}

\def\Dstarp  {{\ensuremath{\D^{*+}}}\xspace}

\def\B       {{\ensuremath{\PB}}\xspace}
\def\Bbar    {{\ensuremath{\kern 0.18em\overline{\kern -0.18em \PB}{}}}\xspace}

\def\BorBbar    {\kern 0.18em\optbar{\kern -0.18em B}{}\xspace}

\def\Bub     {{\ensuremath{\B^-}}\xspace}

\def\Bm      {{\ensuremath{\Bub}}\xspace}

\def\Y#1S{\ensuremath{\PUpsilon{(#1S)}}\xspace}

\def\proton      {{\ensuremath{\Pp}}\xspace}

\def\LorLbar     {\kern 0.18em\optbar{\kern -0.18em \PLambda}{}\xspace}

\newcommand{\decay}[2]{\mbox{\ensuremath{#1\!\to #2}}\xspace}         % {\Pa}{\Pb \Pc}

\def\to                 {\ensuremath{\rightarrow}\xspace}

\def\CP                {{\ensuremath{C\!P}}\xspace}

\def\AT#1     {\ensuremath{A_{\mathrm{T}}^{#1}}\xspace}           % 2

\def\C#1      {\ensuremath{\mathcal{C}_{#1}}\xspace}                       % 9
\def\Cp#1     {\ensuremath{\mathcal{C}_{#1}^{'}}\xspace}                    % 7
\def\Ceff#1   {\ensuremath{\mathcal{C}_{#1}^{\mathrm{(eff)}}}\xspace}        % 9  
\def\Cpeff#1  {\ensuremath{\mathcal{C}_{#1}^{'\mathrm{(eff)}}}\xspace}       % 7
\def\Ope#1    {\ensuremath{\mathcal{O}_{#1}}\xspace}                       % 2
\def\Opep#1   {\ensuremath{\mathcal{O}_{#1}^{'}}\xspace}                    % 7

\newcommand{\tev}{\ifthenelse{\boolean{inbibliography}}{\ensuremath{~T\kern -0.05em eV}}{\ensuremath{\mathrm{\,Te\kern -0.1em V}}}\xspace}
\newcommand{\gev}{\ensuremath{\mathrm{\,Ge\kern -0.1em V}}\xspace}
\newcommand{\mev}{\ensuremath{\mathrm{\,Me\kern -0.1em V}}\xspace}
\newcommand{\kev}{\ensuremath{\mathrm{\,ke\kern -0.1em V}}\xspace}
\newcommand{\ev}{\ensuremath{\mathrm{\,e\kern -0.1em V}}\xspace}
\newcommand{\mevc}{\ensuremath{{\mathrm{\,Me\kern -0.1em V\!/}c}}\xspace}
\newcommand{\gevc}{\ensuremath{{\mathrm{\,Ge\kern -0.1em V\!/}c}}\xspace}
\newcommand{\mevcc}{\ensuremath{{\mathrm{\,Me\kern -0.1em V\!/}c^2}}\xspace}
\newcommand{\gevcc}{\ensuremath{{\mathrm{\,Ge\kern -0.1em V\!/}c^2}}\xspace}
\newcommand{\gevgevcc}{\ensuremath{{\mathrm{\,Ge\kern -0.1em V^2\!/}c^2}}\xspace} % for \pt^2 in CEP
\newcommand{\gevgevcccc}{\ensuremath{{\mathrm{\,Ge\kern -0.1em V^2\!/}c^4}}\xspace} % for q^2

\def\mum  {\ensuremath{{\,\upmu\mathrm{m}}}\xspace}

\def\invfb   {\ensuremath{\mbox{\,fb}^{-1}}\xspace}

\newcommand{\chisq}{\ensuremath{\chi^2}\xspace}
\newcommand{\chisqndf}{\ensuremath{\chi^2/\mathrm{ndf}}\xspace}

\def\gsim{{~\raise.15em\hbox{$>$}\kern-.85em
          \lower.35em\hbox{$\sim$}~}\xspace}
\def\lsim{{~\raise.15em\hbox{$<$}\kern-.85em
          \lower.35em\hbox{$\sim$}~}\xspace}

\def\pt         {\ensuremath{p_{\mathrm{T}}}\xspace}

\def\evtgen     {\mbox{\textsc{EvtGen}}\xspace}

\def\geant      {\mbox{\textsc{Geant4}}\xspace}

\def\photos     {\mbox{\textsc{Photos}}\xspace}

\def\pythia     {\mbox{\textsc{Pythia}}\xspace}

\def\tell1  {TELL1\xspace}
\def\ukl1   {UKL1\xspace}

\newcommand{\eg}{\mbox{\itshape e.g.}\xspace}
\newcommand{\ie}{\mbox{\itshape i.e.}\xspace}

\usepackage{booktabs}
\usepackage{siunitx}
\usepackage{multicol}
\usepackage{multirow}
\usepackage{stackengine}
\usepackage{mathtools}
\usepackage[normalem]{ulem}
\usepackage{xcolor}
\newcommand\redsout{\bgroup\markoverwith{\textcolor{red}{\rule[0.5ex]{2pt}{0.4pt}}}\ULon}
\usepackage{numprint}
\usepackage{bm}

\makeatletter
\g@addto@macro\bfseries{\boldmath}
\makeatother

\def\DztoKKpipi   {\decay{\Dz}{\Kp\Km\pip\pim}}

\def\Kstz  {{\ensuremath{\kaon^*(892)^0}}\xspace}
\def\Kstzb {{\ensuremath{\Kstarb(892)^0}}\xspace}

\def\Kstzone  {{\ensuremath{\kaon^*(1680)^0}}\xspace}

\usepackage{cite} % Allows for ranges in citations
\usepackage{mciteplus}

\begin{document}

%%%%%%%%%%%%%%%%%%%%%%%%%
%%%%% Title     %%%%%%%%%
%%%%%%%%%%%%%%%%%%%%%%%%%
\renewcommand{\thefootnote}{\fnsymbol{footnote}}
\setcounter{footnote}{1}

% %%%%%%% CHOOSE TITLE PAGE--------
%\onecolumn
%\input{title-LHCb-INT}
%\input{title-LHCb-ANA}
%\input{title-LHCb-CONF}
% $Id: title-LHCb-PAPER.tex 122889 2018-08-17 17:59:55Z pkoppenb $
% ===============================================================================
% Purpose: LHCb-PAPER journal paper title page template
% Author:
% Created on: 2010-09-25
% ===============================================================================
% !TEX root = ./ms.tex

%%%%%%%%%%%%%%%%%%%%%%%%%
%%%%%  TITLE PAGE  %%%%%%
%%%%%%%%%%%%%%%%%%%%%%%%%
\begin{titlepage}
\pagenumbering{roman}

% Header ---------------------------------------------------
\vspace*{-1.5cm}
\centerline{\large EUROPEAN ORGANIZATION FOR NUCLEAR RESEARCH (CERN)}
\vspace*{1.5cm}
\noindent
\begin{tabular*}{\linewidth}{lc@{\extracolsep{\fill}}r@{\extracolsep{0pt}}}
\ifthenelse{\boolean{pdflatex}}% Logo format choice
{\vspace*{-1.5cm}\mbox{\!\!\!\includegraphics[width=.14\textwidth]{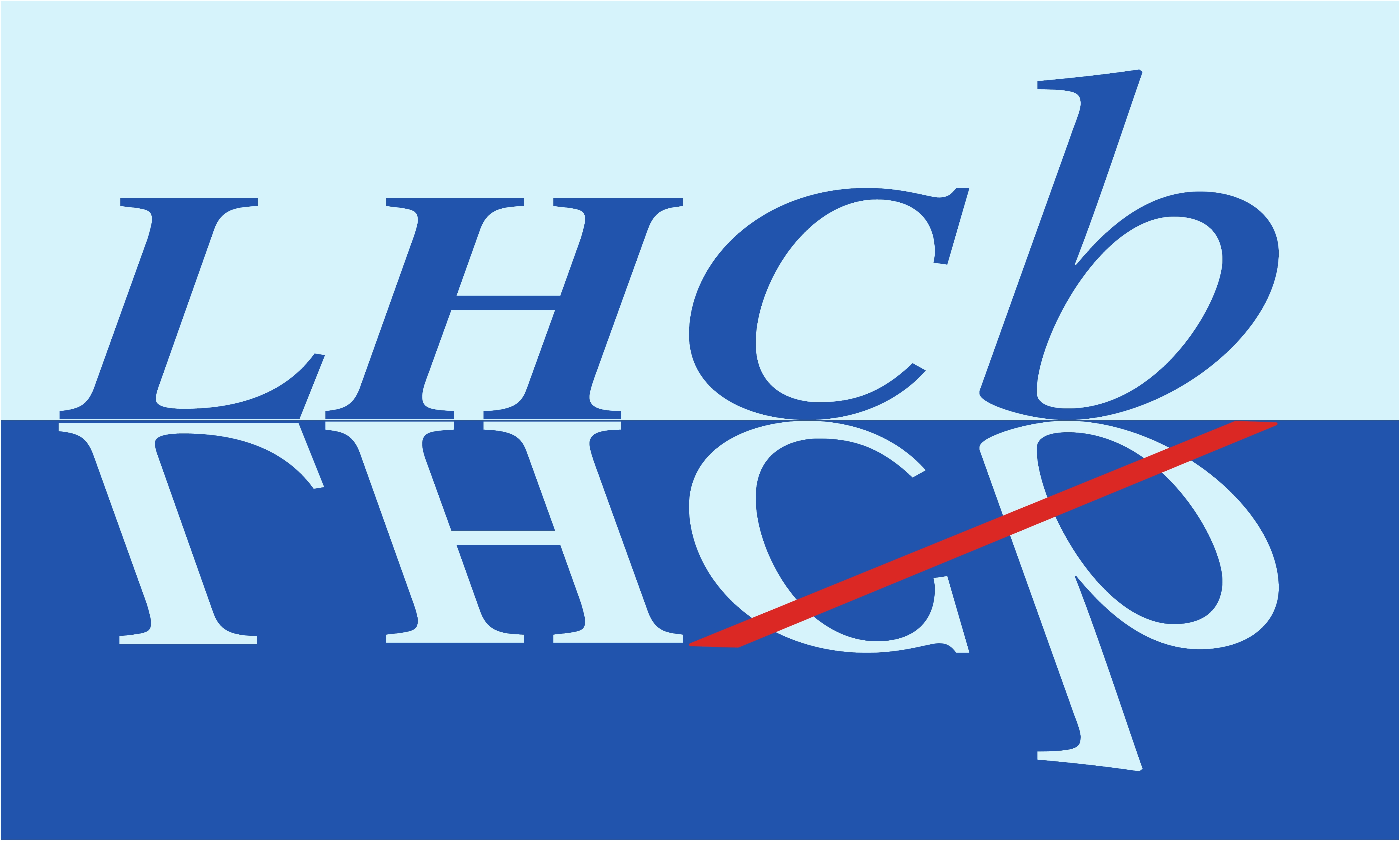}} & &}%
{\vspace*{-1.2cm}\mbox{\!\!\!\includegraphics[width=.12\textwidth]{lhcb-logo.eps}} & &}%
\\
 & & CERN-EP-2018-299 \\  % ID
 & & LHCb-PAPER-2018-041 \\  % ID
 & & 6 March 2019 \\ % Date - Can also hardwire e.g.: 23 March 2010
% & & v3.2\\
% not in paper \hline
\end{tabular*}

\vspace*{4.0cm}

% Title --------------------------------------------------
{\normalfont\bfseries\boldmath\huge
\begin{center}
% DO NOT EDIT HERE. Instead edit macro in main.tex to keep metadata correct
  \papertitle
\end{center}
}

\vspace*{2.0cm}

% Authors -------------------------------------------------
\begin{center}
%In the footnote, replace 'paper' by 'Letter' in case of submission to PRL or PLB
% Edit macro in main.tex to keep metadata correct
\paperauthors\footnote{Authors are listed at the end of this paper.}
\end{center}

\vspace{\fill}

% Abstract -----------------------------------------------
\begin{abstract}
  \noindent
\paperabstract

\end{abstract}

\vspace*{2.0cm}

\begin{center}
  Published in JHEP 02 (2019) 126
\end{center}

\vspace{\fill}

{\footnotesize
% Edit macro in main.tex to keep metadata correct
\centerline{\copyright~\papercopyright. \href{\paperlicenceurl}{\paperlicence}.}}
\vspace*{2mm}

\end{titlepage}

%%%%%%%%%%%%%%%%%%%%%%%%%%%%%%%%
%%%%%  EOD OF TITLE PAGE  %%%%%%
%%%%%%%%%%%%%%%%%%%%%%%%%%%%%%%%

%  empty page follows the title page ----
\newpage
\setcounter{page}{2}
\mbox{~}
%\newpage
%
%% Author List ----------------------------
%%  You need to get a new author list!
%\input{LHCb_authorlist.tex}
%
%The author list for journal publications is provided by the Membership Committee shortly after 'approval to go to paper' has been given.
%%It will be made available on the page
%%\verb!http://www.physik.uzh.ch/~strauman/forMemCo/LHCb-PAPER-XXXX-XXX/! .
%It will be sent to you by email shortly after a paper number has beens assigned.
%The author list should be included already at first circulation,
%to allow new members of the collaboration to verify whether they have been included correctly.
%Occasionally a misspelled name is corrected or associated institutions become full members.
%In that case, a new author list will be sent to you.
%In case line numbering doesn't work well after including the authorlist, try moving the \verb!\bigskip! after the last author to a separate line.
%
%
%The authorship for Conference Reports should be ``The LHCb
%  collaboration'', with a footnote giving the name(s) of the contact
%  author(s), but without the full list of collaboration names.

\cleardoublepage

%\twocolumn
% %%%%%%%%%%%%% ---------

\renewcommand{\thefootnote}{\arabic{footnote}}
\setcounter{footnote}{0}

%%%%%%%%%%%%%%%%%%%%%%%%%%%%%%%%
%%%%%  Table of Content   %%%%%%
%%%%%%%%%%%%%%%%%%%%%%%%%%%%%%%%
%%%% Uncomment next 2 lines if desired
%\tableofcontents
%\cleardoublepage

%%%%%%%%%%%%%%%%%%%%%%%%%
%%%%% Main text %%%%%%%%%
%%%%%%%%%%%%%%%%%%%%%%%%%

\pagestyle{plain} % restore page numbers for the main text
\setcounter{page}{1}
\pagenumbering{arabic}

%% Uncomment during review phase.
%% Comment before a final submission.
%\linenumbers

% You can include short sections directly in the main tex file.
% However, for larger papers it is desirable to split the text into
% several semiautonomous files, which can be revised independently.
% This is especially useful when developing a document in
% collaboration with several people, since then different parts can be
% edited independently.  This type of file organization is shown here.
%

% !TEX root = ./ms.tex
\section{Introduction}
\label{sec:Introduction}

The asymmetry between matter and antimatter in the universe is one of the most compelling questions that awaits explanation in particle physics. One of the three conditions required to explain this asymmetry is charge-parity (\CP) violation~\cite{Sakharov:1967dj}.
The Standard Model (SM) of particle physics allows this violation to arise. This is, however, not sufficient to describe the cosmological observations when taking into account the estimated lifetime of the universe~\cite{Universe_CPV, Bernreuther:2002uj}.

A promising area to search for  \CP violation beyond the SM is the decay of charm hadrons.
In Cabibbo-suppressed charm-hadron decays, the interference between the \decay{c}{q} ($\quark=d,s$) tree diagrams with the \decay{c}{u} loop diagrams is the only potential source of \CP asymmetry in the SM, which is predicted to be smaller than $0.1\%$~\cite{SM_prediction, KHODJAMIRIAN2017235}.
These decays are sensitive to potential new contributions in strong penguin and chromomagnetic dipole operators that could produce effects of up to 1\%~\cite{NP_prediction}.
Multibody decays of charm hadrons have peculiarities that make them particularly interesting for these studies.
They have a rich resonant structure, and the variation of the strong phases over the decay phase space may provide regions with enhanced sensitivity to \CP violation.

This paper presents a search for \CP violation in the individual amplitudes of the decay%
\footnote{Unless specified otherwise, charge conjugation is implied throughout this paper.}
\decay{\Dz}{\Kp\Km\pip\pim}, produced in semileptonic decays of \bquark hadrons into $\Dz\mun X$, where $X$ represents any combination of undetected particles.
The charge of the \mun (\mup) lepton is used to identify \Dz (\Dzb) mesons. The $\Dz\mu^-$ system is  reconstructed in a sample of $\proton\proton$ collisions collected
by the \lhcb experiment in 2011 and 2012 at centre-of-mass energies of 7 and $8\tev$, corresponding to integrated luminosities of 1.0\invfb and 2.0\invfb, respectively. A  \CP-averaged decay model is developed through a full amplitude analysis in the five-dimensional phase space of the four-body \Dz decay, where the \Dz and  \Dzb samples are merged.
This model is then used to search for \CP violation in a simultaneous fit to the distinct \Dz and \Dzb samples.

Such an amplitude analysis has already been performed by the CLEO experiment~\cite{KKpipi_CLEO} and has been recently updated with the CLEO legacy data~\cite{KKpipi_CLEO2017}. The analysis was based on 3000 signal decays and no \CP violation was observed. This paper presents an analysis using a data sample more than 50 times larger. Searches for \CP violation in this decay have also been performed by the BaBar~\cite{delAmoSanchez:2010xj}, LHCb~\cite{LHCb-PAPER-2014-046} and Belle~\cite{Kim:2018mtf} experiments, which are all compatible with \CP symmetry.

These studies use a model-independent technique to search for \CP violation based on triple products~\cite{Durieux:2015zwa}.
Similarly the energy test~\cite{Williams:2011cd} is used to study the \decay{\Dz}{\pip\pim\pip\pim} decays~\cite{LHCb-PAPER-2016-044}.
While these techniques are very powerful in evidencing the presence of \CP violation, they do not provide any direct information on what generated the observed effect.
The technique presented in this paper instead is able to associate any \CP violation effect to a specific source in the resonant structure of the decay.

Further motivation to study the amplitude structure of this decay mode is given by the determination of the Unitarity Triangle angle $\gamma$ in \decay{\Bm}{D\Km} decays, where $D$ stands for \Dz or \Dzb meson decaying to $\Kp\Km\pip\pim$~\cite{Rademacker:2006zx}, for which the amplitude structure of the \Dz decay is a limiting systematic uncertainty.

In Sec.~\ref{sec:Detector}, a description of the \lhcb detector and the simulation is presented. The event selection procedure is described in Sec.~\ref{sec:Selection}. The formalism used for the amplitude analysis is presented in Sec.~\ref{sec:AmplitudeAnalysis}. The systematic uncertainties are discussed in Sec.~\ref{sec:Systematics}. The resulting \CP-averaged model is presented in Sec.~\ref{sec:result}. The \CP-violation results are summarised in Sec.~\ref{sec:CPVResults} and conclusions are discussed in Sec.~\ref{sec:Conclusions}.

% !TEX root = ./ms.tex
\section{Detector and simulation}
\label{sec:Detector}

The \lhcb detector~\cite{Alves:2008zz,LHCb-DP-2014-002} is a single-arm forward
spectrometer covering the \mbox{pseudorapidity} range $2<\eta <5$,
designed for the study of particles containing \bquark or \cquark
quarks. The detector includes a high-precision tracking system
consisting of a silicon-strip vertex detector surrounding the $pp$
interaction region, a large-area silicon-strip detector located
upstream of a dipole magnet with a bending power of about
$4{\mathrm{\,Tm}}$, and three stations of silicon-strip detectors and straw
drift tubes placed downstream of the magnet.
The tracking system provides a measurement of momentum of charged particles with
a relative uncertainty that varies from 0.5\% at low momentum to 1.0\% at 200\gevc.
The polarity of the dipole magnet is reversed periodically throughout data-taking.
The configuration with the magnetic field vertically upwards (downwards) bends positively (negatively) charged particles in the horizontal plane towards the centre of the LHC.
The minimum distance of a track to a primary vertex (PV), the impact parameter (IP),
is measured with a resolution of $(15+29/\pt)\mum$,
where \pt is the component of the momentum transverse to the beam, in\,\gevc.
Different types of charged hadrons are distinguished using information
from two ring-imaging Cherenkov detectors. Photons, electrons and hadrons are identified by a calorimeter system consisting of
scintillating-pad and preshower detectors, an electromagnetic and a hadronic calorimeter. Muons are identified by a
system composed of alternating layers of iron and multiwire
proportional chambers. The online event selection is performed by a trigger consisting of a hardware stage, based on information from the calorimeter and muon
systems, followed by a software stage, which applies a full event
reconstruction.

In the simulation, $pp$ collisions are generated using
\pythia~\cite{Sjostrand:2007gs, *Sjostrand:2006za} with a specific \lhcb
configuration~\cite{LHCb-PROC-2010-056}.  Decays of hadronic particles
are described by \evtgen~\cite{Lange:2001uf}, in which final-state
radiation is generated using \photos~\cite{Golonka:2005pn}. The
interaction of the generated particles with the detector, and its response,
are implemented using the \geant
toolkit~\cite{Allison:2006ve, *Agostinelli:2002hh} as described in
Ref.~\cite{LHCb-PROC-2011-006}.

% !TEX root = ./ms.tex
\section{Event selection}
\label{sec:Selection}

%trigger
At the hardware trigger stage, events are required to have a muon with high \pt or a
  hadron with high transverse energy in the calorimeters.
  For the muon, the \pt threshold is 1.48\gevc in 2011 and 1.76\gevc in 2012. For hadrons,
  the transverse energy threshold is 3.5\gev and 3.62\gev, respectively.
  The software trigger requires a two-, three- or four-track
  secondary vertex with a significant displacement from any $pp$ interaction vertex. At least one charged particle
  must have a transverse momentum $\pt > 1.5\gevc$ and be
  inconsistent with originating from a PV.
  A multivariate algorithm~\cite{BBDT} is used for
  the identification of secondary vertices consistent with the decay
  of a \bquark hadron.

In the offline selection, two kaons and two pions are combined to form a \Dz candidate, which is itself combined with a muon to form a $b$-hadron candidate. These candidates are required to pass preselection criteria, listed below. The kaons and pions are required to have a momentum larger than 2\gevc and a transverse momentum larger than 0.3\gevc. The muon is required to have a momentum larger than 3\gevc and a transverse momentum larger than 1.2\gevc. Requirements on the track quality as well as correct particle identification are also made. The \Dz candidate is required to decay downstream of the $b$-hadron decay vertex and its mass is required to be in the range $[1805,1925]\mevcc$.
The mass of the $b$-hadron candidate is required to be in the range $[2500,6000]\mevcc$.
Furthermore, events are kept only if the positive trigger decision is due to the signal candidate.

%BDT
A boosted decision tree~(BDT)~\cite{Breiman,AdaBoost} is used to separate signal from background.
The description of the signal is taken from simulation while that of the background is taken from the \Dz-mass sidebands in data.
The two datasets are split into two independent subsamples for the training and the testing of the classifier.
Many kinematic and isolation variables are tested, the set of variables being gradually reduced until the performance of the BDT starts to decrease significantly.
The final set of variables used in this analysis contains the mass of the \bquark hadron corrected for the missing particles~\cite{Kodama:1991ij}, the flight distance of the \Dz candidate,
the vertex-fit \chisq of the decay vertex of the \Dz candidate,
particle identification information of the four daughters of the \Dz candidate, the
probability of the tracks being reconstructed from random hits~\cite{LHCb-DP-2014-002}, and isolation variables based on the underlying event. One isolation variable is obtained by considering all tracks inside a cone around the \Dz direction and computing the \pt asymmetry with respect to the \Dz meson. Two additional isolation variables are obtained by adding tracks to the \Dz vertex; the first represents the invariant mass of the new combination when adding the track with the minimum vertex \chisq difference and the second represents the minimum \chisq difference when adding two tracks.
The selection on the BDT outcome is chosen to maximise the figure of merit %$\frac{N_{\rm s}}{\sqrt{N_{\rm s}+N_{\rm b}}}$
$N_{\rm s}/\sqrt{N_{\rm s}+N_{\rm b}}$, where $N_{\rm s}$ and $N_{\rm b}$ are the number of signal and background events in the signal region (SR), where the SR is defined as $\pm2$ standard deviations around the central value of the known \Dz mass~\cite{PDG2018}.

\begin{figure}[t]
\centering
\includegraphics[width=0.75\linewidth]{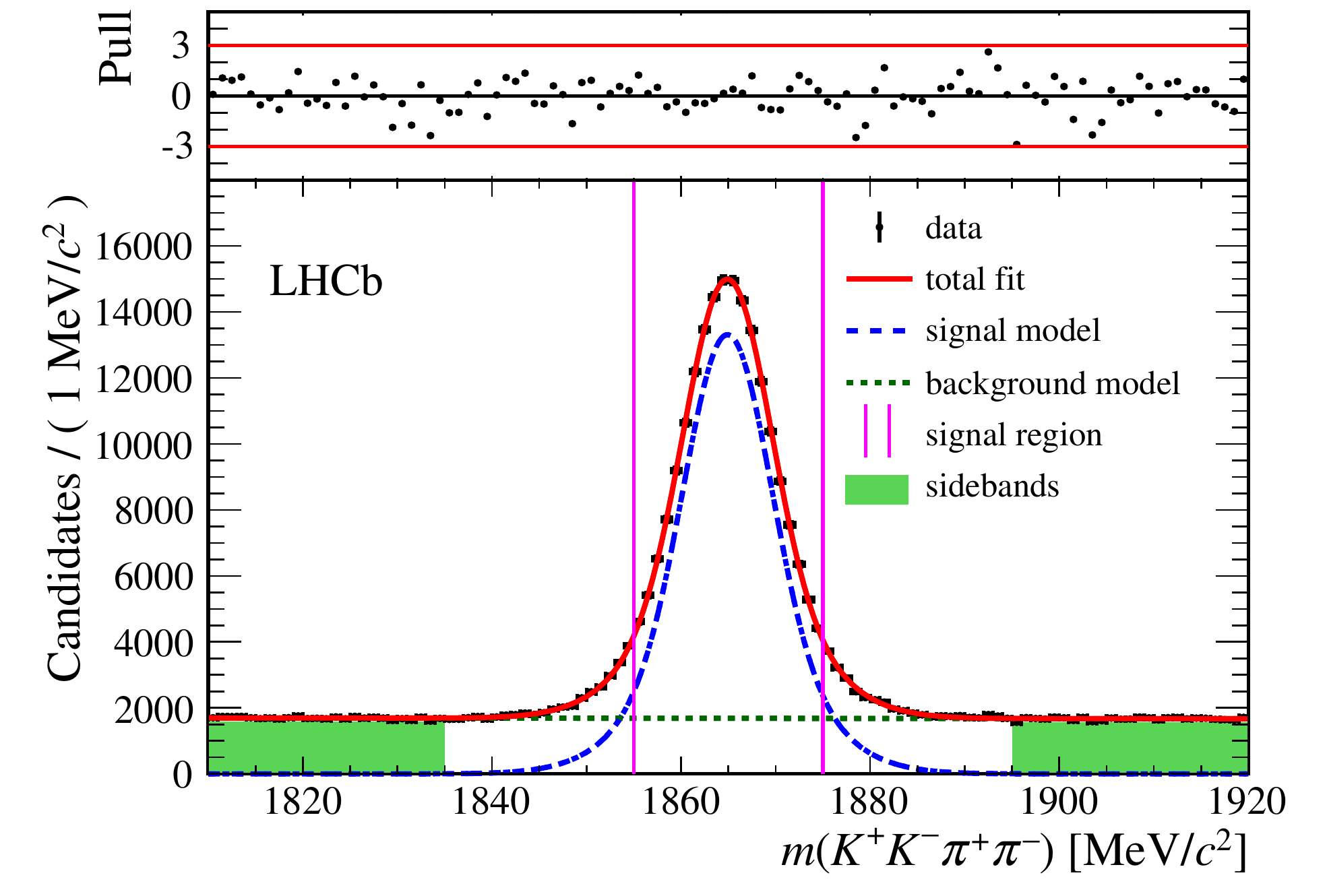}
\vspace{-2mm}
\caption{\label{fig:D0opt}Mass distribution of the \DztoKKpipi candidates after the final selection, with fit result superimposed. The top plot shows the normalised residuals.}
\end{figure}

%Delta m cut
A background component arises when the \Dz meson originates from a \Dstarp resonance,
which is itself coming from a  $b$-hadron decay, and the slow pion from the \Dstarp resonance is used as one of the pions in the \Dz decay.
This misreconstruction generates a structure in the upper end of the $\Kp\Km\pim$ invariant mass spectrum, which is removed with the requirement $m(\Kp\Km\pip\pim)-m(\Kp\Km\pim)>0.18\gevcc$.
%KS veto
A small component of \decay{\Dz}{\KS\Kp\Km} decays is present in the SR. These decays have a different topology than \DztoKKpipi decays and are therefore vetoed by removing all candidates that have a $\pip\pim$ invariant mass in the region $[480.2, 507.2]\mevcc$.
%double mis-ID
Background contributions in which both a kaon is misidentified as a pion and a pion as a kaon are found to be negligible.
%multiple candidates
About $1.7\%$ of the events contain more than one \Dz candidate, in which case one candidate per event is randomly selected.

The resulting \Dz mass distribution, shown in Fig.~\ref{fig:D0opt}, is fitted with a double Gaussian  signal function and an exponential background function. Only the
$N_{\rm data} = 196\,648$  events that are in the SR are kept. According to the fit,
the selected sample contains $162\,909 \pm 516$ signal decays in the SR with a purity of $f_s=82.8\pm0.3\%$.
This selection provides a significant improvement with respect to the cut-based selection
of a previous LHCb analysis of the same decay mode with the same data sample~\cite{LHCb-PAPER-2014-046}: 5\% more signal and 30\% less background are retained.

%D0 mass constraint
In order to improve the resolution on the measured momenta of the \Dz decay products, the tracks and vertices are refitted under the constraint that the invariant mass of the four \Dz daughter particles be equal to the known \Dz mass~\cite{PDG2018}.

%CP transform Dzb
The data sample contains both \decay{\Dz}{\Kp\Km\pip\pim} and \decay{\Dzb}{\Km\Kp\pim\pip} decays, tagged by the sign of the accompanying muon. For the rest of the analysis, the \CP transformation is applied on the \Dzb candidates, \ie on the momentum vectors of the four daughter particles in the \Dzb rest frame. This allows, in absence of \CP violation, the \Dz and \CP-transformed \Dzb candidates to have identical distributions.

% !TEX root = ./ms.tex
\section{Amplitude analysis}
\label{sec:AmplitudeAnalysis}

The amplitude analysis consists of describing the \Dz decay chain as a coherent sum of amplitudes,
each corresponding to a specific decay path (called ``component'' from here on) from the mother particle to the \Kp\Km\pip\pim final state.
These complex amplitudes may interfere.
The main goal of the amplitude analysis is to identify the components that contribute to the decay. Each amplitude $A_k$ is multiplied by a complex coefficient $c_k$, whose modulus $|c_k|$ and phase $\arg(c_k)$ are determined by means of an unbinned maximum likelihood fit (except for one of the complex coefficients, which is fixed to 1).

\begin{figure}[t!]
\centering
\includegraphics[width=0.6\linewidth, clip=true, trim=90 230 140 180]{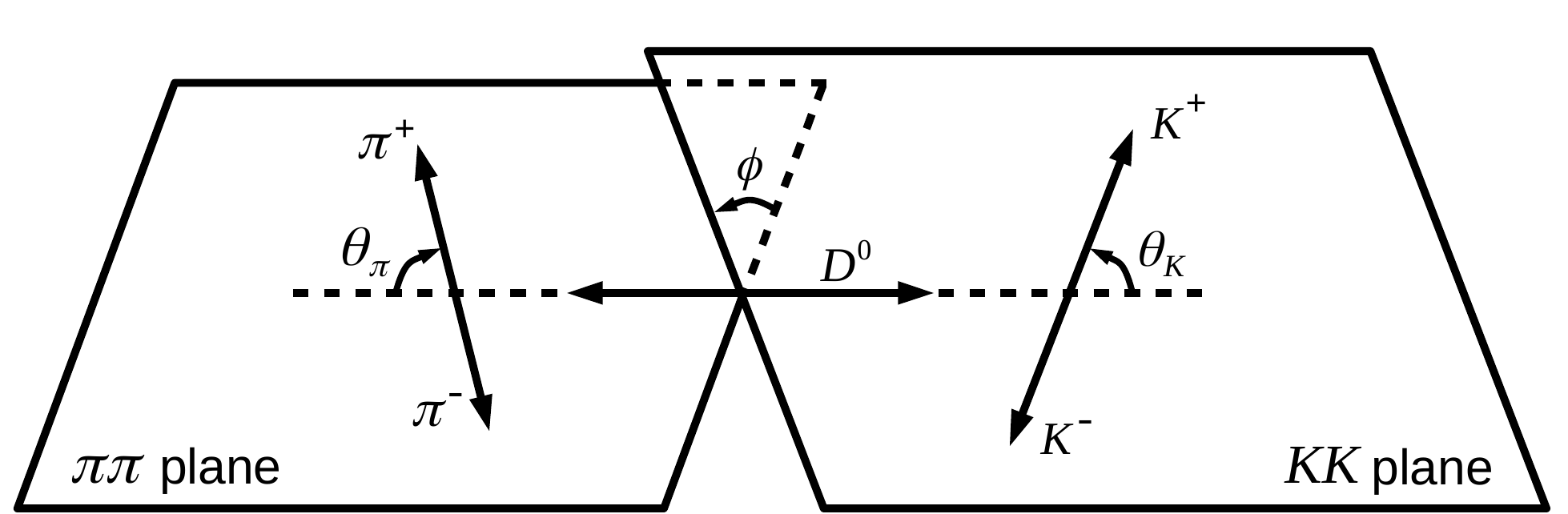}
\caption{\label{fig:CMvars}Definition of the helicity angles $\theta_{K}$ and $\theta_{\pi}$, and the decay-plane angle $\phi$.}
\end{figure}

\subsection{Likelihood fit}
\label{sec:likelihood}

The likelihood function is defined as
\begin{equation}
{\cal L}(\bm{c}) = \prod_j^{N_{\rm data}} \left[ \rule[-.3\baselineskip]{0pt}{\baselineskip} f_s \, a(\bm{x}_j;\bm{c}) + (1-f_s)\, b(\bm{x}_j) \right] \,,
\end{equation}
\noindent where $a(\bm{x};\bm{c})$ and $b(\bm{x})$ are the probability density functions (PDF) of the signal and background components normalized as $\int a(\bm{x};\bm{c}) d^5\bm{x} = \int b(\bm{x}) d^5\bm{x} =1$. Here, $\bm{c}$ represents the fit parameters and $\bm{x}$ represents the five dimensions of the \DztoKKpipi four-body phase space. These five dimensions can be visualised with the five Cabibbo--Maksymowicz (CM) variables~\cite{CM_vars} (see Fig.~\ref{fig:CMvars}):
the invariant mass of the two-kaon system $m(\Kp\Km)$;
the invariant mass of the two-pion system $m(\pip\pim)$;
the cosine of the helicity angle for the two-kaon system $\cos(\theta_{K})$, defined as the angle between the direction of the \Dz momentum and that of one of the kaons in the rest frame of the two kaons;
the cosine of the helicity angle for the two-pion system $\cos(\theta_{\pi})$, defined similarly; and the angle $\phi$ between the plane defined by the directions of the two kaons and the plane defined by the directions of the two pions, in the \Dz rest frame.

The signal PDF is written as
\begin{equation}
a(\bm{x};\bm{c}) = \frac{\epsilon_s(\bm{x}) S(\bm{x};\bm{c}) \mathcal{R}_4(\bm{x})}{I(\bm{c})} ~~ \mbox{with} ~~ I(\bm{c}) = \int \epsilon_s(\bm{x}) S(\bm{x};\bm{c}) \mathcal{R}_4(\bm{x}) d^5\bm{x} \,,
\label{eq:pdf_sig}
\end{equation}
where $\epsilon_s(\bm{x})$ is the signal efficiency, $S(\bm{x};\bm{c})$ is the signal model described in Sec.~\ref{sec:signal_description} and $\mathcal{R}_4(\bm{x})$ is the function representing the four-body phase space.
Instead of explicit parametrisations of the signal efficiency and the four-body phase-space function, the fit
uses a simulated signal sample that encodes the functions $\epsilon_s(\bm{x})$
and $\mathcal{R}_4(\bm{x})$. After the reconstruction and selection are applied, this sample is distributed according to the PDF
$a_{\rm gen}(\bm{x}) = \epsilon_s(\bm{x}) S_{\rm gen}(\bm{x}) \mathcal{R}_4(\bm{x})/I_{\rm gen}$, where $S_{\rm gen}(\bm{x})$ is the signal model used in the generation of the sample and $I_{\rm gen} = \int \epsilon_s(\bm{x}) S_{\rm gen}(\bm{x}) \mathcal{R}_4(\bm{x}) d^5\bm{x}$. This simulated sample contains approximately 10 million signal events, half of which are generated with a pure phase-space decay model (corresponding to a constant function $S_{\rm gen}(\bm{x})$) in order to cover the whole phase space and the other half according to the decay model published by the CLEO collaboration~\cite{KKpipi_CLEO} in order to sufficiently populate the phase-space regions corresponding to the resonances. The normalisation integral of Eq.~\ref{eq:pdf_sig} is then estimated as
\begin{equation}
I(\bm{c}) = \frac{1}{N_{\rm sim}} \sum_{i=1}^{N_{\rm sim}} \frac{S(\bm{x}_i;\bm{c})}{S_{\rm gen}(\bm{x}_i)} \, I_{\rm gen} \,,
\label{eq:MC_integral}
\end{equation}
where $N_{\rm sim}$ is the number of events in the simulated sample.

The background component is described using the same simulated sample. For each simulated event $i$ at position $\bm{x}_i$ in phase space, a weight $w(\bm{x}_i)$ is assigned so that the weighted simulated distribution matches the distribution of the background in five dimensions, as explained in Sec.~\ref{sec:bkg_description}. These weights are obtained by evaluating $w(\bm{x}) = b(\bm{x})/a_{\rm gen}(\bm{x})$ and satisfy $\frac{1}{N_{\rm sim}} \sum_{i=1}^{N_{\rm sim}} w(\bm{x}_i)= 1$. In order to factorise out the signal efficiency and the phase-space function, the background PDF is rewritten as $b(\bm{x}) = \epsilon_s(\bm{x}) B(\bm{x}) \mathcal{R}_4(\bm{x}) / I_{\rm gen}$, where $B(\bm{x})=w(\bm{x}) S_{\rm gen}(\bm{x})$.

Using the above definitions, the likelihood is rewritten as
\begin{equation}
{\cal L}(\bm{c}) = C \prod_{j=1}^{N_{\rm data}} \left(
f_s \,  \frac{S(\bm{x}_j;\bm{c})}{\frac{1}{N_{\rm sim}} \sum_{i=1}^{N_{\rm sim}} \frac{S(\bm{x}_i;\bm{c})}{S_{\rm gen}(\bm{x}_i)} } + (1-f_s) \, B(\bm{x}_j)
\right) \,,
\end{equation}
where $C=\prod_{j=1}^{N_{\rm data}} \left(\epsilon_s(\bm{x}_j) \mathcal{R}_4(\bm{x}_j) / I_{\rm gen}\right)$ is a constant independent of the fit parameters $\bm{c}$. Therefore, this constant does not need to be computed in order to maximize ${\cal L}(\bm{c})$. The value $B(\bm{x}_j)$ of data event $j$ is obtained as the average of the quantities $B(\bm{x}_i)$ over all simulated events with $\bm{x}_i$ falling in the same and neighbouring phase-space bins of $\bm{x}_j$.
For this purpose an adaptive binning is used, which splits five-dimensional rectangular bins along the CM variables until each bin contains between 5 and 9 simulated events, resulting in smaller bin volumes in higher density regions.

\subsection{Signal description}
\label{sec:signal_description}

The formalism\footnote{Additional details on the description of the formalism are available in the supplementary material.} chosen for this amplitude analysis is the so-called isobar model~\cite{isobar,isobar_2}, which assumes that each amplitude can be built as a series of two-body decays. The two allowed patterns for \decay{\Dz}{abcd} decays, both involving two intermediate resonances $r_1$ and $r_2$, are \decay{\Dz}{r_1r_2} followed by \decay{r_1}{ab} and \decay{r_2}{cd}, and \decay{\Dz}{r_1 a} followed by \decay{r_1}{r_2 b} and \decay{r_2}{cd}.
In both cases the $k^\text{th}$ complex amplitude $A_k(\bm{x})$ is computed as the product of the lineshapes for the resonances $r_1$ and $r_2$, the normalised Blatt--Weisskopf barrier factors~\cite{BlattWeisskopf}, and a spin factor defined using the covariant formalism~\cite{CovForm}.

\begin{table}[t!]
\centering
\caption{\label{tab:resonances}Resonances considered in the analysis, classified according to their spin-parity $J^P$ and  decay products.}
% !TEX root = ./ms.tex

\begin{tabular}{lccccc}
\toprule
 & $KK$ & $\pi\pi$  & $K\pi$  & $KK\pi$     & $K\pi\pi$                                                                       \\ \toprule
$J^P=0^+$ & \begin{tabular}[c]{@{}l@{}}$a_0(980)$\\ $f_0(980)$\\ $f_0(1370)$\end{tabular} & \begin{tabular}[c]{@{}l@{}}$f_0(980)$\\ $f_0(1370)$\end{tabular}                   & $K_0^*(1430)$                                                    &             &      \\ \midrule
%$J^P=0^-$ &  &      &  &             & $K(1460)$      \\ \midrule
$J^P=1^+$ &  &  &  & $a_1(1260)$ & \begin{tabular}[c]{@{}l@{}}$K_1(1270)$\\ $K_1(1400)$\end{tabular} \\ \midrule
$J^P=1^-$ & $\phi(1020)$                                                                  & \begin{tabular}[c]{@{}l@{}}$\rho(770)$\\ $\omega(782)$\\ $\rho(1450)$\end{tabular} & \begin{tabular}[c]{@{}l@{}}$K^*(892)$\\ $K^*(1680)$\end{tabular} &  & \begin{tabular}[c]{@{}l@{}}$K^*(1410)$\\ $K^*(1680)$\end{tabular} \\ \midrule
$J^P=2^+$ & \begin{tabular}[c]{@{}l@{}}$f_2(1270)$\\ $a_2(1320)$\end{tabular}             & $f_2(1270)$                                                                        & $K^*_2(1430)$                                                    &             & $K^*_2(1430)$ \\ \bottomrule                                                               
\end{tabular}

\end{table}

The resonances considered in this analysis are listed in Table~\ref{tab:resonances}. The default lineshape, used for most resonances, is the relativistic Breit--Wigner (RBW) function~\cite{BW}. For the $a_1(1260)^+$ and the $K_1(1270)^+$ resonances, a correction is applied to their mass-dependent width according to the formalism described in Ref.~\cite{tau23pinu}, in order to take into account the effect of intermediate resonances. In addition, an exponential form factor derived from Refs.~\cite{tau23pinu,rhoexp} is used for these two three-body resonances instead of the Blatt--Weisskopf factors.
Exceptions are the Flatt\'e parametrisation~\cite{Flatte} used for the $a_0(980)^0$ resonance near the $KK$ threshold and
the Gounaris--Sakurai parametrisation~\cite{GoSa} used for the $\rho(770)^0$ resonance.
The aforementioned lineshapes describe accurately well-separated narrow resonances. In the case of broad overlapping resonances, the above descriptions may fail to account properly for interferences and the K-matrix formalism~\cite{Kmatrix} is used instead.
The $\pi\pi$  and the $KK$ $S$-wave contributions (referred to as $[\pip\pim]_{L=0}$ and $[\Kp\Km]_{L=0}$ in the following) are both described with the K-matrix formalism. They couple to five different channels ($\pi\pi$, $KK$, $\pi\pi\pi\pi$, $\eta\eta$ and $\eta\eta'$) with five different poles ($f_0(980)$, $f_0(1300)$, $f_0(1500)$, $f_0(1750)$ and $f_0(1200-1600)$) and a non-resonant contribution, according to the parametrisation of Ref.~\cite{Anisovich2003}, with parameters taken from Ref.~\cite{Kmatrix_Babar}.
The $K\pi$ $S$-wave contribution (referred to as $[\Kp\pim]_{L=0}$ in the following) is also described with the K-matrix formalism. It couples to the $K\pi$ and $K\eta'$ channels in the isospin-state  $I = \frac{3}{2}$, or only to the $K\pi$ channel if $I = \frac{1}{2}$, with a single pole ($K_0^*(1430)$) and a non-resonant contribution. The parametrisation is taken from Ref.~\cite{KpiKmatrix}.

The mass and the width of the $K_1(1270)^+$ resonance are left floating in the final fit. The parametrisation of the $a_1(1260)^+$ resonance is taken from Ref.~\cite{LHCb-PAPER-2017-040} with mass and width set to 1195\mevcc and 422\mevcc. For all the other resonances, their masses and widths are fixed to the values of Ref.~\cite{PDG2018}. The radii of the normalised Blatt--Weisskopf barrier factors used in the parametrisation of the \Dz, \Kstz and \Kstzone mesons are set to $1.21\gev^{-1}$, $1.13\gev^{-1}$ and $1.93\gev^{-1}$, respectively. Each of these three values is obtained by maximising the likelihood of amplitude fits while fixing the values of the other two. The radius of the $a_1(1260)^+$ resonance is fixed to $1.7\gev^{-1}$ according to Ref.~\cite{LHCb-PAPER-2017-040} and all the other radii are fixed to $1.5\gev^{-1}$.

The total signal function is then described by a coherent sum of $N$ amplitudes,
\begin{equation}
S(\bm{x};\bm{c}) = \left|\sum_{k=1}^N c_k A_k(\bm{x})\right|^2 \,, \label{Pdf_sig}
\end{equation}
where $c_1 = 1$ and the other complex coefficients $c_k$ are defined relative to $c_1$.
The moduli and phases of these other complex coefficients are left floating in the fit. After the fit, in order to quantify the relative importance of each component, the fit fraction
\begin{equation}
\mathcal{F}_k = \frac{\int|c_k A_k(\bm{x})|^2 {\mathcal R}_4(\bm{x}) d^5\bm{x}}{\int |\sum_{\ell=1}^N c_\ell A_\ell(\bm{x})|^2 {\mathcal R}_4(\bm{x}) d^5\bm{x}}
\label{eqt:fitfrac}
\end{equation}
is computed. Interferences may lead to $\sum_{k=1}^N \mathcal{F}_k \ne 1$.

\subsection{Background description}
\label{sec:bkg_description}

The sidebands of the \Dz-mass distribution are used to describe the background, assuming that the sum of the lower
($m(\Kp\Km\pip\pim) \in [1.81, 1.835]\gevcc$)
and upper
($m(\Kp\Km\pip\pim) \in [1.895, 1.92]\gevcc$)
sidebands gives a good description of the background in the SR. Although the background has a nontrivial distribution in the five-dimensional phase space, as it also contains resonances, this assumption is justified by the observation that the five-dimensional distributions of the candidates in the two sidebands are similar and that the contributions of the various resonances vary smoothly as a function of the $\Kp\Km\pip\pim$~invariant mass.

Since the point in phase space of each candidate is recomputed under the \Dz-mass constraint, the phase-space boundaries are the same for
candidates in the SR and in each of the sideband regions. However, as a side effect,
the peaks of the resonances are slightly shifted to larger (smaller) values in the lower (upper) sideband.
This effect is corrected for by weighting the sideband events to shift the resonance peaks back to their correct position while keeping their widths unchanged.
The correction is applied for the $\phi(1020)$,  $\Kstarb(892)^0$ and $K^*(892)^0$ peaks visible in the \Kp\Km, $\Km\pip$ and \Kp\pim invariant mass distributions, respectively. Since the $K^*(892)^0$ peak is much less pronounced than the $\Kstarb(892)^0$ peak, the positions and shapes of these two peaks are constrained to be the same. No correction is applied for the small $\rho(770)^0$ contribution in the \pip\pim invariant mass of the sideband sample.

After these corrections, the sample of fully simulated signal events is reweighted to match the distribution of the sideband data. To obtain the weights $w(\bm{x}_i)$ needed for the amplitude fit (see Sec.~\ref{sec:likelihood}), a multidimensional reweighting is performed simultaneously on 31 different projections of the five-dimensional phase space (10 invariant masses, 18 helicity angles, and 3 decay plane angles) using the package {\tt hep\_ml}~\cite{hepml}.

\subsection{Model building}
\label{sec:model_building}

In order to build a \CP-averaged decay model,  a single model is fitted to the full data sample, containing both \Dz and \CP-transformed \Dzb candidates. This allows the construction of the signal model in a way that is blind to possible \CP-violating effects.

The signal model is built iteratively.
As a starting point, the model is made of the sum of the \decay{\Dz}{\phi(1020)(\rho-\omega)^0} and \decay{\Dz}{\Kstz\Kstzb} amplitudes, where in both cases three different values of the orbital angular momentum ($L=0,1,2$) between the vector resonances from the \Dz decay are allowed.
The $\rho(770)^0$ and $\omega(782)$ mesons, being close in mass, interfere heavily. They are therefore described as a quantum superposition of the two individual states, defined as the ``$(\rho-\omega)^0$'' state, with a free relative complex coefficient~$\tilde{c}$. A long list of other possible amplitudes is defined from all possible combinations of the resonances listed in Table~\ref{tab:resonances}. At each iteration, the amplitude of this list that produces the largest decrease in the minimised value of $-2\ln(\mathcal{L}(\bm{c}))$ is added to the model.

As no \CP violation is expected to arise in strong decays, the two charge states of the same three-body resonance are constrained to have the same decay substructure, implying that the two charged-conjugate states are always added together in the model building procedure. The $K^*(1410)$ and  $K^*(1680)$ resonances give similar contributions and the fit cannot distinguish them. One of the two needs therefore to be removed. The $K^*(1680)$ components are chosen as the fit shows a slightly better \chisq, but the $K^*(1410)$ components are considered in one of the alternative models used to estimate systematic uncertainties.

As iterations proceed, the sum of the fit fractions remains quite stable but diverges at some point. The procedure is then stopped and the model from the previous iteration is taken as the nominal one.

\subsection{\boldmath{\CP}-violating observables}
\label{sec:cpv_description}

For measuring \CP violation, the data is split in two samples according to the charge of the muon, one with \Dz decays, and another one with \CP-transformed \Dzb decays.
This allows the use of the same model in a simultaneous fit to the two samples, where different parameter values are allowed for the two samples.
Any significant difference between the fitted parameters on the two samples will signal \CP violation. For each amplitude in the model, the fit is parametrised with the average modulus $\overline{|c_k|}$, modulus asymmetry $A_{|c_k|}$, average phase $\overline{\arg(c_k)}$ and phase difference $\Delta \arg(c_k)$, defined as
\begin{align}
\overline{|c_k|} &= \frac{|c_k|_{\Dz}+|c_k|_{\Dzb}}{2}\,, & & & A_{|c_k|} &= \frac{|c_k|_{\Dz}-|c_k|_{\Dzb}}{|c_k|_{\Dz}+|c_k|_{\Dzb}}\,, \\
\overline{\arg(c_k)} &= \frac{\arg(c_k)_{\Dz}+\arg(c_k)_{\Dzb}}{2}\,, & & & \Delta \arg(c_k) &= \frac{\arg(c_k)_{\Dz}-\arg(c_k)_{\Dzb}}{2}\,,
\end{align}
\noindent where $|c_k|$ and $\arg(c_k)$ are the polar coordinates (modulus and phase) of the complex fit coefficient multiplying the $k^{\rm th}$ amplitude. The simultaneous fit minimises the sum of the two negative log-likelihoods for the \Dz and \Dzb samples. Since no \CP violation is expected in the strong decays of the three-body resonances, their modulus and phases are therefore simultaneously fitted to common values for the two samples.

An additional information on \CP violation in each amplitude can be obtained from the fit fractions. The  asymmetry
\begin{equation}
A_{\mathcal{F}_k} = \frac{\mathcal{F}_k^\Dz-\mathcal{F}_k^\Dzb}{\mathcal{F}_k^\Dz+\mathcal{F}_k^\Dzb}\,,
\end{equation}
is considered,
where $\mathcal{F}_k^\Dz$ and $\mathcal{F}_k^\Dzb$ are the fit fractions of the $k^{\rm th}$ amplitude for the \Dz and the \Dzb samples respectively.

% !TEX root = ./ms.tex
\section{Systematic uncertainties}
\label{sec:Systematics}
Various systematic uncertainties are considered, most of them  related to the fitting procedure and the model determination, and others to the interaction of particles with the detector.
Among the effects that could directly influence the amplitude model are the determination of the selection efficiency, the background description, the signal fraction, the description of the resonance shapes, alternative components in the model, and any bias intrinsic to the fit procedure. Moreover, the flavour misidentification and the track reconstruction asymmetry may have an effect on the \CP-violation measurements.

The size of these uncertainties is determined either by running pseudoexperiments (fit bias, background description, flavour misidentification, detection asymmetry, alternative models), by performing alternative fits after applying some modifications (background description, selection efficiency, resonance shapes), or by fitting multiple times the data to simulate statistical fluctuations using resampling (background description) or to take into account the uncertainties on the fixed parameters (masses and widths, radii).

The selection efficiency is a source of systematic uncertainty since the detector simulation may not be perfect.
To reduce this effect, a reweighting of the simulated sample is applied to match the kinematical distributions of the data before the BDT selection.
Its effect on the result is tested by performing again the reweighting after the BDT selection, and repeating the fit.
The difference with respect to the nominal fit is taken as a systematic uncertainty. The dependence of the selection efficiencies with respect to the \Dz transverse momentum has also been studied, without a significant effect being found.

There are a couple of sources of systematic uncertainty related to the background description: the finite size of the \Dz sidebands that are used as a proxy, and the technique to describe the background itself.
In the first case, the fit is repeated many times with alternative descriptions of the background obtained by resampling the sidebands using a bootstrapping technique~\cite{efron:1979}. In the second case, an alternative technique is used for the weighting of the simulated sample to match the sidebands. This technique determines the weights by considering seven subsets of the 31 variables defined in Sec.~\ref{sec:bkg_description}. The weighting is performed in turn using each subset, assuming that the variables in that subset are independent. The correlations are taken into account by iterating until the matching is satisfactory. The systematic uncertainties are taken from the differences with respect to the nominal fit.

Alternative parametrisations of the resonance shapes are tested. The Gounaris--Sakurai lineshape, used for the $\rho(770)^0$ resonance, is compared to the RBW lineshape.
Alternative solutions for the $KK$ and $\pi\pi$ K-matrices as reported in Refs.~\cite{Anisovich2003,Kmatrix_Babar} are compared with the nominal parametrisation.
The fit is repeated many times with the values of the masses and widths are randomly drawn from a Gaussian distribution with mean and width taken as their central value and uncertainty. The values of the Blatt--Weisskopf radii for the \Dz, \Kstz and \Kstzone mesons are varied around their nominal value according to a Gaussian distribution with a width defined by the result of the corresponding amplitude fit. The radii of the other resonances are uniformly varied in a range of $\pm0.2\gev^{-1}$ around their central values, similarly to Refs.~\cite{LHCb-PAPER-2015-026,LHCb-PAPER-2017-040}.

The effect of the fit bias is studied by generating samples and fitting them using the same model. Four sets of pseudoexperiments are performed. The first only uses the nominal signal model to test the stability of the fitter. The second uses the signal and background models to test the effect of the background description. The third uses the signal model with the introduction of a wrong flavour assignment of the \Dz candidate. In data, about 0.5\% of the \Dz mesons are associated with the wrong muon~\cite{LHCb-PAPER-2014-046}.
The effect on the measurement is tested by generating pseudoexperiments in which 0.5\% of the sample is generated as \Dzb instead of \Dz. Finally, the fourth uses the signal model where a detection asymmetry between the two kaons is introduced. While some intermediate states are \CP eigenstates (\eg \decay{\Dz}{\phi(1020)\rho(770)^0}), others are not (\eg \decay{\Dz}{K_1(1270)^+K^-}) and could be affected by a kaon detection asymmetry.
The effect is studied with pseudoexperiments in which a momentum-dependent reconstruction asymmetry of the order of 1\% is introduced between the \Kp and \Km mesons.

The model building method chosen in this analysis produces one model, which is the best solution given the amplitudes considered and the criteria used for the selection of the amplitudes.
By slightly varying this method, alternative models can be produced with similar fit qualities. Three alternative models are considered to assign a systematic uncertainty.
The $K^*(1410)^0$ resonance is used as an alternative to the $K^*(1680)^0$ resonance in the first alternative model.
In the second alternative model, five additional amplitudes are added to the nominal model to test the effect of the stopping criteria.
Finally, the \babar collaboration has recently observed the decay of the  $\rho(1450)^0$ meson into two kaons~\cite{Lees:2017ndy}. This contribution is tested by adding to the nominal model the component \decay{\Dz}{\rho(1450)^0\rho(770)^0} in $D$-wave, with \decay{\rho(1450)^0}{\Kp\Km} and \decay{\rho(770)^0}{\pip\pim}.
The results of these three alternative models are described in Appendix~\ref{app:alt_mod}. Pseudoexperiments are performed by generating a signal sample according to the alternative model and fitting it according to the nominal model. For each amplitude, the largest difference  with respect to the nominal fit among the three alternative models is assigned as a systematic uncertainty.

The fraction of signal candidates $f_s$ is fixed in the fit.
To estimate the impact of its statistical uncertainty to the final result, the fit is repeated many times with values of $f_s$ sampled from a Gaussian with mean 0.828 and width of 0.003. No significant effect is found.

The breakdown of all the systematic uncertainties on the model parameters and fit fractions is shown in Appendix~\ref{app:syst_tables} for the \CP-averaged and the \CP-violating fits.
The largest contributions to the systematic uncertainty come from the resonance parameters, the alternative models and the alternative parametrisations used for the description of the S-wave shapes.

In addition, some cross-checks are performed.
In particular it is checked that the choices made during the selection (the cut on $m(\Kp\Km\pip\pim)-m(\Kp\Km\pim)$, the \KS~veto and the treatment of multiple candidates) do not bias the results and that resolution effects are negligible.

% !TEX root = ./ms.tex

\section{\CP-averaged results}
\label{sec:result}

\begin{figure}[t!]
\includegraphics[width=0.49\linewidth]{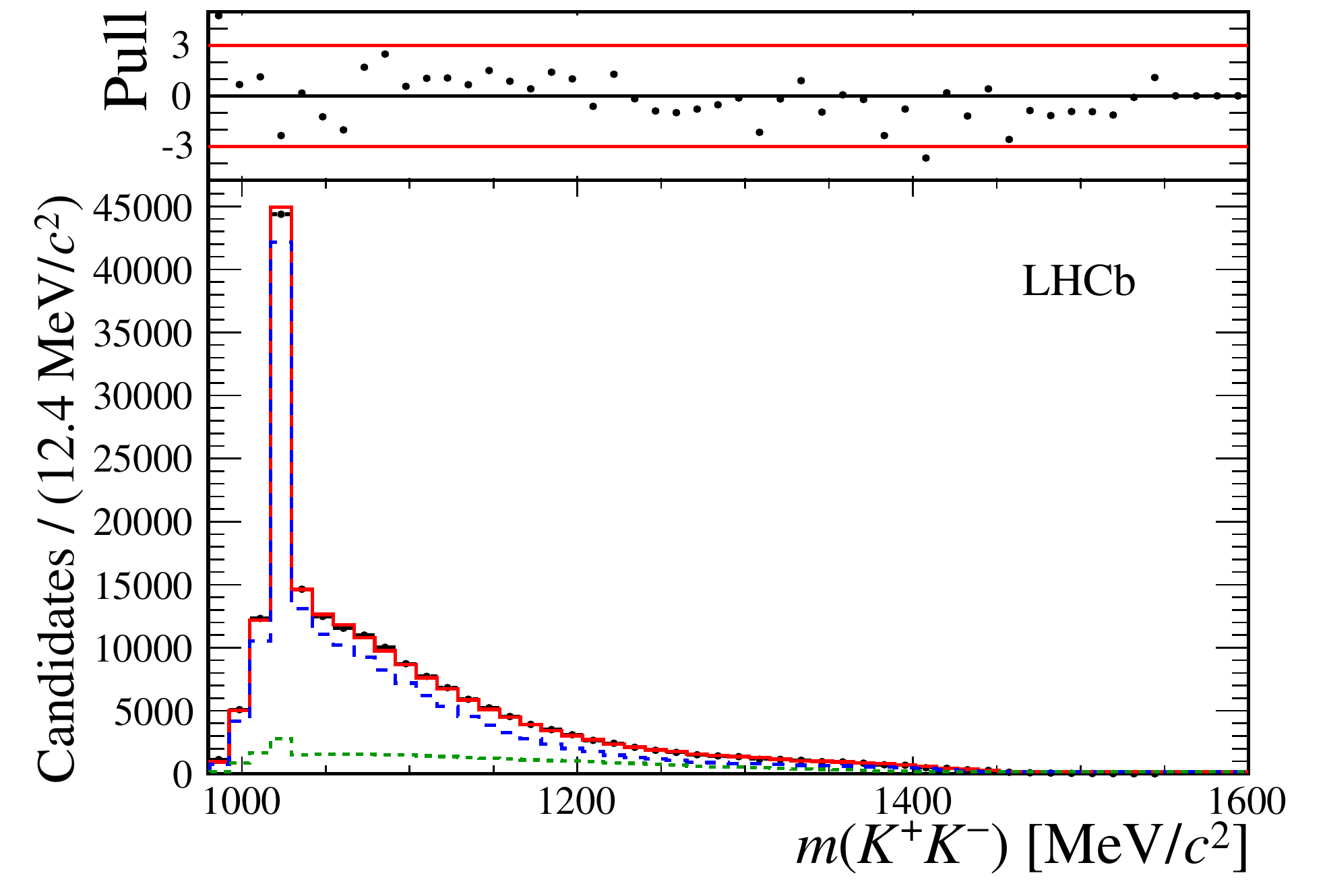}
\includegraphics[width=0.49\linewidth]{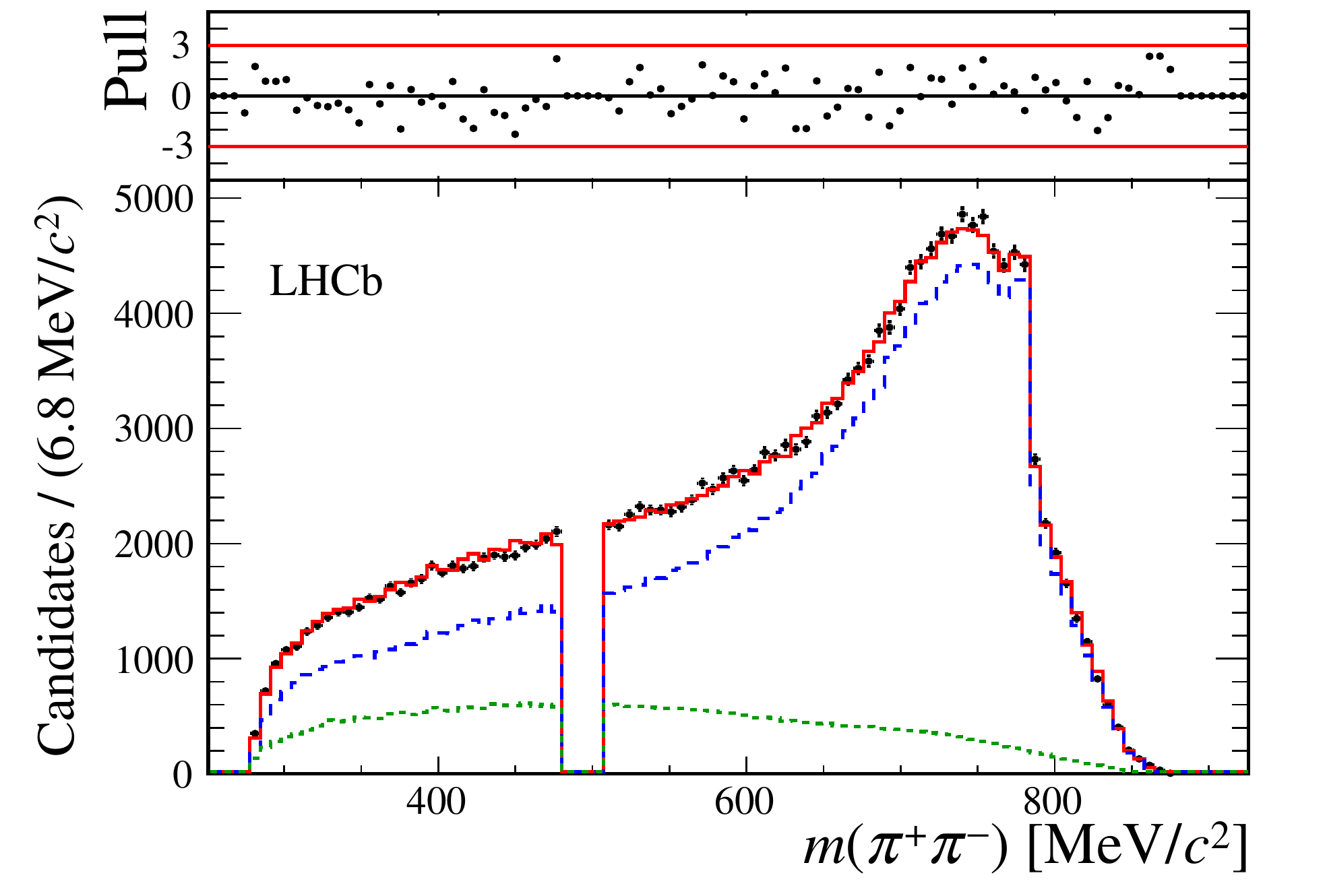}

\includegraphics[width=0.49\linewidth]{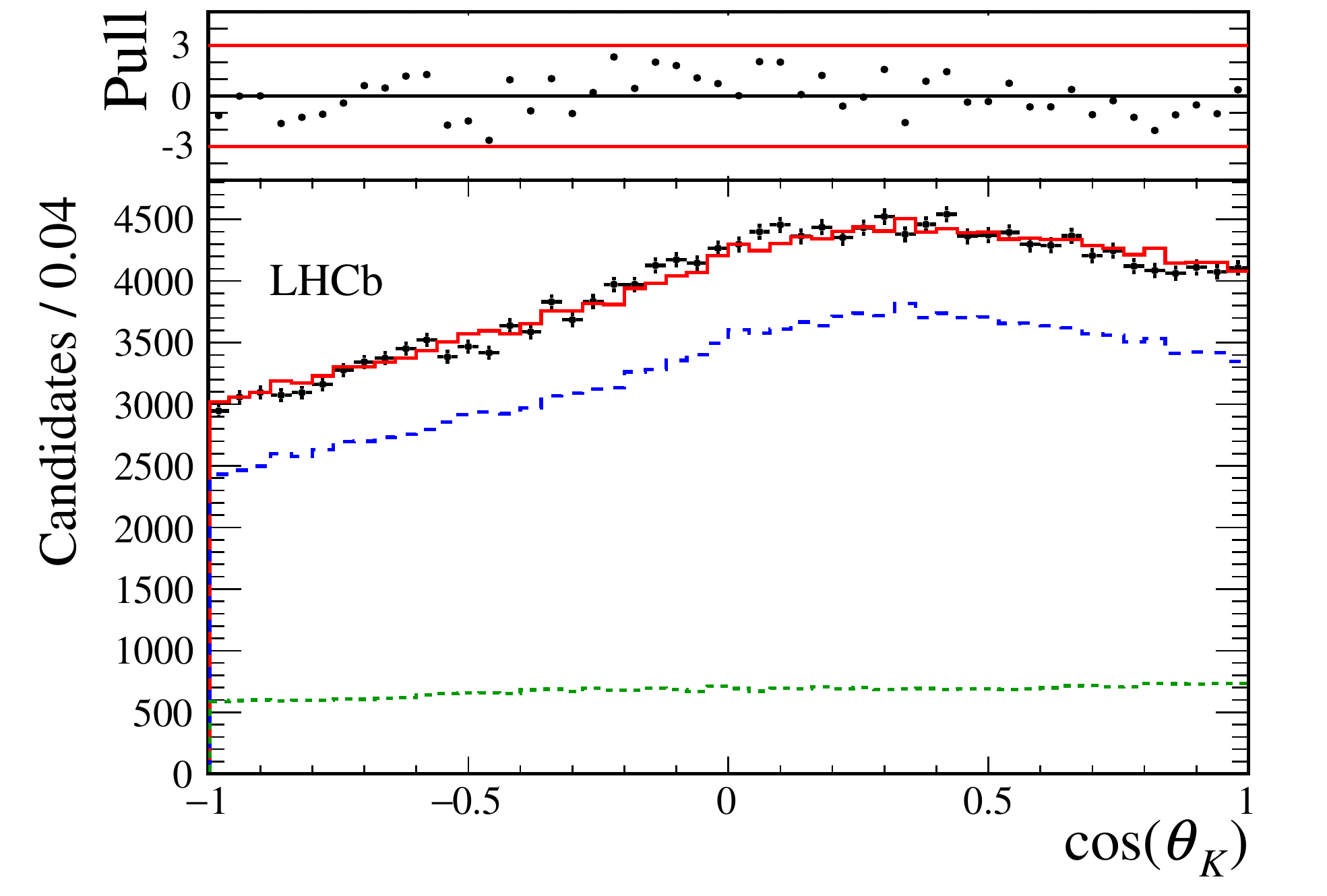}
\includegraphics[width=0.49\linewidth]{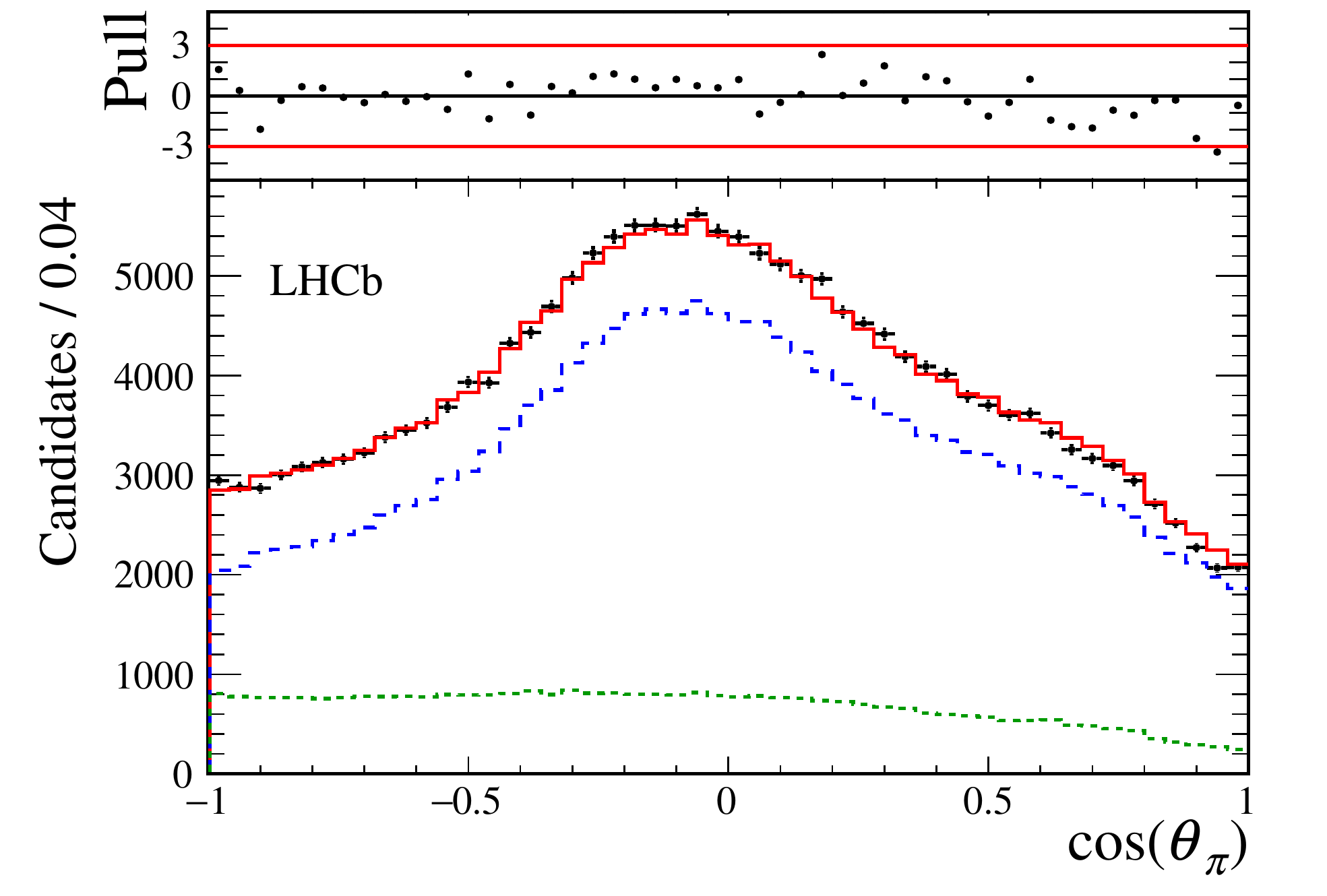}

$\vcenter{\hbox{\includegraphics[width=0.49\linewidth]{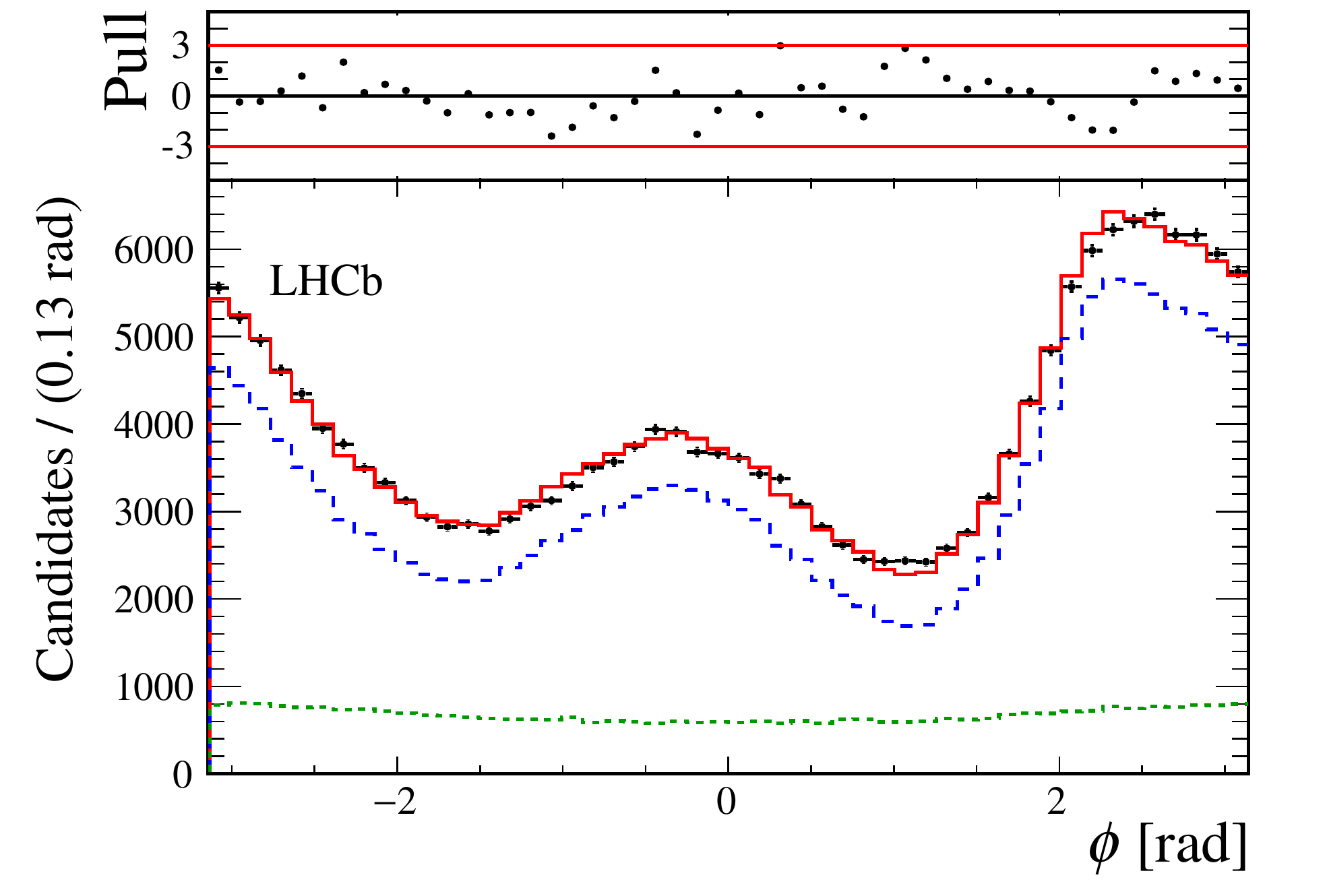}}}$
\hspace{0.09\linewidth}
$\vcenter{\hbox{\includegraphics[width=0.35\linewidth]{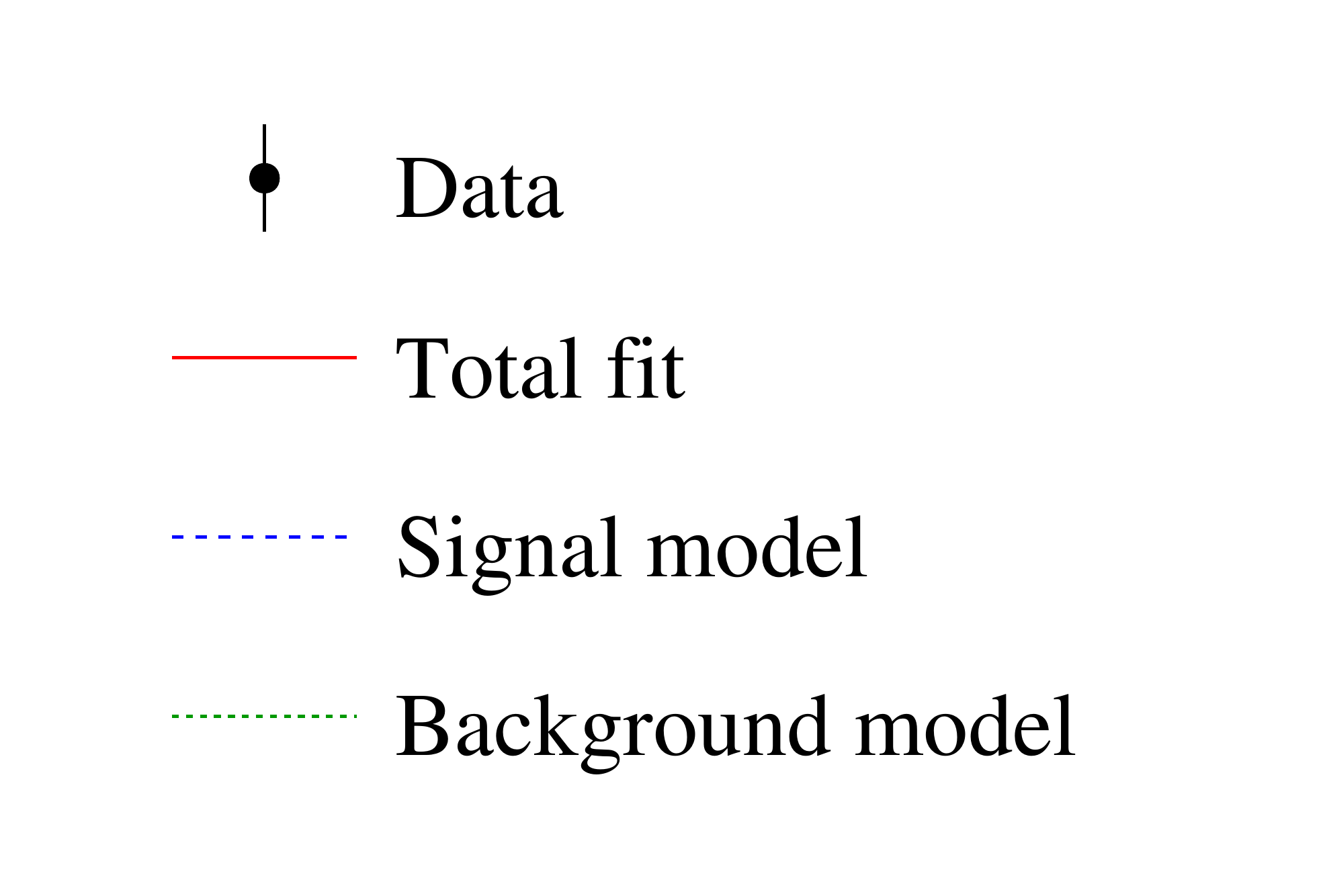}}}$
\caption{\label{fig:result}Distributions of the five CM variables for the selected \Dz and \CP-transformed \Dzb candidates (black points with error bars). The results of the five-dimensional amplitude fit is superimposed with the signal model (dashed blue), the background model (dotted green) and the total fit function (plain red). The plot on top of each distribution shows the normalised residuals, where the error is defined as the quadratic sum of the statistical uncertainties of the data and simulated samples.}
\end{figure}

The fit results are summarised in Table~\ref{tab:Result}, which shows the fit parameters and fit fractions of each component in the model. The fit is performed relative to the fixed component \decay{\Dz}{[\phi(1020)(\rho-\omega)^0]_{L=0}}. The resulting parametrisations of the three-body decays are reported in Table~\ref{tab:K1_1270} and those of the $\rho-\omega$ superposition in Table~\ref{tab:rhoomega}. A visualisation of the fit quality is provided by overlapping the fitted model and data projections on the five CM variables, shown in Fig.~\ref{fig:result}. The fit has also been inspected on 26 other projections, showing similar qualities. In order to quantify the quality of the fit, a \chisq~value is computed between the data and the fit model, using a five-dimensional adaptive binning. The obtained \chisq~value is 9226 for 8123 degrees of freedom, yielding a \chisqndf value of~1.14, not including systematic uncertainties. Such value is typical in comparison to similar analyses~\cite{LHCb-PAPER-2017-040,KKpipi_CLEO2017}.

\begin{table}[t]
\centering
\caption{\label{tab:Result}Modulus and phase of the fit parameters along with the fit fractions of the amplitudes included in the model. The substructures of the three-body resonances are listed in Table~\ref{tab:K1_1270}. The first uncertainty is statistical and the second is systematic.}
% !TEX root = ./ms.tex

\resizebox{\textwidth}{!}{
\begin{tabular}{lS[table-auto-round,table-format=1.3]@{\,\( \pm \)\,}S[table-auto-round,table-format=1.3]@{\,\( \pm \)\,}S[table-auto-round,table-format=1.3]S[table-auto-round,table-format=3.2]@{\,\( \pm \)\,}S[table-auto-round,table-format=1.2]@{\,\( \pm \)\,}S[table-auto-round,table-format=1.2]S[table-auto-round,table-format=3.2]@{\,\( \pm \)\,}S[table-auto-round,table-format=1.2]@{\,\( \pm \)\,}S[table-auto-round,table-format=1.2]}
\toprule
{Amplitude} & \multicolumn{3}{c}{$|c_k|$} & \multicolumn{3}{c}{$\arg(c_k)$ [rad]} & \multicolumn{3}{c}{Fit fraction [\%]}  \\ \midrule
$D^0 \rightarrow [\phi(1020)(\rho-\omega)^0]_{L=0}$ 	 & 	 \multicolumn{3}{c}{1 (fixed)} 	 & 	 \multicolumn{3}{c}{0 (fixed)} 	 & 	 23.820598 	 & 	 0.38241342 	 & 	 0.495364764663 	  \\ 
$D^0 \rightarrow K_1(1400)^+K^-$ 	 & 	 0.613904 	 & 	 0.0112591 	 & 	 0.0311320191384 	 & 	 1.05339 	 & 	 0.0216712 	 & 	 0.0534339481644 	 &  19.076028 	 & 	 0.59997329 	 & 	 1.46285634726 	  \\ 
$D^0 \rightarrow [K^-\pi^+]_{L=0}[K^+\pi^-]_{L=0}$ 	 & 	 0.282145 	 & 	 0.00374617 	 & 	 0.00772225236562 	 & 	 -0.603985 	 & 	 0.0153123 	 & 	 0.103548987199 	 & 	 18.464568 	 & 	 0.34720379 	 & 	 0.935302622459 	 \\ 
$D^0 \rightarrow K_1(1270)^+K^-$ 	 & 	 0.452287 	 & 	 0.010503 	 & 	 0.0172979921721 	 & 	 2.01724 	 & 	 0.0268733 	 & 	 0.0495470190537 	 &  18.052284 	 & 	 0.5214153 	 & 	 0.98195113359 	\\ 
$D^0 \rightarrow [K^*(892)^0\Kstarb(892)^0]_{L=0}$ 	 & 	 0.259205 	 & 	 0.00384906 	 & 	 0.0175184548966 	 & 	 -0.266961 	 & 	 0.0159128 	 & 	 0.0254856381889 	 & 	 9.1774181 	 & 	 0.20728835 	 & 	 0.277486530999 	  \\ 
$D^0 \rightarrow K^*(1680)^0[K^-\pi^+]_{L=0}$ 	 & 	 2.35857 	 & 	 0.0357751 	 & 	 0.623954196071 	 & 	 0.444357 	 & 	 0.016194 	 & 	 0.0296781472689 	 & 	 6.6125505 	 & 	 0.14818318 	 & 	 0.36784715246 	 \\ 
$D^0 \rightarrow [K^*(892)^0\Kstarb(892)^0]_{L=1}$ 	 & 	 0.249344 	 & 	 0.00458924 	 & 	 0.0167248486603 	 & 	 1.22179 	 & 	 0.0208534 	 & 	 0.0268994854706 	 & 	 4.9041923 	 & 	 0.15523346 	 & 	 0.181312757588 	  \\ 
$D^0 \rightarrow K_1(1270)^-K^+$ 	 & 	 0.220373 	 & 	 0.00599977 	 & 	 0.0108564351272 	 & 	 2.08838 	 & 	 0.0292048 	 & 	 0.0747256494412 	 &  4.2865819 	 & 	 0.17953251 	 & 	 0.405211788338 	\\ 
$D^0 \rightarrow [K^+K^-]_{L=0}[\pi^+\pi^-]_{L=0}$ 	 & 	 0.119605 	 & 	 0.00335915 	 & 	 0.0179465148698 	 & 	 -2.49485 	 & 	 0.0301723 	 & 	 0.163164636971 	 & 	 3.1448596 	 & 	 0.16791969 	 & 	 0.722720145746 	 \\ 
$D^0 \rightarrow K_1(1400)^-K^+$ 	 & 	 0.236265 	 & 	 0.008487 	 & 	 0.0182336201564 	 & 	 0.0437416 	 & 	 0.0424829 	 & 	 0.0876130930529 	 &  2.8171095 	 & 	 0.19162083 	 & 	 0.393823990434 	 \\ 
$D^0 \rightarrow [K^*(1680)^0\Kstarb(892)^0]_{L=0}$ 	 & 	 0.822878 	 & 	 0.0233662 	 & 	 0.218363318489 	 & 	 2.99156 	 & 	 0.0294719 	 & 	 0.0470637234954 	 & 	 2.7545096 	 & 	 0.14724509 	 & 	 0.189208895289 	 \\ 
$D^0 \rightarrow [\Kstarb(1680)^0K^*(892)^0]_{L=1}$ 	 & 	 1.00948 	 & 	 0.0218498 	 & 	 0.275797212303 	 & 	 -2.76213 	 & 	 0.021658 	 & 	 0.0291868777704 	 & 	 2.6964275 	 & 	 0.10546369 	 & 	 0.0934871200022 	 \\ 
$D^0 \rightarrow \Kstarb(1680)^0[K^+\pi^-]_{L=0}$ 	 & 	 1.37936 	 & 	 0.0287076 	 & 	 0.37262143691 	 & 	 1.05666 	 & 	 0.0236771 	 & 	 0.0305499657734 	 & 	 2.4112079 	 & 	 0.091089995 	 & 	 0.274605078331 	\\ 
$D^0 \rightarrow [\phi(1020)(\rho-\omega)^0]_{L=2}$ 	 & 	 1.31063 	 & 	 0.0306024 	 & 	 0.0179115133289 	 & 	 0.537725 	 & 	 0.0231829 	 & 	 0.0186097583035 	 & 	 2.2922153 	 & 	 0.076328417 	 & 	 0.0768510690169 	 \\ 
$D^0 \rightarrow [K^*(892)^0\Kstarb(892)^0]_{L=2}$ 	 & 	 0.652249 	 & 	 0.0178521 	 & 	 0.0426943872231 	 & 	 2.84918 	 & 	 0.027302 	 & 	 0.0402496423946 	 & 	 1.8487031 	 & 	 0.094854473 	 & 	 0.0994948005872 	  \\ 
$D^0 \rightarrow \phi(1020)[\pi^+\pi^-]_{L=0}$ 	 & 	 0.0486562 	 & 	 0.00149917 	 & 	 0.00415769868158 	 & 	 -1.71214 	 & 	 0.0403376 	 & 	 0.368436161161 	 & 	 1.4886758 	 & 	 0.08997412 	 & 	 0.325468008229 	 \\ 
$D^0 \rightarrow [K^*(1680)^0\Kstarb(892)^0]_{L=1}$ 	 & 	 0.747139 	 & 	 0.0214379 	 & 	 0.20340319157 	 & 	 0.140611 	 & 	 0.0308149 	 & 	 0.0395668213067 	 & 	 1.4779688 	 & 	 0.079670509 	 & 	 0.0963001180248 	\\ 
$D^0 \rightarrow [\phi(1020)\rho(1450)^0]_{L=1}$ 	 & 	 0.7618 	 & 	 0.035319 	 & 	 0.0679546462904 	 & 	 1.17161 	 & 	 0.0377484 	 & 	 0.0376498306684 	 & 	 0.98468291 	 & 	 0.088579436 	 & 	 0.0454743038681 	 \\ 
$D^0 \rightarrow a_0(980)^0f_2(1270)^0$ 	 & 	 1.52373 	 & 	 0.0575761 	 & 	 0.189018383122 	 & 	 0.213574 	 & 	 0.0379071 	 & 	 0.190443225542 	 & 	 0.69749665 	 & 	 0.051620198 	 & 	 0.0829510790037 	 \\ 
$D^0 \rightarrow a_1(1260)^+\pi^-$ 	 & 	 0.189122 	 & 	 0.011073 	 & 	 0.0418787783575 	 & 	 -2.8356 	 & 	 0.0669081 	 & 	 0.380249425011 	 &  0.46172906 	 & 	 0.054631865 	 & 	 0.22004966406 	 \\ 
$D^0 \rightarrow a_1(1260)^-\pi^+$ 	 & 	 0.187535 	 & 	 0.0137736 	 & 	 0.0306707878321 	 & 	 0.176563 	 & 	 0.0598272 	 & 	 0.431267942009 	 &  0.4525891 	 & 	 0.063138529 	 & 	 0.15623269672 	  \\ 
$D^0 \rightarrow [\phi(1020)(\rho-\omega)^0]_{L=1}$ 	 & 	 0.159697 	 & 	 0.011486 	 & 	 0.00521677533584 	 & 	 0.278105 	 & 	 0.0708505 	 & 	 0.0272336214917 	 & 	 0.42784564 	 & 	 0.049437689 	 & 	 0.0278207712443 	\\ 
$D^0 \rightarrow [K^*(1680)^0\Kstarb(892)^0]_{L=2}$ 	 & 	 1.2177 	 & 	 0.0891501 	 & 	 0.353706035557 	 & 	 -2.443 	 & 	 0.0839784 	 & 	 0.149960784542 	 & 	 0.32987773 	 & 	 0.048007044 	 & 	 0.0588164191526 	 \\ 
$D^0 \rightarrow [K^+K^-]_{L=0}(\rho-\omega)^0$ 	 & 	 0.194997 	 & 	 0.0150836 	 & 	 0.0349825916142 	 & 	 2.95021 	 & 	 0.0843851 	 & 	 0.292029269587 	 & 	 0.27352249 	 & 	 0.035558583 	 & 	 0.0536064667464 	\\ 
$D^0 \rightarrow [\phi(1020)f_2(1270)^0]_{L=1}$ 	 & 	 1.38765 	 & 	 0.0948117 	 & 	 0.256651681056 	 & 	 1.71026 	 & 	 0.0607552 	 & 	 0.372712411442 	 & 	 0.18296895 	 & 	 0.024473849 	 & 	 0.0746132554441 	 \\ 
$D^0 \rightarrow [K^*(892)^0 \Kbar{}^*_2(1430)^0]_{L=1}$ 	 & 	 1.52977 	 & 	 0.0862167 	 & 	 0.130576420361 	 & 	 2.01417 	 & 	 0.0668342 	 & 	 0.0865599838338 	 & 	 0.17834329 	 & 	 0.020169554 	 & 	 0.02436712161 	\\ 
\midrule
 	& \multicolumn{6}{r}{Sum of fit fractions} 	 &  	 129.315 	 & 	 1.086 	 & 	 2.378 	  \\ 
	& \multicolumn{6}{r}{$\chisqndf$} & 	 \multicolumn{3}{r}{$9242/8121=1.14$} 	  	 \\ 
\bottomrule
\end{tabular}
}

\end{table}

\begin{table}[t]
\centering
\caption{\label{tab:K1_1270}Parameters of the amplitudes contributing to the three-body decays of the $a_1(1260)^+$, $K_1(1270)^+$ and $K_1(1400)^+$. The first uncertainty is statistical and the second is systematic. }
% !TEX root = ./ms.tex

\resizebox{\textwidth}{!}{
\begin{tabular}{lS[table-auto-round,table-format=1.3]@{\,\( \pm \)\,}S[table-auto-round,table-format=1.3]@{\,\( \pm \)\,}S[table-auto-round,table-format=1.3]S[table-auto-round,table-format=3.2]@{\,\( \pm \)\,}S[table-auto-round,table-format=1.2]@{\,\( \pm \)\,}S[table-auto-round,table-format=1.2]S[table-auto-round,table-format=3.2]@{\,\( \pm \)\,}S[table-auto-round,table-format=1.2]@{\,\( \pm \)\,}S[table-auto-round,table-format=1.2]}
\toprule
{Amplitude} & \multicolumn{3}{c}{$|c_k|$} & \multicolumn{3}{c}{$\arg(c_k)$ [rad]} & \multicolumn{3}{c}{Fit fraction [\%]} \\ \midrule
$a_1(1260)^+ \rightarrow [\phi(1020)\pi^+]_{L=0}$ 	 & 	 \multicolumn{3}{c}{1 (fixed)} 	 & 	 \multicolumn{3}{c}{0 (fixed)} 	 &  	 \multicolumn{3}{c}{100} 	  \\ 
\midrule
 	%& \multicolumn{6}{r}{Sum of fit fractions} 	 &  	 \multicolumn{3}{c}{100} 	 & \\ 
\vspace{10pt}
\\ \midrule
$K_1(1270)^+ \rightarrow [K^*(892)^0\pi^+]_{L=0}$ 	 & 	 0.583591 	 & 	 0.0161474 	 & 	 0.0395075723128 	 & 	 0.631944 	 & 	 0.0307629 	 & 	 0.0524084546841 	 & 	 51.215637 	 & 	 1.0575526 	 & 	 3.21282463279 	\\ 
$K_1(1270)^+ \rightarrow [(\rho-\omega)^0K^+]_{L=0}$ 	 & 	 \multicolumn{3}{c}{1 (fixed)} 	 & 	 \multicolumn{3}{c}{0 (fixed)} 	 & 	 49.58307 	 & 	 1.9934943 	 & 	 4.3521588311 \\ 
$K_1(1270)^+ \rightarrow [K^+\pi^-]_{L=0}\pi^+$ 	 & 	 0.611561 	 & 	 0.026976 	 & 	 0.094237488354 	 & 	 -1.93863 	 & 	 0.0417981 	 & 	 0.0826215795071 	 & 	 6.2737535 	 & 	 0.48369986 	 & 	 1.6600040234 	 \\ 
$K_1(1270)^+ \rightarrow [K^*(892)^0\pi^+]_{L=2}$ 	 & 	 0.859048 	 & 	 0.0440729 	 & 	 0.0602191738369 	 & 	 -2.53474 	 & 	 0.0409534 	 & 	 0.04530302363 	 & 	 2.0292766 	 & 	 0.16941125 	 & 	 0.195235614515 	  \\ 
$K_1(1270)^+ \rightarrow [\rho(1450)^0K^+]_{L=0}$ 	 & 	 0.482425 	 & 	 0.0683832 	 & 	 0.187061285214 	 & 	 -2.37278 	 & 	 0.0998564 	 & 	 0.445333144672 	 & 	 1.495995 	 & 	 0.47249784 	 & 	 1.04099356456 	  \\ 
\midrule
 	& \multicolumn{6}{r}{Sum of fit fractions} 	 &  	 110.598 	 & 	 2.200 	 & 	 5.757 	  \\ 
\vspace{10pt}
\\ \midrule
$K_1(1400)^+ \rightarrow [K^*(892)^0\pi^+]_{L=0}$ 	 & 	 \multicolumn{3}{c}{1 (fixed)} 	 & 	 \multicolumn{3}{c}{0 (fixed)} 	 & 	 \multicolumn{3}{c}{100}  	  \\ 
\bottomrule
 	%& \multicolumn{6}{r}{Sum of fit fractions} 	 &  	 \multicolumn{3}{c}{100} 	 & \\ 
\end{tabular}
}
\end{table}

\begin{table}[b]
\centering
\caption{\label{tab:rhoomega}Parameters of the $\rho-\omega$ interference for all relevant amplitudes. The first uncertainty is statistical and the second is systematic.}
% !TEX root = ./ms.tex

\resizebox{\textwidth}{!}{
\begin{tabular}{lS[table-auto-round,table-format=1.3]@{\,\( \pm \)\,}S[table-auto-round,table-format=1.3]@{\,\( \pm \)\,}S[table-auto-round,table-format=1.3]S[table-auto-round,table-format=3.2]@{\,\( \pm \)\,}S[table-auto-round,table-format=1.2]@{\,\( \pm \)\,}S[table-auto-round,table-format=1.2]S[table-auto-round,table-format=3.2]@{\,\( \pm \)\,}S[table-auto-round,table-format=1.2]@{\,\( \pm \)\,}S[table-auto-round,table-format=1.2]}
\toprule
{Amplitude}  & \multicolumn{3}{c}{$|\tilde{c}_k|$} & \multicolumn{3}{c}{$\arg(\tilde{c}_k)$ [rad]} & \multicolumn{3}{c}{Fit fraction [\%]} \\ \midrule
$D^0 \rightarrow [\phi(1020)\rho(770)^0]_{L=0}$ 	 	 & \multicolumn{3}{c}{1 (fixed)} 	 & 	 \multicolumn{3}{c}{0 (fixed)} 	 & 	 92.546287 	 & 	 0.46298338& 	 0.275463867603 \\ 
$D^0 \rightarrow [\phi(1020)\omega(782)]_{L=0}$ 	 &0.11355 	 & 	 0.00444098 	 & 	 0.0031618524758 	 & 	 1.30064 	 & 	 0.0422617 	 & 	 0.0360786427055 	 & 	 1.4161069 	 &  0.10606457 & 	 0.0408208402451 \\ \midrule
 	& \multicolumn{6}{r}{Sum of fit fractions} 	 &  	 93.96 	 & 	 0.40 	 & 	 0.28   \\ 
\vspace{10pt}
\\ \midrule
$D^0 \rightarrow [\phi(1020)\rho(770)^0]_{L=1}$ 	  	 & \multicolumn{3}{c}{1 (fixed)} 	 & 	 \multicolumn{3}{c}{0 (fixed)} 	 & 	 83.111722 	 & 	 4.1073466& 	 1.69575270894 \\ 
$D^0 \rightarrow [\phi(1020)\omega(782)]_{L=1}$	 &0.253818 	 & 	 0.0515847 	 & 	 0.0176131132714 	 & 	 1.31639 	 & 	 0.193714 	 & 	 0.0687018754566 	 & 	 4.3336918 	 &  1.5767466 & 	 0.515257237739 \\ \midrule
 	& \multicolumn{6}{r}{Sum of fit fractions} 	 &  	 87.45	 & 	 2.99 	 & 	 1.78   \\ 
\vspace{10pt}
\\ \midrule
$D^0 \rightarrow [\phi(1020)\rho(770)^0]_{L=2}$ 		 & \multicolumn{3}{c}{1 (fixed)} 	 & 	 \multicolumn{3}{c}{0 (fixed)} 	 & 	 94.638866 	 & 	 1.6899223& 	 0.777958073703 \\ 
$D^0 \rightarrow [\phi(1020)\omega(782)]_{L=2}$	 &0.162457 	 & 	 0.0317563 	 & 	 0.0137173328957 	 & 	 1.49845 	 & 	 0.167493 	 & 	 0.0593690439742 	 & 	 0.7120005 	 &  0.2700828 & 	 0.116227863943 \\ \midrule
 	& \multicolumn{6}{r}{Sum of fit fractions} 	 &  	 95.35 	 & 	 1.54 	 & 	 0.79 	 \\ 
\vspace{10pt}
\\ \midrule
$D^0 \rightarrow [K^+K^-]_{L=0}\rho(770)^0$ 	 	 & \multicolumn{3}{c}{1 (fixed)} 	 & 	 \multicolumn{3}{c}{0 (fixed)} 	 & 	 85.411894 	 & 	 5.8949219&  3.49228154062 \\ 
$D^0 \rightarrow [K^+K^-]_{L=0}\omega(782)$	 &0.494178 	 & 	 0.0977038 	 & 	 0.0982925395904 	 & 	 -0.954386 	 & 	 0.185513 	 & 	 0.149180587947 	 & 	 9.2421845 	 &  3.2585626 & 	 3.64153481408 \\ \midrule
 	& \multicolumn{6}{r}{Sum of fit fractions} 	 &  	 94.65 	 & 	 5.03 	 & 	 5.04  \\ 
\vspace{10pt}
\\ \midrule
$K_1(1270)^+ \rightarrow [\rho(770)^0K^+]_{L=0}$ 		 & \multicolumn{3}{c}{1 (fixed)} 	 & 	 \multicolumn{3}{c}{0 (fixed)} 	 & 	 139.02732 	 & 	 1.9808632& 	 3.80580772455 \\ 
$K_1(1270)^+ \rightarrow [\omega(782)K^+]_{L=0}$	 &0.159107 	 & 	 0.012062 	 & 	 0.0108237354399 	 & 	 1.36445 	 & 	 0.0736758 	 & 	 0.0566620657172 	 & 	 1.5182573 	 &  0.22047933 & 	 0.190984352188 \\ \midrule
 	& \multicolumn{6}{r}{Sum of fit fractions} 	 &  	 140.55 	 & 	 1.90 	 & 	 3.81 	 \\ 
\bottomrule
\end{tabular}
}
\end{table}

A few features of the model are worth noting. The \decay{\Dz}{\phi(1020)\rho(1450)^0} and \decay{\Dz}{\Kstarb(1680)^0 K^*(892)^0} components appear only in $P$-wave without their $S$- and \mbox{$D$-wave} counterparts, which are also allowed. The $a_1(1260)^+$ resonance is decaying only to $\phi(1020)\pip$, while a contribution of $\Kstarb(892)^0 \Kp$ is reported by the PDG~\cite{PDG2018}. Finally, the $\rho-\omega$ superposition seems to be different between the decay modes, as shown in Table~\ref{tab:rhoomega}.

The resulting mass and width of the $K_1(1270)^+$ resonance are $1297 \pm 1 \mevcc$ and $148 \pm 4 \mevcc$, respectively, where the quoted uncertainties are statistical only. These values can be compared with the values quoted by the PDG~\cite{PDG2018}, $1272 \pm 7 \mevcc$ and $90 \pm 20 \mevcc$, the latter being an estimate from values ranging from $75\mevcc$ to $260\mevcc$. However, these values are model-dependent and the lineshape of three-body resonances is not as well established as that of two-body resonances.

These results are compared to the CLEO legacy-data model of Ref.~\cite{KKpipi_CLEO2017}. The main components are present in both models. While the \decay{\Dz}{\phi(1020)\rho(770)^0} components are compatible, the \decay{\Dz}{\Kstar(892)^0\Kstarb(892)^0} components do not have the same hierarchy between the three angular momentum configurations. The contributions of the \decay{\Dz}{K_1(1270)^\pm K^\mp} and \decay{\Dz}{K_1(1400)^\pm K^\mp} components are different. One possible explanation is that Ref.~\cite{KKpipi_CLEO2017} did not impose \CP conservation in the strong decays of the kaon resonances. The only component of the CLEO legacy-data model that is not found in this analysis is \decay{\Dz}{K^*(1680)^+K^-}, although it is included in the list of potential amplitudes. Instead, other amplitudes with small fit fractions are uncovered in this analysis, as a consequence of the more than 50 times larger dataset.

% !TEX root = ./ms.tex

\section{\CP-violation results}
\label{sec:CPVResults}

For the \CP-violation fit, the dataset is split into two subsets according to the charge of the muon to separate the \Dz and \CP-transformed \Dzb decays. The \CP-violation fit described in Sec.~\ref{sec:cpv_description} is applied to these two samples. Table~\ref{tab:CPV} shows the resulting \CP-violation parameters. The average moduli and phases are not shown in the table as they are identical to the moduli and phases from the \CP-averaged fit in Table~\ref{tab:Result}.

\begin{table}[t]
\centering
\caption{\label{tab:CPV}\CP-violation parameters fitted simultaneously to the \Dz and (\CP-transformed) \Dzb samples. The first uncertainty is statistical and the second is systematic.}
% !TEX root = ./ms.tex

\resizebox{\textwidth}{!}{
\begin{tabular}{lS[table-auto-round,table-format=1.1]@{\,\( \pm \)\,}S[table-auto-round,table-format=1.1]@{\,\( \pm \)\,}S[table-auto-round,table-format=1.1]S[table-auto-round,table-format=3.1]@{\,\( \pm \)\,}S[table-auto-round,table-format=1.1]@{\,\( \pm \)\,}S[table-auto-round,table-format=1.1]S[table-auto-round,table-format=3.1]@{\,\( \pm \)\,}S[table-auto-round,table-format=2.1]@{\,\( \pm \)\,}S[table-auto-round,table-format=1.1]}
\toprule
{Amplitude} & \multicolumn{3}{c}{$A_{|c_k|}$ [\%]} & \multicolumn{3}{c}{$\Delta\arg(c_k)$ [\%]} &  \multicolumn{3}{c}{$A_{\mathcal{F}_k}$ [\%]}\\ \midrule
$D^0 \rightarrow [\phi(1020)(\rho-\omega)^0]_{L=0}$ & \multicolumn{3}{c}{0 (fixed)} & \multicolumn{3}{c}{0 (fixed)} & -1.82189320616 & 1.49601725228 & 0.192587202406\\ 
$D^0 \rightarrow K_1(1400)^+K^-$ & -1.35914 & 1.08448 & 0.19985445302 & 1.33317 & 1.46873 & 0.251077597125 & -4.53741608437 & 2.08695814523 & 0.280325256178 \\ 
$D^0 \rightarrow [K^-\pi^+]_{L=0}[K^+\pi^-]_{L=0}$ & 1.91145 & 1.13402 & 0.30396365352 & -1.15464 & 1.34911 & 0.29249601139 & 2.00101229622 & 1.81614845282 & 0.658613744639 \\ 
$D^0 \rightarrow K_1(1270)^+K^-$ & -0.394449 & 0.999095 & 0.190299117951 & -1.11886 & 1.44375 & 0.231795002621 & -2.61040206944 & 1.71559957981 & 0.207478793722 \\ 
$D^0 \rightarrow [K^*(892)^0\Kstarb(892)^0]_{L=0}$ & -1.26197 & 1.29878 & 0.2648890709 & -1.69149 & 1.47057 & 0.204761571778 & -4.34342503783 & 2.17207559221 & 0.479811343281 \\ 
$D^0 \rightarrow K^*(1680)^0[K^-\pi^+]_{L=0}$ & 2.20411 & 1.30944 & 0.262353211186 & 1.35699 & 1.49105 & 0.213661045336 & 2.58625393767 & 2.16726936151 & 0.412492921752 \\ 
$D^0 \rightarrow [K^*(892)^0\Kstarb(892)^0]_{L=1}$ & -0.383939 & 1.7132 & 0.220326499405 & 3.74391 & 2.00162 & 0.238990541509 & -2.58939819724 & 3.15235642856 & 0.291575145804 \\ 
$D^0 \rightarrow K_1(1270)^-K^+$ & 2.58568 & 1.70659 & 0.435611428544 & -0.105977 & 2.07411 & 0.310679202873 & 3.34917149124 & 3.52026975696 & 0.540159695851 \\ 
$D^0 \rightarrow [K^+K^-]_{L=0}[\pi^+\pi^-]_{L=0}$ & 3.47357 & 2.48228 & 1.52071580946 & -5.45121 & 2.6465 & 1.55690135454 & 5.12335460958 & 5.05833569378 & 3.11918809664 \\ 
$D^0 \rightarrow K_1(1400)^-K^+$ & 0.247324 & 2.8788 & 0.68681818763 & 2.47408 & 3.54658 & 1.0493761091 & -1.32737015511 & 6.04988781007 & 0.963148958425 \\ 
$D^0 \rightarrow [K^*(1680)^0\Kstarb(892)^0]_{L=0}$ & 4.01797 & 2.6728 & 0.775178349877 & -5.41451 & 2.76328 & 0.821440862921 & 6.21017707447 & 5.24743412282 & 1.51535259026 \\ 
$D^0 \rightarrow [\Kstarb(1680)^0K^*(892)^0]_{L=1}$ & -0.350713 & 2.06322 & 0.255642348155 & 0.411065 & 2.08669 & 0.27826653173 & -2.52298828263 & 3.89599354671 & 0.443738058776 \\ 
$D^0 \rightarrow \Kstarb(1680)^0[K^+\pi^-]_{L=0}$ & 2.09909 & 1.9585 & 0.628295947768 & -1.78967 & 2.24935 & 0.328249049479 & 2.37625374905 & 3.74849066694 & 1.0698970013 \\ 
$D^0 \rightarrow [\phi(1020)(\rho-\omega)^0]_{L=2}$ & 0.836926 & 1.88275 & 0.314183148502 & -1.19725 & 1.99088 & 0.468930060841 & -0.148204208401 & 3.27715479255 & 0.46199139855 \\ 
$D^0 \rightarrow [K^*(892)^0\Kstarb(892)^0]_{L=2}$ & -0.576444 & 2.51026 & 0.418672828703 & 0.55705 & 2.61656 & 0.448760374112 & -2.97411914834 & 4.96321539465 & 0.687022018704 \\ 
$D^0 \rightarrow \phi(1020)[\pi^+\pi^-]_{L=0}$ & 3.823 & 3.07797 & 0.650050412514 & -0.467035 & 3.89983 & 0.676876292394 & 5.82104678359 & 6.07835783622 & 0.802435989261 \\ 
$D^0 \rightarrow [K^*(1680)^0\Kstarb(892)^0]_{L=1}$ & 1.55328 & 2.75233 & 0.465299664782 & 0.69646 & 2.98815 & 0.424942111998 & 1.28464295904 & 5.34212945631 & 0.553385019376 \\ 
$D^0 \rightarrow [\phi(1020)\rho(1450)^0]_{L=1}$ & 4.64409 & 4.12027 & 0.561373841053 & 9.26793 & 3.34236 & 0.593140842281 & 7.45889097723 & 8.52702583941 & 1.10207895335 \\ 
$D^0 \rightarrow a_0(980)^0f_2(1270)^0$ & 1.641 & 3.5638 & 0.693572256913 & -7.33179 & 3.34073 & 0.834141954208 & 1.4600993099 & 7.18978907442 & 1.30464693118 \\ 
$D^0 \rightarrow a_1(1260)^+\pi^-$ & -4.42586 & 5.63956 & 3.6602948082 & 9.2995 & 6.14782 & 1.26804640295 & -10.6391821942 & 11.6997988795 & 7.04194367471 \\ 
$D^0 \rightarrow a_1(1260)^-\pi^+$ & -3.42523 & 7.02477 & 1.92144937398 & -5.80143 & 5.55398 & 4.32691569068 & -8.65354183988 & 13.6724330915 & 2.86005586987 \\ 
$D^0 \rightarrow [\phi(1020)(\rho-\omega)^0]_{L=1}$ & 2.13529 & 5.17338 & 0.759142662492 & -12.2448 & 5.46847 & 0.610763603109 & 2.44864943504 & 10.9989472728 & 1.35691160625 \\ 
$D^0 \rightarrow [K^*(1680)^0\Kstarb(892)^0]_{L=2}$ & 5.15648 & 7.06365 & 1.87195146186 & -5.58651 & 8.12384 & 1.3124366708 & 8.47960495017 & 14.3040260493 & 3.53507144409 \\ 
$D^0 \rightarrow [K^+K^-]_{L=0}(\rho-\omega)^0$ & 11.6681 & 5.99994 & 1.86634597895 & 4.79817 & 6.24957 & 1.12470458457 & 21.2901505323 & 12.5340490167 & 2.79107990599 \\ 
$D^0 \rightarrow [\phi(1020)f_2(1270)^0]_{L=1}$ & 2.72715 & 6.70962 & 1.68618361809 & 0.906099 & 6.03837 & 1.68766810665 & 3.63196549391 & 13.3014218602 & 2.98486529797 \\ 
$D^0 \rightarrow [K^*(892)^0\Kbar{}^*_2(1430)^0]_{L=1}$ & 3.94513 & 5.19447 & 1.03751035648 & 6.75997 & 6.35103 & 1.36385803361 & 6.06481572204 & 10.8049736016 & 1.81010277936 \\ 
\bottomrule
\end{tabular}
}

\end{table}

All the asymmetry parameters are compatible with zero. The most significant deviation, observed for the phase difference for the \decay{\Dz}{[\phi(1020)\rho(1450)^0]_{L=1}} component, corresponds to a $2.8\sigma$ statistical fluctuation. To check how likely such a deviation would be in absence of \CP violation, the fit is repeated many times, where the data is randomly split instead of splitting it according to the flavour of the $D$ candidate. The largest deviation among all the asymmetry parameters exceeds $2.8\sigma$ in $35\%$ of the fits, confirming that the deviation observed in the \CP-violation fit is not significant.

% !TEX root = ./ms.tex
\section{Conclusion}
\label{sec:Conclusions}
An amplitude analysis of the Cabibbo-suppressed decay mode \DztoKKpipi is performed.
The resulting amplitude model provides the most precise description of this decay to date and is used to perform a search for \CP violation.

More than 25 decay amplitudes of the \Dz meson have been identified. The most abundant being \decay{\Dz}{[\phi(1020)(\rho-\omega)^0]_{L=0}}, followed by \decay{\Dz}{K_1(1400)^+K^-}, \decay{\Dz}{[K^-\pi^+]_{L=0}[K^+\pi^-]_{L=0}}, and \decay{\Dz}{K_1(1270)^+K^-}, all together representing about 80\% of the total decay rate (neglecting interference).
This model confirms the main findings of Ref.~\cite{KKpipi_CLEO2017} and provides an improved description of the data. In particular, a $\rho-\omega$ interference is found that does not allow to treat the two resonances separately and the contribution of the $KK$, $\pi\pi$ and $K\pi$ $S$-waves are studied.

For each component of the model, \CP asymmetries related to the amplitude modulus, amplitude phase and fit fraction are measured with a total uncertainty ranging from 1\% to 15\%, dominated by the statistical uncertainty.
At this level of sensitivity, no effect of \CP violation is found. 
This is expected from SM predictions~\cite{SM_prediction} and large effects from beyond the SM processes are ruled out.
It is interesting to report that most of the systematic uncertainties of the \CP-averaged fit are marginally affecting the \CP-violation fit, resulting in a much smaller systematic uncertainty.
Finally, the systematic uncertainties of the \CP-violation measurements are expected to scale down with luminosity, since most of them are estimated with inputs from data.

% !TEX root = ./ms.tex
\section*{Acknowledgements}
%
% These Acknowledgements valid from 14-Aug-2018
%
\noindent We express our gratitude to our colleagues in the CERN
accelerator departments for the excellent performance of the LHC. We
thank the technical and administrative staff at the LHCb
institutes.
We acknowledge support from CERN and from the national agencies:
CAPES, CNPq, FAPERJ and FINEP (Brazil); 
MOST and NSFC (China); 
CNRS/IN2P3 (France); 
BMBF, DFG and MPG (Germany); 
INFN (Italy); 
NWO (Netherlands); 
MNiSW and NCN (Poland); 
MEN/IFA (Romania); 
MSHE (Russia); 
MinECo (Spain); 
SNSF and SER (Switzerland); 
NASU (Ukraine); 
STFC (United Kingdom); 
NSF (USA).
We acknowledge the computing resources that are provided by CERN, IN2P3
(France), KIT and DESY (Germany), INFN (Italy), SURF (Netherlands),
PIC (Spain), GridPP (United Kingdom), RRCKI and Yandex
LLC (Russia), CSCS (Switzerland), IFIN-HH (Romania), CBPF (Brazil),
PL-GRID (Poland) and OSC (USA).
We are indebted to the communities behind the multiple open-source
software packages on which we depend.
Individual groups or members have received support from
AvH Foundation (Germany);
EPLANET, Marie Sk\l{}odowska-Curie Actions and ERC (European Union);
ANR, Labex P2IO and OCEVU, and R\'{e}gion Auvergne-Rh\^{o}ne-Alpes (France);
Key Research Program of Frontier Sciences of CAS, CAS PIFI, and the Thousand Talents Program (China);
RFBR, RSF and Yandex LLC (Russia);
GVA, XuntaGal and GENCAT (Spain);
the Royal Society
and the Leverhulme Trust (United Kingdom);
Laboratory Directed Research and Development program of LANL (USA).

\clearpage
% !TEX root = ./ms.tex

\section*{Appendices}

\appendix

\section{Alternative models}
\label{app:alt_mod}
Three alternative models are considered to assign a systematic uncertainty. The results of the corresponding fits are listed in Tables~\ref{tab:alt_mod_Kst1410}--\ref{tab:alt_mod_rho1450}.

\begin{table}[h!]
\centering
\caption{\label{tab:alt_mod_Kst1410}Modulus and phase of the fit parameters along with the fit fractions of the amplitudes included in the alternative model using the $K^*(1410)^0$ instead of the $K^*(1680)^0$ resonance.}
% !TEX root = ./ms.tex

\fontsize{8pt}{1pt}\selectfont
\npdecimalsign{.}
\nprounddigits{2}
\resizebox{\textwidth}{!}{
\begin{tabular}{ln{2}{3}@{\,\( \pm \)\,}n{2}{2}n{3}{3}@{\,\( \pm \)\,}n{2}{2}n{3}{3}@{\,\( \pm \)\,}n{2}{2}}
\toprule
{Amplitude} & \multicolumn{2}{c}{$|c_k|$} & \multicolumn{2}{c}{$\arg(c_k)$ [rad]} & \multicolumn{2}{c}{Fit fraction [\%]} \\ \midrule
$D^0 \rightarrow [\phi(1020)(\rho-\omega)^0]_{L=0}$ 	 & 	 \multicolumn{2}{c}{1 (fixed)} 	 & 	 \multicolumn{2}{c}{0 (fixed)} 	 & 	 23.781732 	 & 	 0.38158305 	  \\ 
$D^0 \rightarrow K_1(1400)^+K^-$ 	 & 	 0.634156 	 & 	 0.00998808 	 & 	 1.03452 	 & 	 0.0193115 	 & 	 18.930194 	 & 	 0.46380671 	 \\ 
$D^0 \rightarrow [K^-\pi^+]_{L=0}[K^+\pi^-]_{L=0}$ 	 & 	 0.284656 	 & 	 0.0037843 	 & 	 -0.599114 	 & 	 0.015004 	 & 	 18.759375 	 & 	 0.3566133 	 \\ 
$D^0 \rightarrow K_1(1270)^+K^-$ 	 & 	 0.462099 	 & 	 0.00938781 	 & 	 2.02835 	 & 	 0.022449 	 & 	 18.401285 	 & 	 0.37130925 	 \\ 
$D^0 \rightarrow [K^*(892)^0\Kstarb(892)^0]_{L=0}$ 	 & 	 0.27866 	 & 	 0.00412252 	 & 	 -0.280665 	 & 	 0.0155052 	 & 	 9.2333442 	 & 	 0.20878872 	 \\ 
$D^0 \rightarrow [K^*(1410)^0[K^-\pi^+]_{L=0}]_{L=1}$ 	 & 	 1.82136 	 & 	 0.0282207 	 & 	 0.435287 	 & 	 0.0161907 	 & 	 6.5861793 	 & 	 0.15172595 	 \\ 
$D^0 \rightarrow [K^*(892)^0\Kstarb(892)^0]_{L=1}$ 	 & 	 0.269252 	 & 	 0.00504785 	 & 	 1.2121 	 & 	 0.0207554 	 & 	 4.9954975 	 & 	 0.16191011 	 \\ 
$D^0 \rightarrow K_1(1270)^-K^+$ 	 & 	 0.220843 	 & 	 0.00574495 	 & 	 2.12388 	 & 	 0.0267147 	 & 	 4.2035713 	 & 	 0.17130304 	 \\ 
$D^0 \rightarrow [K^+K^-]_{L=0}[\pi^+\pi^-]_{L=0}$ 	 & 	 0.119227 	 & 	 0.0033507 	 & 	 -2.52597 	 & 	 0.0300814 	 & 	 3.1191478 	 & 	 0.16709177 	 \\ 
$D^0 \rightarrow [K^*(1410)^0\Kstarb(892)^0]_{L=0}$ 	 & 	 0.68681 	 & 	 0.0190716 	 & 	 2.94556 	 & 	 0.0285309 	 & 	 2.9707832 	 & 	 0.15496554 	 \\ 
$D^0 \rightarrow K_1(1400)^-K^+$ 	 & 	 0.241115 	 & 	 0.00861341 	 & 	 0.103423 	 & 	 0.0411969 	 & 	 2.7285615 	 & 	 0.18478111 	 \\ 
$D^0 \rightarrow [\Kstarb(1410)^0K^*(892)^0]_{L=1}$ 	 & 	 0.81644 	 & 	 0.0179625 	 & 	 -2.77828 	 & 	 0.0220556 	 & 	 2.6224144 	 & 	 0.10480844 	 \\ 
$D^0 \rightarrow [\Kstarb(1410)^0[K^+\pi^-]_{L=0}]_{L=1}$ 	 & 	 1.0311 	 & 	 0.0220978 	 & 	 1.06897 	 & 	 0.0244519 	 & 	 2.3167453 	 & 	 0.090693574 	 \\ 
$D^0 \rightarrow [\phi(1020)(\rho-\omega)^0]_{L=2}$ 	 & 	 1.30951 	 & 	 0.0305978 	 & 	 0.540571 	 & 	 0.0232351 	 & 	 2.2842368 	 & 	 0.076103218 	 \\ 
$D^0 \rightarrow [K^*(892)^0\Kstarb(892)^0]_{L=2}$ 	 & 	 0.685212 	 & 	 0.0195627 	 & 	 2.86738 	 & 	 0.027524 	 & 	 1.797538 	 & 	 0.096620228 	 \\ 
$D^0 \rightarrow \phi(1020)[\pi^+\pi^-]_{L=0}$ 	 & 	 0.0483265 	 & 	 0.00150808 	 & 	 -1.70434 	 & 	 0.040476 	 & 	 1.4658104 	 & 	 0.089880385 	 \\ 
$D^0 \rightarrow [K^*(1410)^0\Kstarb(892)^0]_{L=1}$ 	 & 	 0.609339 	 & 	 0.0178523 	 & 	 0.0878262 	 & 	 0.0306182 	 & 	 1.4616228 	 & 	 0.080656968 	 \\ 
$D^0 \rightarrow [\phi(1020)\rho(1450)^0]_{L=1}$ 	 & 	 0.757687 	 & 	 0.035357 	 & 	 1.18096 	 & 	 0.0378961 	 & 	 0.97224933 	 & 	 0.088068239 	 \\ 
$D^0 \rightarrow a_0(980)^0f_2(1270)^0$ 	 & 	 1.46469 	 & 	 0.0575177 	 & 	 0.332109 	 & 	 0.0388594 	 & 	 0.64328297 	 & 	 0.049647347 	 \\ 
$D^0 \rightarrow a_1(1260)^-\pi^+$ 	 & 	 0.191478 	 & 	 0.013762 	 & 	 0.184132 	 & 	 0.0587437 	 & 	 0.47093056 	 & 	 0.064244773 	 \\ 
$D^0 \rightarrow a_1(1260)^+\pi^-$ 	 & 	 0.190654 	 & 	 0.0111267 	 & 	 -2.80449 	 & 	 0.0665231 	 & 	 0.4683593 	 & 	 0.055012337 	 \\ 
$D^0 \rightarrow [\phi(1020)(\rho-\omega)^0]_{L=1}$ 	 & 	 0.161023 	 & 	 0.0114749 	 & 	 0.271888 	 & 	 0.0704436 	 & 	 0.4323853 	 & 	 0.049598033 	 \\ 
$D^0 \rightarrow [K^*(1410)^0\Kstarb(892)^0]_{L=2}$ 	 & 	 1.03026 	 & 	 0.0764462 	 & 	 -2.51866 	 & 	 0.080374 	 & 	 0.33284581 	 & 	 0.049074525 	 \\ 
$D^0 \rightarrow [K^+K^-]_{L=0}(\rho-\omega)^0$ 	 & 	 0.205895 	 & 	 0.0150263 	 & 	 3.06228 	 & 	 0.0809116 	 & 	 0.29159403 	 & 	 0.036447753 	 \\ 
$D^0 \rightarrow [\phi(1020)f_2(1270)^0]_{L=1}$ 	 & 	 1.39898 	 & 	 0.095099 	 & 	 1.72451 	 & 	 0.0602075 	 & 	 0.18561767 	 & 	 0.02469805 	 \\ 
$D^0 \rightarrow [K^*(892)^0\Kbar{}^*_2(1430)^0]_{L=1}$ 	 & 	 1.51341 	 & 	 0.0886133 	 & 	 1.98435 	 & 	 0.0702038 	 & 	 0.16353078 	 & 	 0.019200979 	 \\ 
\midrule

 	& \multicolumn{4}{r}{Sum of fit fractions} 	 &  	 129.619 	 & 	 0.953 	 \\ 
	& \multicolumn{6}{r}{$\chisqndf ~~~ 9224/8123=1.14$} 	  	  	 \\ 
%	& \multicolumn{4}{r}{ndf} & 	 \multicolumn{1}{c}{8123} 	 & 	  	 \\ 	

\vspace{10pt} \\ \midrule

$a_1(1260)^+ \rightarrow [\phi(1020)\pi^+]_{L=0}$ 	 & 	 \multicolumn{2}{c}{1 (fixed)} 	 & 	 \multicolumn{2}{c}{0 (fixed)} 	 &  	 \multicolumn{2}{c}{100} 	 \\ 
\midrule
% 	& \multicolumn{4}{r}{Sum of fit fractions} 	 &  	 \multicolumn{2}{c}{100} 	 \\ 
\vspace{10pt}
\\ \midrule
$K_1(1270)^+ \rightarrow [K^*(892)^0\pi^+]_{L=0}$ 	 & 	 0.615408 	 & 	 0.015366 	 & 	 0.591736 	 & 	 0.0261907 	 & 	 51.640326 	 & 	 0.88886642 	  \\ 
$K_1(1270)^+ \rightarrow [(\rho-\omega)^0K^+]_{L=0}$ 	 & 	 \multicolumn{2}{c}{1 (fixed)} 	 & 	 \multicolumn{2}{c}{0 (fixed)} 	 & 	 48.325553 	 & 	 1.8180199 	  \\ 
$K_1(1270)^+ \rightarrow [K^+\pi^-]_{L=0}\pi^+$ 	 & 	 0.576714 	 & 	 0.0263198 	 & 	 -1.8869 	 & 	 0.0434435 	 & 	 5.3450904 	 & 	 0.44008312 	  \\ 
$K_1(1270)^+ \rightarrow [K^*(892)^0\pi^+]_{L=2}$ 	 & 	 0.917374 	 & 	 0.0453398 	 & 	 -2.55509 	 & 	 0.0403779 	 & 	 2.059334 	 & 	 0.16776693 	  \\ 
$K_1(1270)^+ \rightarrow [\rho(1450)^0K^+]_{L=0}$ 	 & 	 0.425259 	 & 	 0.065613 	 & 	 -2.28657 	 & 	 0.110361 	 & 	 1.1166302 	 & 	 0.37955632 	  \\ 
\midrule
 	& \multicolumn{4}{r}{Sum of fit fractions} 	 &  	 108.487 	 & 	 2.079 	  \\ 
\vspace{10pt}
\\ \midrule
$K_1(1400)^+ \rightarrow [K^*(892)^0\pi^+]_{L=0}$ 	 & 	 \multicolumn{2}{c}{1 (fixed)} 	 & 	 \multicolumn{2}{c}{0 (fixed)} 	 & 	 \multicolumn{2}{c}{100}	  \\ 
\bottomrule
 	%& \multicolumn{4}{r}{Sum of fit fractions} 	 &  	 100.000 	 & 	 0.000 	  \\ 
\end{tabular}
}
\npnoround

\end{table}

\begin{table}[h!]
\centering
\caption{\label{tab:alt_mod_MoreAmps}Modulus and phase of the fit parameters along with the fit fractions of the amplitudes included in the alternative model using five additional amplitudes.}
% !TEX root = ./ms.tex

\fontsize{8pt}{1pt}\selectfont
\npdecimalsign{.}
\nprounddigits{2}
\resizebox{\textwidth}{!}{
\begin{tabular}{ln{2}{3}@{\,\( \pm \)\,}n{2}{2}n{3}{3}@{\,\( \pm \)\,}n{2}{2}n{3}{3}@{\,\( \pm \)\,}n{2}{2}}
\toprule
{Amplitude} & \multicolumn{2}{c}{$|c_k|$} & \multicolumn{2}{c}{$\arg(c_k)$ [rad]} & \multicolumn{2}{c}{Fit fraction [\%]} \\ \midrule
$D^0 \rightarrow [\phi(1020)(\rho-\omega)^0]_{L=0}$ 	 & 	 \multicolumn{2}{c}{1 (fixed)} 	 & 	 \multicolumn{2}{c}{0 (fixed)} 	 & 	 24.121617 	 & 	 0.39570579 	  \\ 
$D^0 \rightarrow K_1(1400)^+K^-$ 	 & 	 0.64542 	 & 	 0.010062 	 & 	 0.988042 	 & 	 0.0192851 	 & 	 19.336337 	 & 	 0.46511436 	 \\ 
$D^0 \rightarrow K_1(1270)^+K^-$ 	 & 	 0.45487 	 & 	 0.00981214 	 & 	 2.06958 	 & 	 0.0233382 	 & 	 19.275503 	 & 	 0.38871141 	 \\ 
$D^0 \rightarrow [K^-\pi^+]_{L=0}[K^+\pi^-]_{L=0}$ 	 & 	 0.285882 	 & 	 0.00388071 	 & 	 -0.573021 	 & 	 0.0151434 	 & 	 19.204344 	 & 	 0.36033103 	 \\ 
$D^0 \rightarrow [K^*(892)^0\Kstarb(892)^0]_{L=0}$ 	 & 	 0.278455 	 & 	 0.00411141 	 & 	 -0.291271 	 & 	 0.0156528 	 & 	 9.3576685 	 & 	 0.20827564 	 \\ 
$D^0 \rightarrow K^*(1680)^0[K^-\pi^+]_{L=0}$ 	 & 	 2.17925 	 & 	 0.0343651 	 & 	 0.448685 	 & 	 0.0162693 	 & 	 6.2195171 	 & 	 0.14598972 	 \\ 
$D^0 \rightarrow [K^*(892)^0\Kstarb(892)^0]_{L=1}$ 	 & 	 0.26236 	 & 	 0.00499996 	 & 	 1.20633 	 & 	 0.0209841 	 & 	 4.8139411 	 & 	 0.15794928 	 \\ 
$D^0 \rightarrow K_1(1270)^-K^+$ 	 & 	 0.218546 	 & 	 0.00585617 	 & 	 2.14051 	 & 	 0.0268918 	 & 	 4.4503541 	 & 	 0.17864884 	 \\ 
$D^0 \rightarrow [K^+K^-]_{L=0}[\pi^+\pi^-]_{L=0}$ 	 & 	 0.13705 	 & 	 0.0036339 	 & 	 -2.3931 	 & 	 0.0284166 	 & 	 4.1829895 	 & 	 0.20681014 	 \\ 
$D^0 \rightarrow K_1(1400)^-K^+$ 	 & 	 0.25222 	 & 	 0.0087128 	 & 	 0.0199881 	 & 	 0.0401512 	 & 	 2.9439312 	 & 	 0.19187276 	 \\ 
$D^0 \rightarrow [K^*(1680)^0\Kstarb(892)^0]_{L=0}$ 	 & 	 0.815751 	 & 	 0.0230352 	 & 	 2.99528 	 & 	 0.0293859 	 & 	 2.7894206 	 & 	 0.14824161 	 \\ 
$D^0 \rightarrow [\Kstarb(1680)^0K^*(892)^0]_{L=1}$ 	 & 	 0.989004 	 & 	 0.0220325 	 & 	 -2.76573 	 & 	 0.0218354 	 & 	 2.6419193 	 & 	 0.10595658 	 \\ 
$D^0 \rightarrow \Kstarb(1680)^0[K^+\pi^-]_{L=0}$ 	 & 	 1.31693 	 & 	 0.0281168 	 & 	 1.11764 	 & 	 0.0238593 	 & 	 2.4361647 	 & 	 0.094293284 	 \\ 
$D^0 \rightarrow [\phi(1020)(\rho-\omega)^0]_{L=2}$ 	 & 	 1.29123 	 & 	 0.0312075 	 & 	 0.559396 	 & 	 0.0235695 	 & 	 2.2525715 	 & 	 0.078302885 	 \\ 
$D^0 \rightarrow [K^*(892)^0\Kstarb(892)^0]_{L=2}$ 	 & 	 0.667433 	 & 	 0.0190099 	 & 	 2.82693 	 & 	 0.0281673 	 & 	 1.7309724 	 & 	 0.092491513 	 \\ 
$D^0 \rightarrow \phi(1020)[\pi^+\pi^-]_{L=0}$ 	 & 	 0.0473439 	 & 	 0.00148884 	 & 	 -1.69911 	 & 	 0.0404274 	 & 	 1.4278429 	 & 	 0.087805283 	 \\ 
$D^0 \rightarrow [K^*(1680)^0\Kstarb(892)^0]_{L=1}$ 	 & 	 0.692637 	 & 	 0.0219157 	 & 	 0.153285 	 & 	 0.0331225 	 & 	 1.2965776 	 & 	 0.077407629 	 \\ 
$D^0 \rightarrow [\phi(1020)\rho(1450)^0]_{L=1}$ 	 & 	 0.755969 	 & 	 0.03498 	 & 	 1.2091 	 & 	 0.0378654 	 & 	 0.98231922 	 & 	 0.088263788 	 \\ 
$D^0 \rightarrow a_0(980)^0f_2(1270)^0$ 	 & 	 1.5188 	 & 	 0.0602023 	 & 	 0.327827 	 & 	 0.0387701 	 & 	 0.70202994 	 & 	 0.054142332 	 \\ 
$D^0 \rightarrow a_1(1260)^+\pi^-$ 	 & 	 0.195837 	 & 	 0.0110099 	 & 	 -2.86437 	 & 	 0.065229 	 & 	 0.50156151 	 & 	 0.057279685 	 \\ 
$D^0 \rightarrow a_1(1260)^-\pi^+$ 	 & 	 0.185361 	 & 	 0.0136087 	 & 	 0.216299 	 & 	 0.0608382 	 & 	 0.44792496 	 & 	 0.062566362 	 \\ 
$D^0 \rightarrow [K^*(1680)^0\Kstarb(892)^0]_{L=2}$ 	 & 	 1.41216 	 & 	 0.0904412 	 & 	 -2.44563 	 & 	 0.0717729 	 & 	 0.44710176 	 & 	 0.056829311 	 \\ 
$D^0 \rightarrow [\phi(1020)(\rho-\omega)^0]_{L=1}$ 	 & 	 0.161418 	 & 	 0.0113299 	 & 	 0.252539 	 & 	 0.0702882 	 & 	 0.43847659 	 & 	 0.049586023 	 \\ 
$D^0 \rightarrow [K^+K^-]_{L=0}(\rho-\omega)^0$ 	 & 	 0.206486 	 & 	 0.0150238 	 & 	 3.01701 	 & 	 0.0816987 	 & 	 0.30245776 	 & 	 0.037516553 	 \\ 
$D^0 \rightarrow [\phi(1020)f_2(1270)^0]_{L=1}$ 	 & 	 1.41473 	 & 	 0.0944078 	 & 	 1.70307 	 & 	 0.0593527 	 & 	 0.19266141 	 & 	 0.025211476 	 \\ 
$D^0 \rightarrow [K^*(892)^0\Kbar{}^*_2(1430)^0]_{L=1}$ 	 & 	 1.44297 	 & 	 0.0902868 	 & 	 2.0998 	 & 	 0.0733804 	 & 	 0.15088584 	 & 	 0.019001292 	 \\ 
$D^0 \rightarrow [K_2^*(1430)^0\Kbar{}^*_2(1430)^0]_{L=0}$ 	 & 	 6.26565 	 & 	 0.571855 	 & 	 1.65749 	 & 	 0.0958601 	 & 	 0.10631329 	 & 	 0.019221276 	 \\ 
$D^0 \rightarrow [f_2(1270)^0f_2(1270)^0]_{L=0}$ 	 & 	 0.781431 	 & 	 0.083163 	 & 	 -1.55211 	 & 	 0.106368 	 & 	 0.079093553 	 & 	 0.016818499 	 \\ 
$D^0 \rightarrow [\Kstarb(892)^0K_2^*(1430)^0]_{L=2}$ 	 & 	 0.910558 	 & 	 0.111148 	 & 	 -0.435158 	 & 	 0.100306 	 & 	 0.048994601 	 & 	 0.011930036 	 \\ 
$D^0 \rightarrow [\phi(1020)f_2(1270)^0]_{L=2}$ 	 & 	 0.676108 	 & 	 0.0785204 	 & 	 -1.14287 	 & 	 0.101539 	 & 	 0.041301864 	 & 	 0.0095908973 	 \\ 
\midrule

 	& \multicolumn{4}{r}{Sum of fit fractions} 	 &  	 132.923 	 & 	 0.976 \\ 
	& \multicolumn{6}{r}{$\chisqndf ~~~ 9092/8113=1.12$} 	  	  	 \\

\vspace{10pt} \\ \midrule

$a_1(1260)^+ \rightarrow [\phi(1020)\pi^+]_{L=0}$ 	 & 	 \multicolumn{2}{c}{1 (fixed)} 	 & 	 \multicolumn{2}{c}{0 (fixed)} 	 &  	 \multicolumn{2}{c}{100} 	 \\ 
\midrule
 	%& \multicolumn{4}{r}{Sum of fit fractions} 	 &  	 \multicolumn{2}{c}{100} 	 \\ 
\vspace{10pt}
\\ \midrule
$K_1(1270)^+ \rightarrow [K^*(892)^0\pi^+]_{L=0}$ 	 & 	 0.645704 	 & 	 0.0172668 	 & 	 0.562136 	 & 	 0.0268185 	 & 	 53.373236 	 & 	 0.89269383 	  \\ 
$K_1(1270)^+ \rightarrow [(\rho-\omega)^0K^+]_{L=0}$ 	 & 	 \multicolumn{2}{c}{1 (fixed)} 	 & 	 \multicolumn{2}{c}{0 (fixed)} 	 & 	 45.449703 	 & 	 1.8840616 	  \\ 
$K_1(1270)^+ \rightarrow [K^+\pi^-]_{L=0}\pi^+$ 	 & 	 0.709746 	 & 	 0.031046 	 & 	 -1.82847 	 & 	 0.0433408 	 & 	 7.6003563 	 & 	 0.55818059 	  \\ 
$K_1(1270)^+ \rightarrow [K^*(892)^0\pi^+]_{L=2}$ 	 & 	 0.967603 	 & 	 0.0479223 	 & 	 -2.64775 	 & 	 0.0399825 	 & 	 2.15091 	 & 	 0.16706805 	  \\ 
$K_1(1270)^+ \rightarrow [\rho(1450)^0K^+]_{L=0}$ 	 & 	 0.474556 	 & 	 0.0701173 	 & 	 -1.78963 	 & 	 0.125271 	 & 	 1.3054797 	 & 	 0.42211521 	  \\ 
\midrule
 	& \multicolumn{4}{r}{Sum of fit fractions} 	 &  	 109.880 	 & 	 2.186 	  \\ 
\vspace{10pt}
\\ \midrule
$K_1(1400)^+ \rightarrow [K^*(892)^0\pi^+]_{L=0}$ 	 & 	 \multicolumn{2}{c}{1 (fixed)} 	 & 	 \multicolumn{2}{c}{0 (fixed)} 	 & 	 102.92474 	 & 	 0.27336112 	  \\ 
$K_1(1400)^+ \rightarrow [\rho(1450)^0K^+]_{L=2}$ 	 & 	 1.82033 	 & 	 0.164718 	 & 	 2.75758 	 & 	 0.0910822 	 & 	 0.84038346 	 & 	 0.15369444 	  \\ 
\midrule
 	& \multicolumn{4}{r}{Sum of fit fractions} 	 &  	 103.765 	 & 	 0.397 	  \\ \bottomrule
\end{tabular}
}
\npnoround

\end{table}

\begin{table}[h!]
\centering
\caption{\label{tab:alt_mod_rho1450}Modulus and phase of the fit parameters along with the fit fractions of the amplitudes of the alternative model that includes the amplitude \decay{\Dz}{\rho(1450)^0\rho(770)^0} in $D$-wave.}
% !TEX root = ./ms.tex

\fontsize{8pt}{1pt}\selectfont
\npdecimalsign{.}
\nprounddigits{2}
\resizebox{\textwidth}{!}{
\begin{tabular}{ln{2}{3}@{\,\( \pm \)\,}n{2}{2}n{3}{3}@{\,\( \pm \)\,}n{2}{2}n{3}{3}@{\,\( \pm \)\,}n{2}{2}}
\toprule
{Amplitude} & \multicolumn{2}{c}{$|c_k|$} & \multicolumn{2}{c}{$\arg(c_k)$ [rad]} & \multicolumn{2}{c}{Fit fraction [\%]} \\ \midrule
$D^0 \rightarrow [\phi(1020)(\rho-\omega)^0]_{L=0}$ 	 & 	 \multicolumn{2}{c}{1 (fixed)} 	 & 	 \multicolumn{2}{c}{0 (fixed)} 	 & 	 23.83373 	 & 	 0.1570876 	  \\ 
$D^0 \rightarrow K_1(1270)^+K^-$ 	 & 	 0.459828 	 & 	 0.00721685 	 & 	 2.02197 	 & 	 0.0203846 	 & 	 19.260998 	 & 	 0.29765105 	 \\ 
$D^0 \rightarrow K_1(1400)^+K^-$ 	 & 	 0.638726 	 & 	 0.00485141 	 & 	 1.01648 	 & 	 0.0132119 	 & 	 19.228129 	 & 	 0.31479691 	 \\ 
$D^0 \rightarrow [K^-\pi^+]_{L=0}[K^+\pi^-]_{L=0}$ 	 & 	 0.276141 	 & 	 0.00217809 	 & 	 -0.579083 	 & 	 0.0117324 	 & 	 17.676118 	 & 	 0.25995292 	 \\ 
$D^0 \rightarrow [K^*(892)^0\Kstarb(892)^0]_{L=0}$ 	 & 	 0.279819 	 & 	 0.00194039 	 & 	 -0.269624 	 & 	 0.0110004 	 & 	 9.3220149 	 & 	 0.12593606 	 \\ 
$D^0 \rightarrow K^*(1680)^0[K^-\pi^+]_{L=0}$ 	 & 	 2.28269 	 & 	 0.0234841 	 & 	 0.438632 	 & 	 0.0112372 	 & 	 6.7318435 	 & 	 0.13134757 	 \\ 
$D^0 \rightarrow [K^*(892)^0\Kstarb(892)^0]_{L=1}$ 	 & 	 0.266119 	 & 	 0.00230014 	 & 	 1.22111 	 & 	 0.0144543 	 & 	 4.886039 	 & 	 0.0831931 	 \\ 
$D^0 \rightarrow K_1(1270)^-K^+$ 	 & 	 0.218223 	 & 	 0.00384952 	 & 	 2.13142 	 & 	 0.0225981 	 & 	 4.3386926 	 & 	 0.14704533 	 \\ 
$D^0 \rightarrow [K^*(1680)^0\Kstarb(892)^0]_{L=0}$ 	 & 	 0.859036 	 & 	 0.0135059 	 & 	 2.98693 	 & 	 0.0189078 	 & 	 3.0515438 	 & 	 0.093420856 	 \\ 
$D^0 \rightarrow [K^+K^-]_{L=0}[\pi^+\pi^-]_{L=0}$ 	 & 	 0.116964 	 & 	 0.00213362 	 & 	 -2.42375 	 & 	 0.025567 	 & 	 3.0056334 	 & 	 0.11282555 	 \\ 
$D^0 \rightarrow K_1(1400)^-K^+$ 	 & 	 0.242355 	 & 	 0.00670476 	 & 	 0.0997417 	 & 	 0.0217981 	 & 	 2.7601389 	 & 	 0.15745544 	 \\ 
$D^0 \rightarrow [\Kstarb(1680)^0K^*(892)^0]_{L=1}$ 	 & 	 0.997146 	 & 	 0.0155353 	 & 	 -2.75103 	 & 	 0.0144808 	 & 	 2.649346 	 & 	 0.081801345 	 \\ 
$D^0 \rightarrow \Kstarb(1680)^0[K^+\pi^-]_{L=0}$ 	 & 	 1.34282 	 & 	 0.0235172 	 & 	 1.05045 	 & 	 0.0185758 	 & 	 2.4986823 	 & 	 0.083747368 	 \\ 
$D^0 \rightarrow [\phi(1020)(\rho-\omega)^0]_{L=2}$ 	 & 	 1.24993 	 & 	 0.0211213 	 & 	 0.492133 	 & 	 0.0201977 	 & 	 2.0750573 	 & 	 0.06354699 	 \\ 
$D^0 \rightarrow [K^*(892)^0\Kstarb(892)^0]_{L=2}$ 	 & 	 0.665104 	 & 	 0.0114796 	 & 	 2.79096 	 & 	 0.0198521 	 & 	 1.6957114 	 & 	 0.058342811 	 \\ 
$D^0 \rightarrow \phi(1020)[\pi^+\pi^-]_{L=0}$ 	 & 	 0.0491378 	 & 	 0.000785439 	 & 	 -1.71805 	 & 	 0.025292 	 & 	 1.517342 	 & 	 0.045309872 	 \\ 
$D^0 \rightarrow [K^*(1680)^0\Kstarb(892)^0]_{L=1}$ 	 & 	 0.732719 	 & 	 0.0152526 	 & 	 0.146393 	 & 	 0.020329 	 & 	 1.4313971 	 & 	 0.058982489 	 \\ 
$D^0 \rightarrow [\phi(1020)\rho(1450)^0]_{L=1}$ 	 & 	 0.771399 	 & 	 0.014963 	 & 	 1.17598 	 & 	 0.0213875 	 & 	 1.0090239 	 & 	 0.038060585 	 \\ 
$D^0 \rightarrow a_0(980)^0f_2(1270)^0$ 	 & 	 1.4907 	 & 	 0.0493767 	 & 	 0.295585 	 & 	 0.0323211 	 & 	 0.66717221 	 & 	 0.044777404 	 \\ 
$D^0 \rightarrow [K^*(1680)^0\Kstarb(892)^0]_{L=2}$ 	 & 	 1.48923 	 & 	 0.0681076 	 & 	 -2.40331 	 & 	 0.0489528 	 & 	 0.4905224 	 & 	 0.043981589 	 \\ 
$D^0 \rightarrow a_1(1260)^+\pi^-$ 	 & 	 0.187356 	 & 	 0.006489 	 & 	 -2.79225 	 & 	 0.0304505 	 & 	 0.4528627 	 & 	 0.031410587 	 \\ 
$D^0 \rightarrow a_1(1260)^-\pi^+$ 	 & 	 0.182257 	 & 	 0.00725351 	 & 	 0.168101 	 & 	 0.034762 	 & 	 0.42720476 	 & 	 0.034229808 	 \\ 
$D^0 \rightarrow [\phi(1020)(\rho-\omega)^0]_{L=1}$ 	 & 	 0.157656 	 & 	 0.0040949 	 & 	 0.277188 	 & 	 0.0347511 	 & 	 0.41832676 	 & 	 0.021598674 	 \\ 
$D^0 \rightarrow [K^+K^-]_{L=0}(\rho-\omega)^0$ 	 & 	 0.189941 	 & 	 0.0114971 	 & 	 3.03342 	 & 	 0.0622037 	 & 	 0.25891986 	 & 	 0.029985324 	 \\ 
$D^0 \rightarrow [\rho(1450)^0(\rho-\omega)^0]_{L=2}$ 	 & 	 -0.662026 	 & 	 0.0673026 	 & 	 0.331396 	 & 	 0.119496 	 & 	 0.21550782 	 & 	 0.028814209 	 \\ 
$D^0 \rightarrow [\phi(1020)f_2(1270)^0]_{L=1}$ 	 & 	 1.40232 	 & 	 0.0717628 	 & 	 1.71387 	 & 	 0.0442583 	 & 	 0.18674137 	 & 	 0.019173645 	 \\ 
$D^0 \rightarrow [K^*(892)^0\Kbar{}^*_2(1430)^0]_{L=1}$ 	 & 	 1.49871 	 & 	 0.0788389 	 & 	 2.01865 	 & 	 0.0621575 	 & 	 0.1605708 	 & 	 0.016858313 	 \\ 
\midrule
 	& \multicolumn{4}{r}{Sum of fit fractions} 	 &  	 130.249 	 & 	 0.652 \\ 
	& \multicolumn{6}{r}{$\chisqndf ~~~ 9099/8119=1.12$} 	  	  	 \\

\vspace{10pt} \\ \midrule

$a_1(1260)^+ \rightarrow [\phi(1020)\pi^+]_{L=0}$ 	 & 	 \multicolumn{2}{c}{1 (fixed)} 	 & 	 \multicolumn{2}{c}{0 (fixed)} 	 &  	 \multicolumn{2}{c}{100} 	 \\ 
\midrule
 	%& \multicolumn{4}{r}{Sum of fit fractions} 	 &  	 \multicolumn{2}{c}{100} 	 \\ 
\vspace{10pt}
\\ \midrule
$K_1(1270)^+ \rightarrow [K^*(892)^0\pi^+]_{L=0}$ 	 & 	 0.628364 	 & 	 0.0121309 	 & 	 0.607753 	 & 	 0.0222927 	 & 	 50.994063 	 & 	 0.65968289 	  \\ 
$K_1(1270)^+ \rightarrow [(\rho-\omega)^0K^+]_{L=0}$ 	 & 	 \multicolumn{2}{c}{1 (fixed)} 	 & 	 \multicolumn{2}{c}{0 (fixed)} 	 & 	 45.766688 	 & 	 1.423116 	  \\ 
$K_1(1270)^+ \rightarrow [K^+\pi^-]_{L=0}\pi^+$ 	 & 	 0.57795 	 & 	 0.0184997 	 & 	 -1.69534 	 & 	 0.0417205 	 & 	 5.0845156 	 & 	 0.30337485 	  \\ 
$K_1(1270)^+ \rightarrow [K^*(892)^0\pi^+]_{L=2}$ 	 & 	 0.955885 	 & 	 0.0377619 	 & 	 -2.54948 	 & 	 0.0335158 	 & 	 2.1177756 	 & 	 0.13811833 	  \\ 
$K_1(1270)^+ \rightarrow [\rho(1450)^0K^+]_{L=0}$ 	 & 	 0.229975 	 & 	 0.0619254 	 & 	 -2.54266 	 & 	 0.175243 	 & 	 0.30931245 	 & 	 0.17548856 	  \\ 
\midrule
 	& \multicolumn{4}{r}{Sum of fit fractions} 	 &  	 104.272 	 & 	 1.539 	  \\ 
\vspace{10pt}
\\ \midrule
$K_1(1400)^+ \rightarrow [K^*(892)^0\pi^+]_{L=0}$ 	 & 	 \multicolumn{2}{c}{1 (fixed)} 	 & 	 \multicolumn{2}{c}{0 (fixed)} 	 &  	 \multicolumn{2}{c}{100} 	  \\ 
\bottomrule
 %	& \multicolumn{4}{r}{Sum of fit fractions} 	 &  	 100.000 	 & 	 0.000 	  \\ 
\end{tabular}
}
\npnoround

\end{table}

\clearpage
\section{Systematic uncertainties}
\label{app:syst_tables}
All the systematic uncertainties from all the sources considered on every amplitudes for the nominal and \CP-violation fits are listed in Tables~\ref{tab:syst:fitfracs}--\ref{tab:syst:fitpars_cpv}.

\begin{table}[ht!]
  \caption{\label{tab:syst:fitfracs}Statistical and systematic uncertainties (in \%) on the fit fractions. Values smaller than 0.0005\% are displayed as ``0.000''. The sources of systematic uncertainty are described in the text in the same order as shown in this table.}
  % !TEX root = ./ms.tex

\resizebox{\textwidth}{!}{
\begin{tabular}{l*{14}{S[table-auto-round,table-format=1.3]}}
\toprule
& &                     {Total} & {Sel.} & {Alt.} & {Alt.} & {RBW} & {Alt.} &  {Mass \&} & {Res.}  & {Sig.} & {Bkg} & {Mistag} & {Det.\,as.} & {Alt.}\\ 
{Amplitude} & {Stat.} &  {syst.} & {eff.} & {bkg 1} & {bkg 2} & {$\rho(770)^0$} & {$S$-wave} & {width}  & {radius} & {bias} & {bias} & {bias} & {bias} & {models} \\ \midrule
$D^0 \rightarrow [\phi(1020)(\rho-\omega)^0]_{L=0}$ 	 &  	 0.382413 	 & 	 0.495365 	 & 	 0.183290 	 & 	 0.054488 	 & 	 0.045825 	 & 	 0.225826 	 & 	 0.095411 	 & 	 0.110020 	 & 	 0.074870 	 & 	 0.270781 	 & 	 0.121652 	 & 	 0.020015 	  & 	 0.017667 	  & 	 0.200194  \\
$D^0 \rightarrow K_1(1400)^+K^-$ 	 &  	 0.599973 	 & 	 1.462856 	 & 	 0.136514 	 & 	 0.053065 	 & 	 0.316164 	 & 	 0.033253 	 & 	 0.160676 	 & 	 1.236382 	 & 	 0.286321 	 & 	 0.047652 	 & 	 0.101928 	 & 	 0.062447 	  & 	 0.030789 	  & 	 0.602898  \\
$D^0 \rightarrow [K^-\pi^+]_{L=0}[K^+\pi^-]_{L=0}$ 	 &  	 0.347204 	 & 	 0.935303 	 & 	 0.034312 	 & 	 0.065493 	 & 	 0.459436 	 & 	 0.040086 	 &  0.246267 	 & 	 0.192936 	 & 	 0.050883 	 & 	 0.051497 	 & 	 0.234954 	 & 	 0.060955 	  & 	 0.017941 	  & 	 0.703053  \\
$D^0 \rightarrow K_1(1270)^+K^-$ 	 &  	 0.521415 	 & 	 0.981951 	 & 	 0.031107 	 & 	 0.044791 	 & 	 0.085008 	 & 	 0.016184 	 & 	 0.139279 	 & 	 0.733875 	 & 	 0.116168 	 & 	 0.113109 	 & 	 0.150753 	 & 	 0.082895 	  & 	 0.029925 	  & 	 0.582248  \\
$D^0 \rightarrow [K^*(892)^0\Kstarb(892)^0]_{L=0}$ 	 &  	 0.207288 	 & 	 0.277487 	 & 	 0.023028 	 & 	 0.026655 	 & 	 0.037050 	 & 	 0.000524 	 & 	 0.126906 	 & 	 0.086931 	 & 	 0.047259 	 & 	 0.063634 	 & 	 0.045592 	 & 	 0.010697 	  & 	 0.013184 	  & 	 0.205118  \\
$D^0 \rightarrow K^*(1680)^0[K^-\pi^+]_{L=0}$ 	 &  	 0.148183 	 & 	 0.367847 	 & 	 0.009514 	 & 	 0.030830 	 & 	 0.023380 	 & 	 0.014072 	 & 	 0.216520 	 & 	 0.068787 	 & 	 0.050128 	 & 	 0.020511 	 & 	 0.071870 	 & 	 0.007079 	  & 	 0.007364 	  & 	 0.271496  \\
$D^0 \rightarrow [K^*(892)^0\Kstarb(892)^0]_{L=1}$ 	 &  	 0.155233 	 & 	 0.181313 	 & 	 0.008750 	 & 	 0.014643 	 & 	 0.062854 	 & 	 0.001433 	 &  0.043397 	 & 	 0.064177 	 & 	 0.057651 	 & 	 0.027975 	 & 	 0.075901 	 & 	 0.008923 	  & 	 0.007902 	  & 	 0.112336  \\
$D^0 \rightarrow K_1(1270)^-K^+$ 	 &  	 0.179533 	 & 	 0.405212 	 & 	 0.001392 	 & 	 0.021080 	 & 	 0.088954 	 & 	 0.016252 	 & 	 0.112200 	 & 	 0.147278 	 & 	 0.023013 	 & 	 0.018019 	 & 	 0.061136 	 & 	 0.064526 	  & 	 0.010524 	  & 	 0.335302  \\
$D^0 \rightarrow [K^+K^-]_{L=0}[\pi^+\pi^-]_{L=0}$ 	 &  	 0.167920 	 & 	 0.722720 	 & 	 0.002069 	 & 	 0.035669 	 & 	 0.033702 	 & 	 0.031319 	 &  0.293249 	 & 	 0.073700 	 & 	 0.037806 	 & 	 0.006595 	 & 	 0.046813 	 & 	 0.020598 	  & 	 0.008026 	  & 	 0.650655  \\
$D^0 \rightarrow K_1(1400)^-K^+$ 	 &  	 0.191621 	 & 	 0.393824 	 & 	 0.073669 	 & 	 0.029162 	 & 	 0.029023 	 & 	 0.003434 	 & 	 0.024741 	 & 	 0.157546 	 & 	 0.018308 	 & 	 0.011335 	 & 	 0.031487 	 & 	 0.055802 	  & 	 0.009334 	  & 	 0.343332  \\
$D^0 \rightarrow [K^*(1680)^0\Kstarb(892)^0]_{L=0}$ 	 &  	 0.147245 	 & 	 0.189209 	 & 	 0.095972 	 & 	 0.037451 	 & 	 0.019904 	 & 	 0.004034 	 & 	 0.108526 	 & 	 0.058931 	 & 	 0.024604 	 & 	 0.011484 	 & 	 0.013944 	 & 	 0.007656 	  & 	 0.008387 	  & 	 0.091994  \\
$D^0 \rightarrow [\Kstarb(1680)^0K^*(892)^0]_{L=1}$ 	 &  	 0.105464 	 & 	 0.093487 	 & 	 0.016460 	 & 	 0.012036 	 & 	 0.026446 	 & 	 0.000918 	 &  0.032176 	 & 	 0.035315 	 & 	 0.045688 	 & 	 0.013408 	 & 	 0.021230 	 & 	 0.006050 	  & 	 0.009169 	  & 	 0.050028  \\
$D^0 \rightarrow \Kstarb(1680)^0[K^+\pi^-]_{L=0}$ 	 &  	 0.091090 	 & 	 0.274605 	 & 	 0.004018 	 & 	 0.018391 	 & 	 0.071736 	 & 	 0.000252 	 &  0.174421 	 & 	 0.042111 	 & 	 0.013589 	 & 	 0.005433 	 & 	 0.056496 	 & 	 0.014725 	  & 	 0.006953 	  & 	 0.184500  \\
$D^0 \rightarrow [\phi(1020)(\rho-\omega)^0]_{L=2}$ 	 &  	 0.076328 	 & 	 0.076851 	 & 	 0.023065 	 & 	 0.008154 	 & 	 0.013033 	 & 	 0.062228 	 &  0.004788 	 & 	 0.012857 	 & 	 0.018625 	 & 	 0.005648 	 & 	 0.014328 	 & 	 0.004880 	  & 	 0.007116 	  & 	 0.020461  \\
$D^0 \rightarrow [K^*(892)^0\Kstarb(892)^0]_{L=2}$ 	 &  	 0.094854 	 & 	 0.099495 	 & 	 0.030484 	 & 	 0.009661 	 & 	 0.017186 	 & 	 0.013849 	 &  0.068873 	 & 	 0.052564 	 & 	 0.015660 	 & 	 0.015472 	 & 	 0.013228 	 & 	 0.004990 	  & 	 0.004577 	  & 	 0.013322  \\
$D^0 \rightarrow \phi(1020)[\pi^+\pi^-]_{L=0}$ 	 &  	 0.089974 	 & 	 0.325468 	 & 	 0.004739 	 & 	 0.009112 	 & 	 0.001628 	 & 	 0.010352 	 &  0.312876 	 & 	 0.021660 	 & 	 0.013648 	 & 	 0.012204 	 & 	 0.021582 	 & 	 0.005387 	  & 	 0.004735 	  & 	 0.080629  \\
$D^0 \rightarrow [K^*(1680)^0\Kstarb(892)^0]_{L=1}$ 	 &  	 0.079671 	 & 	 0.096300 	 & 	 0.000274 	 & 	 0.008398 	 & 	 0.043305 	 & 	 0.002867 	 &  0.022476 	 & 	 0.013189 	 & 	 0.020566 	 & 	 0.009233 	 & 	 0.029060 	 & 	 0.014352 	  & 	 0.004350 	  & 	 0.071154  \\
$D^0 \rightarrow [\phi(1020)\rho(1450)^0]_{L=1}$ 	 &  	 0.088579 	 & 	 0.045474 	 & 	 0.006742 	 & 	 0.007484 	 & 	 0.007488 	 & 	 0.019949 	 &  0.023289 	 & 	 0.013251 	 & 	 0.016698 	 & 	 0.005028 	 & 	 0.016692 	 & 	 0.005596 	  & 	 0.005705 	  & 	 0.012161  \\
$D^0 \rightarrow a_0(980)^0f_2(1270)^0$ 	 &  	 0.051620 	 & 	 0.082951 	 & 	 0.004846 	 & 	 0.008491 	 & 	 0.026220 	 & 	 0.001558 	 & 	 0.048195 	 & 	 0.027003 	 & 	 0.010249 	 & 	 0.002421 	 & 	 0.014882 	 & 	 0.012798 	  & 	 0.004157 	  & 	 0.050300  \\
$D^0 \rightarrow a_1(1260)^+\pi^-$ 	 &  	 0.054632 	 & 	 0.220050 	 & 	 0.001849 	 & 	 0.005410 	 & 	 0.002670 	 & 	 0.065900 	 & 	 0.206270 	 & 	 0.015847 	 & 	 0.018708 	 & 	 0.004226 	 & 	 0.007396 	 & 	 0.004638 	  & 	 0.004165 	  & 	 0.027918  \\
$D^0 \rightarrow a_1(1260)^-\pi^+$ 	 &  	 0.063139 	 & 	 0.156233 	 & 	 0.001824 	 & 	 0.008417 	 & 	 0.000092 	 & 	 0.050796 	 & 	 0.138631 	 & 	 0.018079 	 & 	 0.013034 	 & 	 0.002593 	 & 	 0.020934 	 & 	 0.006656 	  & 	 0.003275 	  & 	 0.039229  \\
$D^0 \rightarrow [\phi(1020)(\rho-\omega)^0]_{L=1}$ 	 &  	 0.049438 	 & 	 0.027821 	 & 	 0.001454 	 & 	 0.003342 	 & 	 0.003844 	 & 	 0.013837 	 &  0.010977 	 & 	 0.008515 	 & 	 0.013865 	 & 	 0.001812 	 & 	 0.005604 	 & 	 0.008646 	  & 	 0.002910 	  & 	 0.007165  \\
$D^0 \rightarrow [K^*(1680)^0\Kstarb(892)^0]_{L=2}$ 	 &  	 0.048007 	 & 	 0.058816 	 & 	 0.010238 	 & 	 0.006364 	 & 	 0.029849 	 & 	 0.003073 	 &  0.009383 	 & 	 0.017688 	 & 	 0.007999 	 & 	 0.002460 	 & 	 0.003589 	 & 	 0.003158 	  & 	 0.002549 	  & 	 0.043742  \\
$D^0 \rightarrow [K^+K^-]_{L=0}(\rho-\omega)^0$ 	 &  	 0.035559 	 & 	 0.053606 	 & 	 0.002885 	 & 	 0.006620 	 & 	 0.021616 	 & 	 0.000203 	 &  0.038600 	 & 	 0.014504 	 & 	 0.004333 	 & 	 0.003652 	 & 	 0.009700 	 & 	 0.003296 	  & 	 0.001754 	  & 	 0.022666  \\
$D^0 \rightarrow [\phi(1020)f_2(1270)^0]_{L=1}$ 	 &  	 0.024474 	 & 	 0.074613 	 & 	 0.002278 	 & 	 0.002632 	 & 	 0.007606 	 & 	 0.002300 	 & 	 0.073525 	 & 	 0.003309 	 & 	 0.004723 	 & 	 0.002224 	 & 	 0.001279 	 & 	 0.001520 	  & 	 0.001613 	  & 	 0.006418  \\
$D^0 \rightarrow [K^*(892)^0\Kbar{}^*_2(1430)^0]_{L=1}$ 	 &  	 0.020170 	 & 	 0.024367 	 & 	 0.002853 	 & 	 0.002331 	 & 	 0.004831 	 & 	 0.000579 	 & 	 0.001670 	 & 	 0.017451 	 & 	 0.008607 	 & 	 0.001273 	 & 	 0.005418 	 & 	 0.001458 	  & 	 0.000999 	  & 	 0.011874  \\ \midrule
$K_1(1270)^+ \rightarrow [K^*(892)^0\pi^+]_{L=0}$ 	 &  	 1.057553 	 & 	 3.212825 	 & 	 0.050447 	 & 	 0.099521 	 & 	 0.000206 	 & 	 0.022143 	 & 	 0.443024 	 & 	 1.916530 	 & 	 0.558919 	 & 	 0.129007 	 & 	 0.298561 	 & 	 0.054065 	  & 	 0.060078 	  & 	 2.452598  \\
$K_1(1270)^+ \rightarrow [(\rho-\omega)^0K^+]_{L=0}$ 	 &  	 1.993494 	 & 	 4.352159 	 & 	 0.299896 	 & 	 0.253483 	 & 	 1.152413 	 & 	 1.780015 	 & 	 2.301399 	 & 	 2.109990 	 & 	 0.961102 	 & 	 0.145024 	 & 	 0.757928 	 & 	 0.095461 	  & 	 0.098756 	  & 	 1.733211  \\
$K_1(1270)^+ \rightarrow [K^+\pi^-]_{L=0}\pi^+$ 	 &  	 0.483700 	 & 	 1.660004 	 & 	 0.015461 	 & 	 0.073413 	 & 	 0.040613 	 & 	 0.045251 	 &  0.188008 	 & 	 0.302506 	 & 	 0.148413 	 & 	 0.078238 	 & 	 0.023917 	 & 	 0.029418 	  & 	 0.070254 	  & 	 1.607766  \\
$K_1(1270)^+ \rightarrow [K^*(892)^0\pi^+]_{L=2}$ 	 &  	 0.169411 	 & 	 0.195236 	 & 	 0.034679 	 & 	 0.014641 	 & 	 0.036897 	 & 	 0.000041 	 &  0.041415 	 & 	 0.138981 	 & 	 0.025871 	 & 	 0.005941 	 & 	 0.023786 	 & 	 0.009470 	  & 	 0.008429 	  & 	 0.113475  \\
$K_1(1270)^+ \rightarrow [\rho(1450)^0K^+]_{L=0}$ 	 &  	 0.472498 	 & 	 1.040994 	 & 	 0.138234 	 & 	 0.071870 	 & 	 0.365289 	 & 	 0.243777 	 & 	 0.446240 	 & 	 0.334906 	 & 	 0.199380 	 & 	 0.041636 	 & 	 0.168825 	 & 	 0.025552 	  & 	 0.024595 	  & 	 0.695696  \\ \midrule
$D^0 \rightarrow [\phi(1020)\rho(770)^0]_{L=0}$ 	 &  	 0.462983 	 & 	 0.275464 	 & 	 0.024474 	 & 	 0.018774 	 & 	 0.085708 	 &  0.160310 	 & 	 0.022576 	 & 	 0.159224 	 & 	 0.058518 	 & 	 0.036443 	 & 	 0.043164 	 & 	 0.024022 	  & 	 0.026000 	  & 	 0.090294  \\
$D^0 \rightarrow [\phi(1020)\omega(782)]_{L=0}$ 	 &  	 0.106065 	 & 	 0.040821 	 & 	 0.001592 	 & 	 0.005423 	 & 	 0.007608 	 &  0.021115 	 & 	 0.001275 	 & 	 0.015392 	 & 	 0.013978 	 & 	 0.020292 	 & 	 0.006292 	 & 	 0.007482 	  & 	 0.005603 	  & 	 0.012572  \\ \midrule
$D^0 \rightarrow [\phi(1020)\rho(770)^0]_{L=1}$ 	 &  	 4.107347 	 & 	 1.695753 	 & 	 0.332794 	 & 	 0.122894 	 & 	 0.436097 	 & 	 0.727823 	 & 	 0.745942 	 & 	 0.367198 	 & 	 0.728110 	 & 	 0.306365 	 & 	 0.450393 	 & 	 0.326052 	  & 	 0.455936 	  & 	 0.444417  \\
$D^0 \rightarrow [\phi(1020)\omega(782)]_{L=1}$ 	 &  	 1.576747 	 & 	 0.515257 	 & 	 0.053052 	 & 	 0.049453 	 & 	 0.104101 	 & 	 0.138004 	 & 	 0.166027 	 & 	 0.116458 	 & 	 0.184117 	 & 	 0.112442 	 & 	 0.179050 	 & 	 0.131131 	  & 	 0.239695 	  & 	 0.189658  \\ \midrule
$D^0 \rightarrow [\phi(1020)\rho(770)^0]_{L=2}$ 	 &  	 1.689922 	 & 	 0.777958 	 & 	 0.038489 	 & 	 0.068474 	 & 	 0.076026 	 & 	 0.424433 	 & 	 0.309921 	 & 	 0.254823 	 & 	 0.162158 	 & 	 0.060134 	 & 	 0.092764 	 & 	 0.124036 	  & 	 0.121440 	  & 	 0.428358  \\
$D^0 \rightarrow [\phi(1020)\omega(782)]_{L=2}$ 	 &  	 0.270083 	 & 	 0.116228 	 & 	 0.004706 	 & 	 0.009609 	 & 	 0.015410 	 & 	 0.038336 	 & 	 0.019125 	 & 	 0.023988 	 & 	 0.020466 	 & 	 0.041594 	 & 	 0.017663 	 & 	 0.018363 	  & 	 0.024598 	  & 	 0.085691  \\ \midrule
$D^0 \rightarrow [K^+K^-]_{L=0}\rho(770)^0$ 	 &  	 5.894922 	 & 	 3.492282 	 & 	 0.054122 	 & 	 0.529524 	 & 	 0.665094 	 & 	 0.237542 	 & 	 1.470694 	 & 	 1.560871 	 & 	 0.680870 	 & 	 0.368447 	 & 	 0.891844 	 & 	 0.309575 	  & 	 0.317384 	  & 	 2.285471  \\
$D^0 \rightarrow [K^+K^-]_{L=0}\omega(782)$ 	 &  	 3.258563 	 & 	 3.641535 	 & 	 0.799345 	 & 	 0.272267 	 & 	 0.712889 	 & 	 0.056485 	 & 	 1.000465 	 & 	 0.735991 	 & 	 0.201015 	 & 	 0.261199 	 & 	 0.415823 	 & 	 0.202018 	  & 	 0.174299 	  & 	 3.184486  \\ \midrule
$K_1(1270)^+ \rightarrow [\rho(770)^0K^+]_{L=0}$ 	 &  	 1.980863 	 & 	 3.805808 	 & 	 0.188730 	 & 	 0.229254 	 & 	 0.729180 	 &  0.630120 	 & 	 0.215810 	 & 	 1.007494 	 & 	 0.585576 	 & 	 0.065231 	 & 	 0.152097 	 & 	 0.574402 	  & 	 0.106358 	  & 	 3.419661  \\
$K_1(1270)^+ \rightarrow [\omega(782)K^+]_{L=0}$ 	 &  	 0.220479 	 & 	 0.190984 	 & 	 0.014988 	 & 	 0.014623 	 & 	 0.040151 	 & 	 0.081778 	 & 	 0.148930 	 & 	 0.035278 	 & 	 0.034738 	 & 	 0.011079 	 & 	 0.047184 	 & 	 0.011277 	  & 	 0.010776 	  & 	 0.022649  \\
\bottomrule
\end{tabular}
}

\end{table}

\begin{table}
  \caption{\label{tab:syst:fitpars}Statistical and systematic uncertainties on the fit parameters shown for all floating components. Values smaller than 0.0005 are displayed as ``0.000''. The sources of systematic uncertainty are described in the text in the same order as shown in this table. For each amplitude, the first value quoted is the modulus and the second is the phase of the complex fit parameter.}
  % !TEX root = ./ms.tex

\resizebox{\textwidth}{!}{
\begin{tabular}{l*{14}{S[table-auto-round,table-format=1.3]}}
  \toprule
  & &                     {Total} & {Sel.} & {Alt.} & {Alt.} & {RBW} & {Alt.} &  {Mass \&} & {Res.}  & {Sig.} & {Bkg} & {Mistag} & {Det.\,as.} & {Alt.}\\
  {Amplitude} & {Stat.} &  {syst.} & {eff.} & {bkg 1} & {bkg 2} & {$\rho(770)^0$} & {$S$-wave} & {width}  & {radius} & {bias} & {bias} & {bias} & {bias} & {models} \\ \midrule
\multirow{ 2}{*}{$D^0 \rightarrow K_1(1400)^+K^-$} 	  	 & 	 0.011259 	 & 	 0.031132 	 & 	 0.004791 	 & 	 0.001259 	 & 	 0.006264 	 &  0.002081 	 & 	 0.001534 	 & 	 0.018360 	 & 	 0.022466 	 & 	 0.000616 	 & 	 0.002436 	 & 	 0.000787 	 & 	 0.000561 	 & 	 0.007049  \\
 	 & 	 0.021671 	 & 	 0.053434 	 & 	 0.006560 	 & 	 0.001904 	 & 	 0.001100 	 & 	 0.014370 	 & 	 0.011450 	 & 	 0.047069 	 & 	 0.007097 	 & 	 0.000749 	 & 	 0.004037 	 & 	 0.001075 	 & 	 0.001351 	 & 	 0.013567  \\
\multirow{ 2}{*}{$D^0 \rightarrow [K^-\pi^+]_{L=0}[K^+\pi^-]_{L=0}$} 	  	 & 	 0.003746 	 & 	 0.007722 	 & 	 0.000861 	 & 	 0.000560 	 & 	 0.003074 	 & 	 0.000991 	 & 	 0.001337 	 & 	 0.001738 	 & 	 0.005021 	 & 	 0.000247 	 & 	 0.002001 	 & 	 0.000487 	 & 	 0.000176 	 &  0.003714  \\
 	 & 	 0.015312 	 & 	 0.103549 	 & 	 0.002655 	 & 	 0.001958 	 & 	 0.006925 	 & 	 0.014222 	 & 	 0.101559 	 & 	 0.007741 	 & 	 0.005546 	 & 	 0.000859 	 & 	 0.001432 	 & 	 0.001603 	 & 	 0.001036 	 & 	 0.007068  \\
\multirow{ 2}{*}{$D^0 \rightarrow K_1(1270)^+K^-$} 	  	 & 	 0.010503 	 & 	 0.017298 	 & 	 0.000373 	 & 	 0.001321 	 & 	 0.006142 	 &  0.005164 	 & 	 0.011704 	 & 	 0.006367 	 & 	 0.004332 	 & 	 0.000405 	 & 	 0.001838 	 & 	 0.000727 	 & 	 0.000494 	 & 	 0.005687  \\
 	 & 	 0.026873 	 & 	 0.049547 	 & 	 0.005850 	 & 	 0.003096 	 & 	 0.005710 	 & 	 0.004240 	 & 	 0.011440 	 & 	 0.022312 	 & 	 0.006714 	 & 	 0.001071 	 & 	 0.005150 	 & 	 0.003626 	 & 	 0.001277 	 & 	 0.040550  \\
\multirow{ 2}{*}{$D^0 \rightarrow [K^*(892)^0\Kstarb(892)^0]_{L=0}$} 	  	 & 	 0.003849 	 & 	 0.017518 	 & 	 0.001451 	 & 	 0.000551 	 & 	 0.000956 	 & 	 0.000682 	 & 	 0.002441 	 & 	 0.001949 	 & 	 0.016664 	 & 	 0.000231 	 & 	 0.000826 	 & 	 0.000227 	 & 	 0.000204 	 & 	 0.003851  \\
 	 & 	 0.015913 	 & 	 0.025486 	 & 	 0.000277 	 & 	 0.001772 	 & 	 0.002792 	 & 	 0.014314 	 & 	 0.012448 	 & 	 0.013723 	 & 	 0.008444 	 & 	 0.000635 	 & 	 0.003090 	 & 	 0.000767 	 & 	 0.000872 	 & 	 0.002781  \\
\multirow{ 2}{*}{$D^0 \rightarrow K^*(1680)^0[K^-\pi^+]_{L=0}$} 	  	 & 	 0.035775 	 & 	 0.623954 	 & 	 0.010530 	 & 	 0.006106 	 & 	 0.007230 	 & 	 0.003000 	 & 	 0.041800 	 & 	 0.617655 	 & 	 0.067233 	 & 	 0.001212 	 & 	 0.015316 	 & 	 0.001867 	 & 	 0.001807 	 &  0.033186  \\
 	 & 	 0.016194 	 & 	 0.029678 	 & 	 0.000239 	 & 	 0.002432 	 & 	 0.006138 	 & 	 0.013400 	 & 	 0.017794 	 & 	 0.016213 	 & 	 0.006893 	 & 	 0.000564 	 & 	 0.000795 	 & 	 0.002971 	 & 	 0.000767 	 & 	 0.004496  \\
\multirow{ 2}{*}{$D^0 \rightarrow [K^*(892)^0\Kstarb(892)^0]_{L=1}$} 	  	 & 	 0.004589 	 & 	 0.016725 	 & 	 0.001298 	 & 	 0.000514 	 & 	 0.001311 	 & 	 0.000610 	 & 	 0.000666 	 & 	 0.002130 	 & 	 0.016259 	 & 	 0.000353 	 & 	 0.001777 	 & 	 0.000283 	 & 	 0.000230 	 &  0.001709  \\
 	 & 	 0.020853 	 & 	 0.026899 	 & 	 0.000060 	 & 	 0.001760 	 & 	 0.000740 	 & 	 0.013620 	 & 	 0.006410 	 & 	 0.013883 	 & 	 0.012440 	 & 	 0.000729 	 & 	 0.003714 	 & 	 0.002102 	 & 	 0.001020 	 & 	 0.011227  \\
\multirow{ 2}{*}{$D^0 \rightarrow K_1(1270)^-K^+$} 	  	 & 	 0.006000 	 & 	 0.010856 	 & 	 0.000028 	 & 	 0.000767 	 & 	 0.000113 	 &  0.001976 	 & 	 0.003689 	 & 	 0.004349 	 & 	 0.001970 	 & 	 0.000253 	 & 	 0.000296 	 & 	 0.001819 	 & 	 0.000293 	 & 	 0.008568  \\
 	 & 	 0.029205 	 & 	 0.074726 	 & 	 0.007640 	 & 	 0.003372 	 & 	 0.009550 	 & 	 0.003650 	 & 	 0.002890 	 & 	 0.016271 	 & 	 0.006929 	 & 	 0.002204 	 & 	 0.006042 	 & 	 0.002087 	 & 	 0.001457 	 & 	 0.070998  \\
\multirow{ 2}{*}{$D^0 \rightarrow [K^+K^-]_{L=0}[\pi^+\pi^-]_{L=0}$} 	  	 & 	 0.003359 	 & 	 0.017947 	 & 	 0.000514 	 & 	 0.000649 	 & 	 0.000467 	 & 	 0.000885 	 & 	 0.013213 	 & 	 0.001469 	 & 	 0.002136 	 & 	 0.000203 	 & 	 0.001010 	 & 	 0.000422 	 & 	 0.000161 	 &  0.011740  \\
 	 & 	 0.030172 	 & 	 0.163165 	 & 	 0.004920 	 & 	 0.005247 	 & 	 0.012810 	 & 	 0.013630 	 & 	 0.158290 	 & 	 0.013568 	 & 	 0.007796 	 & 	 0.002135 	 & 	 0.010829 	 & 	 0.002653 	 & 	 0.001426 	 & 	 0.028100  \\
\multirow{ 2}{*}{$D^0 \rightarrow K_1(1400)^-K^+$} 	  	 & 	 0.008487 	 & 	 0.018234 	 & 	 0.002185 	 & 	 0.001285 	 & 	 0.001609 	 &  0.000445 	 & 	 0.000750 	 & 	 0.005346 	 & 	 0.008304 	 & 	 0.000318 	 & 	 0.001449 	 & 	 0.002401 	 & 	 0.000406 	 & 	 0.014732  \\
 	 & 	 0.042483 	 & 	 0.087613 	 & 	 0.018529 	 & 	 0.008496 	 & 	 0.000340 	 & 	 0.015603 	 & 	 0.011795 	 & 	 0.041209 	 & 	 0.012535 	 & 	 0.002107 	 & 	 0.017696 	 & 	 0.005100 	 & 	 0.002220 	 & 	 0.068367  \\
\multirow{ 2}{*}{$D^0 \rightarrow [K^*(1680)^0\Kstarb(892)^0]_{L=0}$} 	  	 & 	 0.023366 	 & 	 0.218363 	 & 	 0.017455 	 & 	 0.005579 	 & 	 0.001750 	 & 	 0.001393 	 & 	 0.017650 	 & 	 0.213995 	 & 	 0.033858 	 & 	 0.001422 	 & 	 0.001797 	 & 	 0.001353 	 & 	 0.001361 	 & 	 0.009002  \\
 	 & 	 0.029472 	 & 	 0.047064 	 & 	 0.003590 	 & 	 0.004817 	 & 	 0.000480 	 & 	 0.013300 	 & 	 0.031070 	 & 	 0.019693 	 & 	 0.006293 	 & 	 0.001030 	 & 	 0.004859 	 & 	 0.003908 	 & 	 0.001848 	 & 	 0.023784  \\
\multirow{ 2}{*}{$D^0 \rightarrow [\Kstarb(1680)^0K^*(892)^0]_{L=1}$} 	  	 & 	 0.021850 	 & 	 0.275797 	 & 	 0.007060 	 & 	 0.002450 	 & 	 0.006370 	 & 	 0.002610 	 & 	 0.004080 	 & 	 0.271812 	 & 	 0.042851 	 & 	 0.000921 	 & 	 0.005060 	 & 	 0.001109 	 & 	 0.001656 	 &  0.014003  \\
 	 & 	 0.021658 	 & 	 0.029187 	 & 	 0.001350 	 & 	 0.001810 	 & 	 0.011340 	 & 	 0.014010 	 & 	 0.009450 	 & 	 0.013408 	 & 	 0.006457 	 & 	 0.000855 	 & 	 0.005937 	 & 	 0.001532 	 & 	 0.001201 	 & 	 0.013091  \\
\multirow{ 2}{*}{$D^0 \rightarrow \Kstarb(1680)^0[K^+\pi^-]_{L=0}$} 	  	 & 	 0.028708 	 & 	 0.372621 	 & 	 0.004080 	 & 	 0.005263 	 & 	 0.017750 	 & 	 0.003100 	 & 	 0.051410 	 & 	 0.361271 	 & 	 0.041472 	 & 	 0.001380 	 & 	 0.017543 	 & 	 0.004470 	 & 	 0.002114 	 &  0.057128  \\
 	 & 	 0.023677 	 & 	 0.030550 	 & 	 0.006410 	 & 	 0.003929 	 & 	 0.007920 	 & 	 0.015620 	 & 	 0.008790 	 & 	 0.015102 	 & 	 0.008379 	 & 	 0.000806 	 & 	 0.002161 	 & 	 0.001171 	 & 	 0.001164 	 & 	 0.013656  \\
\multirow{ 2}{*}{$D^0 \rightarrow [\phi(1020)(\rho-\omega)^0]_{L=2}$} 	  	 & 	 0.030602 	 & 	 0.017912 	 & 	 0.001630 	 & 	 0.002634 	 &  0.005080 	 & 	 0.009500 	 & 	 0.002470 	 & 	 0.004625 	 & 	 0.008243 	 & 	 0.001043 	 & 	 0.002134 	 & 	 0.002383 	 & 	 0.001823 	 & 	 0.009224  \\
 	 & 	 0.023183 	 & 	 0.018610 	 & 	 0.001100 	 & 	 0.003257 	 & 	 0.002352 	 & 	 0.010817 	 & 	 0.010989 	 & 	 0.004566 	 & 	 0.002801 	 & 	 0.001337 	 & 	 0.004652 	 & 	 0.001136 	 & 	 0.001168 	 & 	 0.006036  \\
\multirow{ 2}{*}{$D^0 \rightarrow [K^*(892)^0\Kstarb(892)^0]_{L=2}$} 	  	 & 	 0.017852 	 & 	 0.042694 	 & 	 0.002950 	 & 	 0.001980 	 & 	 0.002238 	 & 	 0.000944 	 & 	 0.011531 	 & 	 0.010041 	 & 	 0.039364 	 & 	 0.000672 	 & 	 0.002664 	 & 	 0.001007 	 & 	 0.000859 	 &  0.003415  \\
 	 & 	 0.027302 	 & 	 0.040250 	 & 	 0.002120 	 & 	 0.003328 	 & 	 0.010320 	 & 	 0.013180 	 & 	 0.010150 	 & 	 0.018102 	 & 	 0.010841 	 & 	 0.000973 	 & 	 0.002601 	 & 	 0.001404 	 & 	 0.001376 	 & 	 0.027649  \\
\multirow{ 2}{*}{$D^0 \rightarrow \phi(1020)[\pi^+\pi^-]_{L=0}$} 	  	 & 	 0.001499 	 & 	 0.004158 	 & 	 0.000115 	 & 	 0.000159 	 & 	 0.000042 	 & 	 0.000287 	 & 	 0.003727 	 & 	 0.000348 	 & 	 0.001202 	 & 	 0.000070 	 & 	 0.000327 	 & 	 0.000094 	 & 	 0.000077 	 &  0.001256  \\
 	 & 	 0.040338 	 & 	 0.368436 	 & 	 0.001970 	 & 	 0.004923 	 & 	 0.008350 	 & 	 0.005650 	 & 	 0.367500 	 & 	 0.010909 	 & 	 0.005539 	 & 	 0.001698 	 & 	 0.005058 	 & 	 0.002164 	 & 	 0.004262 	 & 	 0.018926  \\
\multirow{ 2}{*}{$D^0 \rightarrow [K^*(1680)^0\Kstarb(892)^0]_{L=1}$} 	  	 & 	 0.021438 	 & 	 0.203403 	 & 	 0.002860 	 & 	 0.002352 	 & 	 0.012119 	 & 	 0.001057 	 & 	 0.004288 	 & 	 0.200503 	 & 	 0.026748 	 & 	 0.000731 	 & 	 0.006970 	 & 	 0.003809 	 & 	 0.001208 	 &  0.014520  \\
 	 & 	 0.030815 	 & 	 0.039567 	 & 	 0.001539 	 & 	 0.002945 	 & 	 0.007164 	 & 	 0.013903 	 & 	 0.000590 	 & 	 0.021779 	 & 	 0.014857 	 & 	 0.001071 	 & 	 0.009136 	 & 	 0.001499 	 & 	 0.001810 	 & 	 0.022898  \\
\multirow{ 2}{*}{$D^0 \rightarrow [\phi(1020)\rho(1450)^0]_{L=1}$} 	  	 & 	 0.035319 	 & 	 0.067955 	 & 	 0.000415 	 & 	 0.003400 	 & 	 0.003990 	 & 	 0.005955 	 & 	 0.007534 	 & 	 0.061645 	 & 	 0.024509 	 & 	 0.001514 	 & 	 0.005451 	 & 	 0.002206 	 & 	 0.002152 	 &  0.007461  \\
 	 & 	 0.037748 	 & 	 0.037650 	 & 	 0.003680 	 & 	 0.002443 	 & 	 0.002650 	 & 	 0.026740 	 & 	 0.014730 	 & 	 0.014181 	 & 	 0.013372 	 & 	 0.001427 	 & 	 0.005473 	 & 	 0.001877 	 & 	 0.001840 	 & 	 0.006339  \\
\multirow{ 2}{*}{$D^0 \rightarrow a_0(980)^0f_2(1270)^0$} 	  	 & 	 0.057576 	 & 	 0.189018 	 & 	 0.000750 	 & 	 0.009482 	 & 	 0.026250 	 & 	 0.005390 	 & 	 0.048970 	 & 	 0.169338 	 & 	 0.034717 	 & 	 0.003031 	 & 	 0.014282 	 & 	 0.014492 	 & 	 0.004785 	 & 	 0.046847  \\
 	 & 	 0.037907 	 & 	 0.190443 	 & 	 0.007629 	 & 	 0.006319 	 & 	 0.021157 	 & 	 0.015237 	 & 	 0.062870 	 & 	 0.175738 	 & 	 0.014679 	 & 	 0.003562 	 & 	 0.006980 	 & 	 0.002755 	 & 	 0.003307 	 & 	 0.018942  \\
\multirow{ 2}{*}{$D^0 \rightarrow a_1(1260)^+\pi^-$} 	  	 & 	 0.011073 	 & 	 0.041879 	 & 	 0.000376 	 & 	 0.001097 	 & 	 0.000274 	 &  0.013563 	 & 	 0.037874 	 & 	 0.005010 	 & 	 0.008370 	 & 	 0.000385 	 & 	 0.001659 	 & 	 0.000917 	 & 	 0.000850 	 & 	 0.005866  \\
 	 & 	 0.066908 	 & 	 0.380249 	 & 	 0.008710 	 & 	 0.007835 	 & 	 0.006580 	 & 	 0.010710 	 & 	 0.375260 	 & 	 0.032760 	 & 	 0.011233 	 & 	 0.002641 	 & 	 0.004867 	 & 	 0.003198 	 & 	 0.003373 	 & 	 0.047145  \\
\multirow{ 2}{*}{$D^0 \rightarrow a_1(1260)^-\pi^+$} 	  	 & 	 0.013774 	 & 	 0.030671 	 & 	 0.000380 	 & 	 0.001857 	 & 	 0.000252 	 &  0.010629 	 & 	 0.026164 	 & 	 0.005617 	 & 	 0.004546 	 & 	 0.000483 	 & 	 0.004362 	 & 	 0.001492 	 & 	 0.000735 	 & 	 0.008078  \\
 	 & 	 0.059827 	 & 	 0.431268 	 & 	 0.017573 	 & 	 0.008464 	 & 	 0.013398 	 & 	 0.045308 	 & 	 0.425896 	 & 	 0.030807 	 & 	 0.015298 	 & 	 0.002817 	 & 	 0.015299 	 & 	 0.005786 	 & 	 0.004019 	 & 	 0.022740  \\
\multirow{ 2}{*}{$D^0 \rightarrow [\phi(1020)(\rho-\omega)^0]_{L=1}$} 	  	 & 	 0.011486 	 & 	 0.005217 	 & 	 0.000045 	 & 	 0.000655 	 &  0.000907 	 & 	 0.002018 	 & 	 0.003069 	 & 	 0.001678 	 & 	 0.001869 	 & 	 0.000872 	 & 	 0.001012 	 & 	 0.001564 	 & 	 0.000850 	 & 	 0.001100  \\
 	 & 	 0.070850 	 & 	 0.027234 	 & 	 0.010611 	 & 	 0.003020 	 & 	 0.006723 	 & 	 0.002456 	 & 	 0.006242 	 & 	 0.011600 	 & 	 0.015907 	 & 	 0.003884 	 & 	 0.004014 	 & 	 0.006881 	 & 	 0.006586 	 & 	 0.004499  \\
\multirow{ 2}{*}{$D^0 \rightarrow [K^*(1680)^0\Kstarb(892)^0]_{L=2}$} 	  	 & 	 0.089150 	 & 	 0.353706 	 & 	 0.024120 	 & 	 0.012127 	 & 	 0.059210 	 & 	 0.008690 	 & 	 0.020020 	 & 	 0.335298 	 & 	 0.038653 	 & 	 0.004038 	 & 	 0.007888 	 & 	 0.005305 	 & 	 0.004633 	 &  0.079686  \\
 	 & 	 0.083978 	 & 	 0.149961 	 & 	 0.021890 	 & 	 0.019505 	 & 	 0.012990 	 & 	 0.013380 	 & 	 0.045810 	 & 	 0.027899 	 & 	 0.006501 	 & 	 0.007195 	 & 	 0.004595 	 & 	 0.005896 	 & 	 0.004929 	 & 	 0.135018  \\
\multirow{ 2}{*}{$D^0 \rightarrow [K^+K^-]_{L=0}(\rho-\omega)^0$} 	  	 & 	 0.015084 	 & 	 0.034983 	 & 	 0.001756 	 & 	 0.002723 	 & 	 0.007501 	 & 	 0.003284 	 & 	 0.032500 	 & 	 0.005596 	 & 	 0.002336 	 & 	 0.001360 	 & 	 0.002487 	 & 	 0.001096 	 & 	 0.000796 	 &  0.006584  \\
 	 & 	 0.084385 	 & 	 0.292029 	 & 	 0.028870 	 & 	 0.014756 	 & 	 0.052180 	 & 	 0.010320 	 & 	 0.164590 	 & 	 0.036042 	 & 	 0.015819 	 & 	 0.007136 	 & 	 0.044806 	 & 	 0.014493 	 & 	 0.004121 	 & 	 0.224669  \\
\multirow{ 2}{*}{$D^0 \rightarrow [\phi(1020)f_2(1270)^0]_{L=1}$} 	  	 & 	 0.094812 	 & 	 0.256652 	 & 	 0.002960 	 & 	 0.009783 	 & 	 0.026290 	 & 	 0.005310 	 & 	 0.250540 	 & 	 0.016529 	 & 	 0.032495 	 & 	 0.003318 	 & 	 0.004589 	 & 	 0.005503 	 & 	 0.006057 	 & 	 0.029116  \\
 	 & 	 0.060755 	 & 	 0.372712 	 & 	 0.006650 	 & 	 0.007172 	 & 	 0.011440 	 & 	 0.011770 	 & 	 0.371300 	 & 	 0.012193 	 & 	 0.014400 	 & 	 0.002228 	 & 	 0.016791 	 & 	 0.003669 	 & 	 0.003507 	 & 	 0.004134  \\
\multirow{ 2}{*}{$D^0 \rightarrow [K^*(892)^0\Kbar{}^*_2(1430)^0]_{L=1}$} 	  	 & 	 0.086217 	 & 	 0.130576 	 & 	 0.019160 	 & 	 0.010547 	 & 	 0.019870 	 & 	 0.001010 	 & 	 0.004790 	 & 	 0.089444 	 & 	 0.073680 	 & 	 0.005869 	 & 	 0.023122 	 & 	 0.006678 	 & 	 0.004234 	 &  0.045743  \\
 	 & 	 0.066834 	 & 	 0.086560 	 & 	 0.022740 	 & 	 0.007892 	 & 	 0.012340 	 & 	 0.013070 	 & 	 0.043100 	 & 	 0.053536 	 & 	 0.013127 	 & 	 0.002800 	 & 	 0.009692 	 & 	 0.007388 	 & 	 0.004759 	 & 	 0.038924  \\ \midrule
\multirow{ 2}{*}{$K_1(1270)^+ \rightarrow [K^*(892)^0\pi^+]_{L=0}$} 	  	 & 	 0.016147 	 & 	 0.039508 	 & 	 0.002105 	 & 	 0.001848 	 & 	 0.007507 	 & 	 0.005169 	 & 	 0.014379 	 & 	 0.024046 	 & 	 0.022870 	 & 	 0.000565 	 & 	 0.006856 	 & 	 0.000795 	 & 	 0.000828 	 & 	 0.010644  \\
 	 & 	 0.030763 	 & 	 0.052408 	 & 	 0.000824 	 & 	 0.003233 	 & 	 0.008594 	 & 	 0.019554 	 & 	 0.003290 	 & 	 0.042694 	 & 	 0.012906 	 & 	 0.001403 	 & 	 0.004870 	 & 	 0.003852 	 & 	 0.001504 	 & 	 0.015375  \\
\multirow{ 2}{*}{$K_1(1270)^+ \rightarrow [K^+\pi^-]_{L=0}\pi^+$} 	  	 & 	 0.026976 	 & 	 0.094237 	 & 	 0.003153 	 & 	 0.004023 	 & 	 0.009427 	 & 	 0.002605 	 & 	 0.026681 	 & 	 0.013780 	 & 	 0.013650 	 & 	 0.002708 	 & 	 0.005629 	 & 	 0.001480 	 & 	 0.003363 	 &  0.087283  \\
 	 & 	 0.041798 	 & 	 0.082622 	 & 	 0.015180 	 & 	 0.007712 	 & 	 0.042390 	 & 	 0.019220 	 & 	 0.012880 	 & 	 0.022853 	 & 	 0.010955 	 & 	 0.001396 	 & 	 0.013916 	 & 	 0.003191 	 & 	 0.002222 	 & 	 0.057890  \\
\multirow{ 2}{*}{$K_1(1270)^+ \rightarrow [K^*(892)^0\pi^+]_{L=2}$} 	  	 & 	 0.044073 	 & 	 0.060219 	 & 	 0.011373 	 & 	 0.004403 	 & 	 0.019353 	 & 	 0.007464 	 & 	 0.015680 	 & 	 0.029704 	 & 	 0.037354 	 & 	 0.001676 	 & 	 0.002625 	 & 	 0.002254 	 & 	 0.002178 	 &  0.022457  \\
 	 & 	 0.040953 	 & 	 0.045303 	 & 	 0.000190 	 & 	 0.004381 	 & 	 0.004760 	 & 	 0.012200 	 & 	 0.017280 	 & 	 0.017040 	 & 	 0.009002 	 & 	 0.002425 	 & 	 0.003735 	 & 	 0.002205 	 & 	 0.003215 	 & 	 0.034009  \\
\multirow{ 2}{*}{$K_1(1270)^+ \rightarrow [\rho(1450)^0K^+]_{L=0}$} 	  	 & 	 0.068383 	 & 	 0.187061 	 & 	 0.022539 	 & 	 0.011655 	 & 	 0.055687 	 & 	 0.038172 	 & 	 0.079969 	 & 	 0.076003 	 & 	 0.022866 	 & 	 0.004728 	 & 	 0.026432 	 & 	 0.004529 	 & 	 0.003695 	 & 	 0.127829  \\
 	 & 	 0.099856 	 & 	 0.445333 	 & 	 0.051180 	 & 	 0.019067 	 & 	 0.006910 	 & 	 0.103880 	 & 	 0.174850 	 & 	 0.154396 	 & 	 0.051109 	 & 	 0.010111 	 & 	 0.027416 	 & 	 0.006056 	 & 	 0.004637 	 & 	 0.355760  \\ \midrule
\multirow{ 2}{*}{$D^0 \rightarrow [\phi(1020)\omega(782)]_{L=0}$} 	  	 & 	 0.004441 	 & 	 0.003162 	 & 	 0.000079 	 & 	 0.000224 	 & 	 0.000358 	 & 	 0.001663 	 & 	 0.000061 	 & 	 0.000576 	 & 	 0.002477 	 & 	 0.000199 	 & 	 0.000303 	 & 	 0.000309 	 &  0.000236 	 & 	 0.000543  \\
 	 & 	 0.042262 	 & 	 0.036079 	 & 	 0.002340 	 & 	 0.001822 	 & 	 0.007450 	 & 	 0.025050 	 & 	 0.002940 	 & 	 0.019980 	 & 	 0.011434 	 & 	 0.002677 	 & 	 0.003056 	 & 	 0.002047 	 & 	 0.002188 	 & 	 0.006768  \\ \midrule
\multirow{ 2}{*}{$D^0 \rightarrow [\phi(1020)\omega(782)]_{L=1}$} 	  	 & 	 0.051585 	 & 	 0.017613 	 & 	 0.002062 	 &  0.001611 	 & 	 0.003724 	 & 	 0.002639 	 & 	 0.005987 	 & 	 0.003664 	 & 	 0.008864 	 & 	 0.005513 	 & 	 0.004701 	 & 	 0.004171 	 & 	 0.007245 	 & 	 0.005687  \\
 	 & 	 0.193714 	 & 	 0.068702 	 & 	 0.016620 	 & 	 0.004784 	 & 	 0.015640 	 & 	 0.029530 	 & 	 0.029250 	 & 	 0.029520 	 & 	 0.026202 	 & 	 0.011142 	 & 	 0.011250 	 & 	 0.011117 	 & 	 0.013021 	 & 	 0.018623  \\ \midrule
\multirow{ 2}{*}{$D^0 \rightarrow [\phi(1020)\omega(782)]_{L=2}$} 	  	 & 	 0.031756 	 & 	 0.013717 	 & 	 0.000576 	 &  0.001141 	 & 	 0.001851 	 & 	 0.008214 	 & 	 0.002480 	 & 	 0.003043 	 & 	 0.003951 	 & 	 0.004551 	 & 	 0.001577 	 & 	 0.001536 	 & 	 0.001793 	 & 	 0.007473  \\
 	 & 	 0.167493 	 & 	 0.059369 	 & 	 0.002350 	 & 	 0.006777 	 & 	 0.001380 	 & 	 0.036500 	 & 	 0.028870 	 & 	 0.021008 	 & 	 0.010423 	 & 	 0.006182 	 & 	 0.008843 	 & 	 0.014088 	 & 	 0.009623 	 & 	 0.018658  \\ \midrule
\multirow{ 2}{*}{$D^0 \rightarrow [K^+K^-]_{L=0}\omega(782)$} 	  	 & 	 0.097704 	 & 	 0.098293 	 & 	 0.020292 	 &  0.007747 	 & 	 0.016314 	 & 	 0.008374 	 & 	 0.013273 	 & 	 0.021296 	 & 	 0.012837 	 & 	 0.003320 	 & 	 0.005403 	 & 	 0.005989 	 & 	 0.005060 	 & 	 0.089203  \\
 	 & 	 0.185513 	 & 	 0.149181 	 & 	 0.032670 	 & 	 0.017417 	 & 	 0.049630 	 & 	 0.011260 	 & 	 0.016690 	 & 	 0.055883 	 & 	 0.020618 	 & 	 0.016927 	 & 	 0.042473 	 & 	 0.013452 	 & 	 0.009241 	 & 	 0.110049  \\ \midrule
\multirow{ 2}{*}{$K_1(1270)^+ \rightarrow [\omega(782)K^+]_{L=0}$} 	  	 & 	 0.012062 	 & 	 0.010824 	 & 	 0.000877 	 &  0.000772 	 & 	 0.001643 	 & 	 0.002619 	 & 	 0.007572 	 & 	 0.003364 	 & 	 0.005122 	 & 	 0.000897 	 & 	 0.002230 	 & 	 0.000772 	 & 	 0.000573 	 & 	 0.002154  \\
 	 & 	 0.073676 	 & 	 0.056662 	 & 	 0.015210 	 & 	 0.004425 	 & 	 0.001030 	 & 	 0.036730 	 & 	 0.022430 	 & 	 0.023203 	 & 	 0.018507 	 & 	 0.003231 	 & 	 0.004280 	 & 	 0.005042 	 & 	 0.003512 	 & 	 0.012609  \\
\bottomrule
\end{tabular}
}

\end{table}

\begin{table}
  \caption{\label{tab:syst:fitfracs_cpv}Statistical and systematic uncertainties (in \%) on $A_{\mathcal{F}_k}$. Values smaller than 0.0005\% are displayed as ``0.000''. The sources of systematic uncertainty are described in the text in the same order as shown in this table.}
  % !TEX root = ./ms.tex

\resizebox{\textwidth}{!}{
\begin{tabular}{l*{14}{S[table-auto-round,table-format=1.3]}}
\toprule
& &                     {Total} & {Sel.} & {Alt.} & {Alt.} & {RBW} & {Alt.} &  {Mass \&} & {Res.}  & {Sig.} & {Bkg} & {Mistag} & {Det.\,as.} & {Alt.}\\
{Amplitude} & {Stat.} &  {syst.} & {eff.} & {bkg 1} & {bkg 2} & {$\rho(770)^0$} & {$S$-wave} & {width}  & {radius} & {bias} & {bias} & {bias} & {bias} & {models} \\ \midrule
$D^0 \rightarrow [\phi(1020)(\rho-\omega)^0]_{L=0}$ 	 &  	 1.496017 	 & 	 0.192587 	 & 	 0.007632 	 & 	 0.027354 	 & 	 0.032490 	 & 	 0.034086 	 & 	 0.101658 	 & 	 0.028229 	 & 	 0.029313 	 & 	 0.045591 	 & 	 0.048091 	 & 	 0.071180 	 & 	 0.044604 	 & 	 0.103092 	 \\
$D^0 \rightarrow K_1(1400)^+K^-$ 	 &  	 2.086958 	 & 	 0.280325 	 & 	 0.000985 	 & 	 0.032384 	 & 	 0.111233 	 & 	 0.026675 	 & 	 0.129325 	 & 	 0.140063 	 & 	 0.037385 	 & 	 0.085468 	 & 	 0.056228 	 & 	 0.072862 	 & 	 0.077163 	 & 	 0.070558 	 \\
$D^0 \rightarrow [K^-\pi^+]_{L=0}[K^+\pi^-]_{L=0}$ 	 &  	 1.816148 	 & 	 0.658614 	 & 	 0.015393 	 & 	 0.042818 	 & 	 0.112745 	 & 	 0.019694 	 &  0.627751 	 & 	 0.057531 	 & 	 0.028599 	 & 	 0.064511 	 & 	 0.053392 	 & 	 0.055216 	 & 	 0.083352 	 & 	 0.058259 	 \\
$D^0 \rightarrow K_1(1270)^+K^-$ 	 &  	 1.715600 	 & 	 0.207479 	 & 	 0.036093 	 & 	 0.026097 	 & 	 0.081676 	 & 	 0.015122 	 & 	 0.049842 	 & 	 0.060694 	 & 	 0.020942 	 & 	 0.128284 	 & 	 0.049824 	 & 	 0.061182 	 & 	 0.046049 	 & 	 0.052483 	 \\
$D^0 \rightarrow [K^*(892)^0\Kstarb(892)^0]_{L=0}$ 	 &  	 2.172076 	 & 	 0.479811 	 & 	 0.010452 	 & 	 0.039553 	 & 	 0.374299 	 & 	 0.022291 	 & 	 0.232947 	 & 	 0.033765 	 & 	 0.024796 	 & 	 0.066538 	 & 	 0.077444 	 & 	 0.070226 	 & 	 0.097496 	 & 	 0.084066 	 \\
$D^0 \rightarrow K^*(1680)^0[K^-\pi^+]_{L=0}$ 	 &  	 2.167269 	 & 	 0.412493 	 & 	 0.052298 	 & 	 0.046454 	 & 	 0.346305 	 & 	 0.020447 	 & 	 0.055677 	 & 	 0.058281 	 & 	 0.032185 	 & 	 0.076403 	 & 	 0.113960 	 & 	 0.063170 	 & 	 0.064165 	 & 	 0.102214 	 \\
$D^0 \rightarrow [K^*(892)^0\Kstarb(892)^0]_{L=1}$ 	 &  	 3.152356 	 & 	 0.291575 	 & 	 0.015991 	 & 	 0.030750 	 & 	 0.009261 	 & 	 0.014836 	 &  0.083708 	 & 	 0.037762 	 & 	 0.030600 	 & 	 0.109558 	 & 	 0.097618 	 & 	 0.145515 	 & 	 0.116265 	 & 	 0.133847 	 \\
$D^0 \rightarrow K_1(1270)^-K^+$ 	 &  	 3.520270 	 & 	 0.540160 	 & 	 0.037392 	 & 	 0.051185 	 & 	 0.255078 	 & 	 0.000353 	 & 	 0.040465 	 & 	 0.112818 	 & 	 0.075746 	 & 	 0.280684 	 & 	 0.100294 	 & 	 0.228284 	 & 	 0.129868 	 & 	 0.211578 	 \\
$D^0 \rightarrow [K^+K^-]_{L=0}[\pi^+\pi^-]_{L=0}$ 	 &  	 5.058336 	 & 	 3.119188 	 & 	 0.244713 	 & 	 0.140047 	 & 	 0.100503 	 & 	 0.099825 	 &  3.066365 	 & 	 0.182199 	 & 	 0.081411 	 & 	 0.161858 	 & 	 0.180897 	 & 	 0.151306 	 & 	 0.184200 	 & 	 0.267594 	 \\
$D^0 \rightarrow K_1(1400)^-K^+$ 	 &  	 6.049888 	 & 	 0.963149 	 & 	 0.054630 	 & 	 0.170919 	 & 	 0.071419 	 & 	 0.014000 	 & 	 0.209030 	 & 	 0.626928 	 & 	 0.131255 	 & 	 0.441806 	 & 	 0.275469 	 & 	 0.224227 	 & 	 0.171642 	 & 	 0.292212 	 \\
$D^0 \rightarrow [K^*(1680)^0\Kstarb(892)^0]_{L=0}$ 	 &  	 5.247434 	 & 	 1.515353 	 & 	 0.085885 	 & 	 0.169662 	 & 	 0.337492 	 & 	 0.064507 	 & 	 1.304197 	 & 	 0.288832 	 & 	 0.159708 	 & 	 0.244018 	 & 	 0.230094 	 & 	 0.161407 	 & 	 0.266946 	 & 	 0.349871 	 \\
$D^0 \rightarrow [\Kstarb(1680)^0K^*(892)^0]_{L=1}$ 	 &  	 3.895994 	 & 	 0.443738 	 & 	 0.013830 	 & 	 0.043732 	 & 	 0.164827 	 & 	 0.008226 	 &  0.006635 	 & 	 0.033250 	 & 	 0.040676 	 & 	 0.133534 	 & 	 0.213817 	 & 	 0.137114 	 & 	 0.131070 	 & 	 0.255404 	 \\
$D^0 \rightarrow \Kstarb(1680)^0[K^+\pi^-]_{L=0}$ 	 &  	 3.748491 	 & 	 1.069897 	 & 	 0.026771 	 & 	 0.078006 	 & 	 0.907590 	 & 	 0.025825 	 &  0.233876 	 & 	 0.129079 	 & 	 0.045698 	 & 	 0.265782 	 & 	 0.218865 	 & 	 0.192235 	 & 	 0.155694 	 & 	 0.245577 	 \\
$D^0 \rightarrow [\phi(1020)(\rho-\omega)^0]_{L=2}$ 	 &  	 3.277155 	 & 	 0.461991 	 & 	 0.050645 	 & 	 0.061366 	 & 	 0.219730 	 & 	 0.066257 	 &  0.158424 	 & 	 0.047666 	 & 	 0.051141 	 & 	 0.219375 	 & 	 0.163034 	 & 	 0.101720 	 & 	 0.128015 	 & 	 0.151683 	 \\
$D^0 \rightarrow [K^*(892)^0\Kstarb(892)^0]_{L=2}$ 	 &  	 4.963215 	 & 	 0.687022 	 & 	 0.220236 	 & 	 0.095402 	 & 	 0.042297 	 & 	 0.033542 	 &  0.374563 	 & 	 0.131667 	 & 	 0.049276 	 & 	 0.162144 	 & 	 0.178757 	 & 	 0.188330 	 & 	 0.246567 	 & 	 0.311303 	 \\
$D^0 \rightarrow \phi(1020)[\pi^+\pi^-]_{L=0}$ 	 &  	 6.078358 	 & 	 0.802436 	 & 	 0.291068 	 & 	 0.180072 	 & 	 0.229546 	 & 	 0.162606 	 &  0.302357 	 & 	 0.109412 	 & 	 0.120943 	 & 	 0.338901 	 & 	 0.183396 	 & 	 0.189676 	 & 	 0.257154 	 & 	 0.281094 	 \\
$D^0 \rightarrow [K^*(1680)^0\Kstarb(892)^0]_{L=1}$ 	 &  	 5.342129 	 & 	 0.553385 	 & 	 0.007992 	 & 	 0.064557 	 & 	 0.017464 	 & 	 0.019179 	 &  0.296671 	 & 	 0.084360 	 & 	 0.094557 	 & 	 0.211701 	 & 	 0.156443 	 & 	 0.154857 	 & 	 0.166234 	 & 	 0.276320 	 \\
$D^0 \rightarrow [\phi(1020)\rho(1450)^0]_{L=1}$ 	 &  	 8.527026 	 & 	 1.102079 	 & 	 0.132984 	 & 	 0.129105 	 & 	 0.178188 	 & 	 0.144058 	 &  0.177927 	 & 	 0.240061 	 & 	 0.359865 	 & 	 0.344024 	 & 	 0.432484 	 & 	 0.517898 	 & 	 0.347156 	 & 	 0.463467 	 \\
$D^0 \rightarrow a_0(980)^0f_2(1270)^0$ 	 &  	 7.189789 	 & 	 1.304647 	 & 	 0.211676 	 & 	 0.195918 	 & 	 0.359212 	 & 	 0.012164 	 & 	 0.688585 	 & 	 0.733392 	 & 	 0.190142 	 & 	 0.264688 	 & 	 0.217127 	 & 	 0.310649 	 & 	 0.301891 	 & 	 0.369756 	 \\
$D^0 \rightarrow a_1(1260)^+\pi^-$ 	 &  	 11.699799 	 & 	 7.041944 	 & 	 0.436700 	 & 	 0.256130 	 & 	 0.256805 	 & 	 1.030703 	 & 	 6.833420 	 & 	 0.419227 	 & 	 0.366711 	 & 	 0.524499 	 & 	 0.642121 	 & 	 0.344421 	 & 	 0.498473 	 & 	 0.379462 	 \\
$D^0 \rightarrow a_1(1260)^-\pi^+$ 	 &  	 13.672433 	 & 	 2.860056 	 & 	 0.741774 	 & 	 0.348605 	 & 	 0.188891 	 & 	 0.879210 	 & 	 2.052026 	 & 	 0.428533 	 & 	 0.431521 	 & 	 0.981899 	 & 	 0.518822 	 & 	 0.404655 	 & 	 0.646402 	 & 	 0.551298 	 \\
$D^0 \rightarrow [\phi(1020)(\rho-\omega)^0]_{L=1}$ 	 &  	 10.998947 	 & 	 1.356912 	 & 	 0.329812 	 & 	 0.148629 	 & 	 0.358527 	 & 	 0.006057 	 &  0.125462 	 & 	 0.565437 	 & 	 0.493795 	 & 	 0.344132 	 & 	 0.338787 	 & 	 0.578182 	 & 	 0.554009 	 & 	 0.357837 	 \\
$D^0 \rightarrow [K^*(1680)^0\Kstarb(892)^0]_{L=2}$ 	 &  	 14.304026 	 & 	 3.535071 	 & 	 0.548572 	 & 	 0.646090 	 & 	 2.042198 	 & 	 0.114816 	 &  2.266215 	 & 	 0.346539 	 & 	 0.242669 	 & 	 0.642246 	 & 	 0.469495 	 & 	 0.478951 	 & 	 0.430349 	 & 	 1.110135 	 \\
$D^0 \rightarrow [K^+K^-]_{L=0}(\rho-\omega)^0$ 	 &  	 12.534049 	 & 	 2.791080 	 & 	 0.018495 	 & 	 0.316192 	 & 	 2.014176 	 & 	 0.032155 	 &  1.181790 	 & 	 0.577962 	 & 	 0.127688 	 & 	 0.942554 	 & 	 0.630252 	 & 	 0.513656 	 & 	 0.380173 	 & 	 0.436915 	 \\
$D^0 \rightarrow [\phi(1020)f_2(1270)^0]_{L=1}$ 	 &  	 13.301422 	 & 	 2.984865 	 & 	 0.124228 	 & 	 0.355177 	 & 	 0.583569 	 & 	 0.284157 	 & 	 2.659817 	 & 	 0.209532 	 & 	 0.298338 	 & 	 0.460416 	 & 	 0.435587 	 & 	 0.391637 	 & 	 0.394519 	 & 	 0.654414 	 \\
$D^0 \rightarrow [K^*(892)^0\Kbar{}^*_2(1430)^0]_{L=1}$ 	 &  	 10.804974 	 & 	 1.810103 	 & 	 0.476823 	 & 	 0.170970 	 & 	 0.830287 	 & 	 0.065245 	 & 	 1.066403 	 & 	 0.588098 	 & 	 0.236658 	 & 	 0.332159 	 & 	 0.379588 	 & 	 0.327052 	 & 	 0.324797 	 & 	 0.565945 	 \\
\bottomrule
\end{tabular}
}

\end{table}

\begin{table}
  \caption{\label{tab:syst:fitpars_cpv}Statistical and systematic uncertainties (in \%) on the \CP-violation parameters shown for all floating components. Values smaller than 0.0005\% are displayed as ``0.000''. The sources of systematic uncertainty are described in the text in the same order as shown in this table. For each amplitude, the first value quoted is the modulus asymmetry and the second is the phase difference.}
  % !TEX root = ./ms.tex

\resizebox{\textwidth}{!}{
\begin{tabular}{l*{14}{S[table-auto-round,table-format=1.3]}}
  \toprule
  & &                     {Total} & {Sel.} & {Alt.} & {Alt.} & {RBW} & {Alt.} &  {Mass \&} & {Res.}  & {Sig.} & {Bkg} & {Mistag} & {Det.\,as.} & {Alt.}\\
  {Amplitude} & {Stat.} &  {syst.} & {eff.} & {bkg 1} & {bkg 2} & {$\rho(770)^0$} & {$S$-wave} & {width}  & {radius} & {bias} & {bias} & {bias} & {bias} & {models} \\ \midrule
\multirow{ 2}{*}{$D^0 \rightarrow K_1(1400)^+K^-$} 	  	 & 	 1.084480 	 & 	 0.199854 	 & 	 0.004310 	 & 	 0.020681 	 & 	 0.071880 	 &  0.030380 	 & 	 0.095420 	 & 	 0.074092 	 & 	 0.026258 	 & 	 0.039401 	 & 	 0.070954 	 & 	 0.050191 	 & 	 0.057393 	 & 	 0.075642 	 \\
 	 & 	 1.468730 	 & 	 0.251078 	 & 	 0.007760 	 & 	 0.031940 	 & 	 0.026090 	 & 	 0.021500 	 & 	 0.154630 	 &  0.067080 	 & 	 0.022410 	 & 	 0.048717 	 & 	 0.099048 	 & 	 0.068540 	 & 	 0.069045 	 & 	 0.101269 	 \\
\multirow{ 2}{*}{$D^0 \rightarrow [K^-\pi^+]_{L=0}[K^+\pi^-]_{L=0}$} 	  	 & 	 1.134020 	 & 	 0.303964 	 & 	 0.003882 	 & 	 0.024095 	 & 	 0.040136 	 & 	 0.007195 	 & 	 0.263071 	 & 	 0.025499 	 & 	 0.019403 	 & 	 0.034891 	 & 	 0.064795 	 & 	 0.061723 	 & 	 0.053184 	 &  0.088592 	 \\
 	 & 	 1.349110 	 & 	 0.292496 	 & 	 0.003842 	 & 	 0.031080 	 & 	 0.081292 	 & 	 0.003916 	 & 	 0.133742 	 &  0.039424 	 & 	 0.020374 	 & 	 0.083804 	 & 	 0.122997 	 & 	 0.106970 	 & 	 0.115273 	 & 	 0.105885 	 \\
\multirow{ 2}{*}{$D^0 \rightarrow K_1(1270)^+K^-$} 	  	 & 	 0.999095 	 & 	 0.190299 	 & 	 0.014230 	 & 	 0.019010 	 & 	 0.024600 	 &  0.024600 	 & 	 0.075750 	 & 	 0.034789 	 & 	 0.014079 	 & 	 0.062733 	 & 	 0.045063 	 & 	 0.109784 	 & 	 0.044946 	 & 	 0.085172 	 \\
 	 & 	 1.443750 	 & 	 0.231795 	 & 	 0.033400 	 & 	 0.027154 	 & 	 0.055570 	 & 	 0.003570 	 & 	 0.114030 	 &  0.036708 	 & 	 0.020686 	 & 	 0.045464 	 & 	 0.076010 	 & 	 0.066847 	 & 	 0.065249 	 & 	 0.132010 	 \\
\multirow{ 2}{*}{$D^0 \rightarrow [K^*(892)^0\Kstarb(892)^0]_{L=0}$} 	  	 & 	 1.298780 	 & 	 0.264889 	 & 	 0.001411 	 & 	 0.024349 	 & 	 0.170904 	 & 	 0.028189 	 & 	 0.065644 	 & 	 0.021726 	 & 	 0.022162 	 & 	 0.041028 	 & 	 0.114312 	 & 	 0.081554 	 & 	 0.076257 	 & 	 0.084139 	 \\
 	 & 	 1.470570 	 & 	 0.204762 	 & 	 0.000271 	 & 	 0.033062 	 & 	 0.066035 	 & 	 0.000669 	 & 	 0.063269 	 &  0.029977 	 & 	 0.019293 	 & 	 0.051157 	 & 	 0.100899 	 & 	 0.069686 	 & 	 0.083233 	 & 	 0.081350 	 \\
\multirow{ 2}{*}{$D^0 \rightarrow K^*(1680)^0[K^-\pi^+]_{L=0}$} 	  	 & 	 1.309440 	 & 	 0.262353 	 & 	 0.029976 	 & 	 0.027329 	 & 	 0.189466 	 & 	 0.027270 	 & 	 0.074489 	 & 	 0.030310 	 & 	 0.025392 	 & 	 0.041896 	 & 	 0.063162 	 & 	 0.058077 	 & 	 0.064247 	 &  0.100923 	 \\
 	 & 	 1.491050 	 & 	 0.213661 	 & 	 0.051721 	 & 	 0.031873 	 & 	 0.078955 	 & 	 0.011789 	 & 	 0.031728 	 &  0.056000 	 & 	 0.018051 	 & 	 0.048325 	 & 	 0.068352 	 & 	 0.067279 	 & 	 0.066459 	 & 	 0.123159 	 \\
\multirow{ 2}{*}{$D^0 \rightarrow [K^*(892)^0\Kstarb(892)^0]_{L=1}$} 	  	 & 	 1.713200 	 & 	 0.220326 	 & 	 0.004190 	 & 	 0.019210 	 & 	 0.020880 	 & 	 0.009600 	 & 	 0.092740 	 & 	 0.023914 	 & 	 0.023420 	 & 	 0.064643 	 & 	 0.082303 	 & 	 0.082824 	 & 	 0.108387 	 &  0.091366 	 \\
 	 & 	 2.001620 	 & 	 0.238991 	 & 	 0.036680 	 & 	 0.025909 	 & 	 0.027280 	 & 	 0.004670 	 & 	 0.068310 	 &  0.032717 	 & 	 0.014947 	 & 	 0.087179 	 & 	 0.104009 	 & 	 0.093055 	 & 	 0.094883 	 & 	 0.110876 	 \\
\multirow{ 2}{*}{$D^0 \rightarrow K_1(1270)^-K^+$} 	  	 & 	 1.706590 	 & 	 0.435611 	 & 	 0.022513 	 & 	 0.030053 	 & 	 0.143798 	 &  0.016866 	 & 	 0.030594 	 & 	 0.055779 	 & 	 0.046492 	 & 	 0.134312 	 & 	 0.135434 	 & 	 0.213484 	 & 	 0.178954 	 & 	 0.217242 	 \\
 	 & 	 2.074110 	 & 	 0.310679 	 & 	 0.022980 	 & 	 0.043917 	 & 	 0.063734 	 & 	 0.019095 	 & 	 0.143553 	 &  0.063169 	 & 	 0.018519 	 & 	 0.065054 	 & 	 0.092373 	 & 	 0.101253 	 & 	 0.117434 	 & 	 0.167001 	 \\
\multirow{ 2}{*}{$D^0 \rightarrow [K^+K^-]_{L=0}[\pi^+\pi^-]_{L=0}$} 	  	 & 	 2.482280 	 & 	 1.520716 	 & 	 0.118544 	 & 	 0.069467 	 & 	 0.066498 	 & 	 0.032872 	 & 	 1.482500 	 & 	 0.087044 	 & 	 0.042098 	 & 	 0.077795 	 & 	 0.110177 	 & 	 0.114070 	 & 	 0.167084 	 &  0.148055 	 \\
 	 & 	 2.646500 	 & 	 1.556901 	 & 	 0.001600 	 & 	 0.072299 	 & 	 0.664900 	 & 	 0.025990 	 & 	 1.340750 	 &  0.106282 	 & 	 0.083941 	 & 	 0.135851 	 & 	 0.241097 	 & 	 0.132882 	 & 	 0.134759 	 & 	 0.218155 	 \\
\multirow{ 2}{*}{$D^0 \rightarrow K_1(1400)^-K^+$} 	  	 & 	 2.878800 	 & 	 0.686818 	 & 	 0.023540 	 & 	 0.088578 	 & 	 0.051990 	 &  0.024050 	 & 	 0.155450 	 & 	 0.317435 	 & 	 0.066711 	 & 	 0.211143 	 & 	 0.206813 	 & 	 0.292469 	 & 	 0.273697 	 & 	 0.287848 	 \\
 	 & 	 3.546580 	 & 	 1.049376 	 & 	 0.160828 	 & 	 0.155483 	 & 	 0.823376 	 & 	 0.031581 	 & 	 0.247875 	 &  0.122328 	 & 	 0.112777 	 & 	 0.137367 	 & 	 0.229567 	 & 	 0.171681 	 & 	 0.251904 	 & 	 0.344353 	 \\
\multirow{ 2}{*}{$D^0 \rightarrow [K^*(1680)^0\Kstarb(892)^0]_{L=0}$} 	  	 & 	 2.672800 	 & 	 0.775178 	 & 	 0.046860 	 & 	 0.084028 	 & 	 0.152900 	 & 	 0.015300 	 & 	 0.602520 	 & 	 0.141057 	 & 	 0.085288 	 & 	 0.114277 	 & 	 0.270041 	 & 	 0.127521 	 & 	 0.160478 	 & 	 0.223237 	 \\
 	 & 	 2.763280 	 & 	 0.821441 	 & 	 0.140120 	 & 	 0.063992 	 & 	 0.650140 	 & 	 0.022820 	 & 	 0.179020 	 &  0.119276 	 & 	 0.067825 	 & 	 0.153121 	 & 	 0.202596 	 & 	 0.149431 	 & 	 0.198920 	 & 	 0.224875 	 \\
\multirow{ 2}{*}{$D^0 \rightarrow [\Kstarb(1680)^0K^*(892)^0]_{L=1}$} 	  	 & 	 2.063220 	 & 	 0.255642 	 & 	 0.003100 	 & 	 0.025094 	 & 	 0.098698 	 & 	 0.012930 	 & 	 0.048658 	 & 	 0.019616 	 & 	 0.028631 	 & 	 0.068914 	 & 	 0.094714 	 & 	 0.106000 	 & 	 0.108917 	 &  0.120060 	 \\
 	 & 	 2.086690 	 & 	 0.278267 	 & 	 0.023790 	 & 	 0.030143 	 & 	 0.118140 	 & 	 0.016390 	 & 	 0.059680 	 &  0.034597 	 & 	 0.019717 	 & 	 0.067307 	 & 	 0.095262 	 & 	 0.101905 	 & 	 0.149078 	 & 	 0.101835 	 \\
\multirow{ 2}{*}{$D^0 \rightarrow \Kstarb(1680)^0[K^+\pi^-]_{L=0}$} 	  	 & 	 1.958500 	 & 	 0.628296 	 & 	 0.009580 	 & 	 0.040910 	 & 	 0.437710 	 & 	 0.004130 	 & 	 0.167810 	 & 	 0.060698 	 & 	 0.026969 	 & 	 0.137911 	 & 	 0.093800 	 & 	 0.202247 	 & 	 0.127246 	 &  0.289652 	 \\
 	 & 	 2.249350 	 & 	 0.328249 	 & 	 0.022731 	 & 	 0.045167 	 & 	 0.058129 	 & 	 0.040801 	 & 	 0.125255 	 &  0.044516 	 & 	 0.030880 	 & 	 0.072810 	 & 	 0.195717 	 & 	 0.105877 	 & 	 0.102011 	 & 	 0.127672 	 \\
\multirow{ 2}{*}{$D^0 \rightarrow [\phi(1020)(\rho-\omega)^0]_{L=2}$} 	  	 & 	 1.882750 	 & 	 0.314183 	 & 	 0.021550 	 & 	 0.034082 	 &  0.093800 	 & 	 0.050210 	 & 	 0.130140 	 & 	 0.026689 	 & 	 0.021338 	 & 	 0.121481 	 & 	 0.095658 	 & 	 0.149332 	 & 	 0.092815 	 & 	 0.113308 	 \\
 	 & 	 1.990880 	 & 	 0.468930 	 & 	 0.076560 	 & 	 0.054594 	 & 	 0.205640 	 & 	 0.086360 	 & 	 0.146290 	 &  0.036512 	 & 	 0.015057 	 & 	 0.159115 	 & 	 0.147391 	 & 	 0.183795 	 & 	 0.122149 	 & 	 0.206408 	 \\
\multirow{ 2}{*}{$D^0 \rightarrow [K^*(892)^0\Kstarb(892)^0]_{L=2}$} 	  	 & 	 2.510260 	 & 	 0.418673 	 & 	 0.114490 	 & 	 0.051313 	 & 	 0.005040 	 & 	 0.033880 	 & 	 0.238990 	 & 	 0.066661 	 & 	 0.029019 	 & 	 0.083324 	 & 	 0.127121 	 & 	 0.117859 	 & 	 0.113590 	 &  0.214650 	 \\
 	 & 	 2.616560 	 & 	 0.448760 	 & 	 0.054760 	 & 	 0.083798 	 & 	 0.014600 	 & 	 0.001490 	 & 	 0.071450 	 &  0.046094 	 & 	 0.061728 	 & 	 0.104961 	 & 	 0.159345 	 & 	 0.291004 	 & 	 0.121496 	 & 	 0.210374 	 \\
\multirow{ 2}{*}{$D^0 \rightarrow \phi(1020)[\pi^+\pi^-]_{L=0}$} 	  	 & 	 3.077970 	 & 	 0.650050 	 & 	 0.141781 	 & 	 0.088287 	 & 	 0.098577 	 & 	 0.064285 	 & 	 0.100416 	 & 	 0.058891 	 & 	 0.059940 	 & 	 0.180862 	 & 	 0.342842 	 & 	 0.205475 	 & 	 0.332165 	 &  0.246717 	 \\
 	 & 	 3.899830 	 & 	 0.676876 	 & 	 0.107310 	 & 	 0.127975 	 & 	 0.235320 	 & 	 0.183300 	 & 	 0.254810 	 &  0.082103 	 & 	 0.083089 	 & 	 0.133909 	 & 	 0.282105 	 & 	 0.181501 	 & 	 0.200070 	 & 	 0.303704 	 \\
\multirow{ 2}{*}{$D^0 \rightarrow [K^*(1680)^0\Kstarb(892)^0]_{L=1}$} 	  	 & 	 2.752330 	 & 	 0.465300 	 & 	 0.000180 	 & 	 0.034068 	 & 	 0.007514 	 & 	 0.026633 	 & 	 0.199193 	 & 	 0.048305 	 & 	 0.048967 	 & 	 0.120529 	 & 	 0.203811 	 & 	 0.238440 	 & 	 0.136350 	 &  0.196614 	 \\
 	 & 	 2.988150 	 & 	 0.424942 	 & 	 0.034080 	 & 	 0.039237 	 & 	 0.007770 	 & 	 0.007030 	 & 	 0.023510 	 &  0.050671 	 & 	 0.030110 	 & 	 0.119710 	 & 	 0.176333 	 & 	 0.168323 	 & 	 0.235552 	 & 	 0.210942 	 \\
\multirow{ 2}{*}{$D^0 \rightarrow [\phi(1020)\rho(1450)^0]_{L=1}$} 	  	 & 	 4.120270 	 & 	 0.561374 	 & 	 0.062900 	 & 	 0.064147 	 & 	 0.073160 	 & 	 0.055260 	 & 	 0.058970 	 & 	 0.126992 	 & 	 0.173933 	 & 	 0.160677 	 & 	 0.258403 	 & 	 0.195157 	 & 	 0.183229 	 &  0.290788 	 \\
 	 & 	 3.342360 	 & 	 0.593141 	 & 	 0.005130 	 & 	 0.037288 	 & 	 0.193910 	 & 	 0.062170 	 & 	 0.085160 	 &  0.087935 	 & 	 0.145437 	 & 	 0.179706 	 & 	 0.244895 	 & 	 0.201234 	 & 	 0.193361 	 & 	 0.320383 	 \\
\multirow{ 2}{*}{$D^0 \rightarrow a_0(980)^0f_2(1270)^0$} 	  	 & 	 3.563800 	 & 	 0.693572 	 & 	 0.110000 	 & 	 0.099218 	 & 	 0.196420 	 & 	 0.010920 	 & 	 0.294800 	 & 	 0.361448 	 & 	 0.095022 	 & 	 0.130440 	 & 	 0.208492 	 & 	 0.185731 	 & 	 0.201802 	 & 	 0.241068 	 \\
 	 & 	 3.340730 	 & 	 0.834142 	 & 	 0.104810 	 & 	 0.105302 	 & 	 0.225550 	 & 	 0.022920 	 & 	 0.614030 	 &  0.232831 	 & 	 0.073329 	 & 	 0.115903 	 & 	 0.188864 	 & 	 0.182936 	 & 	 0.223097 	 & 	 0.230997 	 \\
\multirow{ 2}{*}{$D^0 \rightarrow a_1(1260)^+\pi^-$} 	  	 & 	 5.639560 	 & 	 3.660295 	 & 	 0.223350 	 & 	 0.129541 	 & 	 0.145340 	 &  0.535700 	 & 	 3.475808 	 & 	 0.214023 	 & 	 0.193913 	 & 	 0.255005 	 & 	 0.591074 	 & 	 0.311717 	 & 	 0.427405 	 & 	 0.405099 	 \\
 	 & 	 6.147820 	 & 	 1.268046 	 & 	 0.031320 	 & 	 0.124781 	 & 	 0.485880 	 & 	 0.175680 	 & 	 0.594990 	 &  0.139551 	 & 	 0.131039 	 & 	 0.205241 	 & 	 0.664578 	 & 	 0.287667 	 & 	 0.293265 	 & 	 0.530324 	 \\
\multirow{ 2}{*}{$D^0 \rightarrow a_1(1260)^-\pi^+$} 	  	 & 	 7.024770 	 & 	 1.921449 	 & 	 0.378420 	 & 	 0.182949 	 & 	 0.079130 	 &  0.460420 	 & 	 1.088180 	 & 	 0.222324 	 & 	 0.228921 	 & 	 0.459959 	 & 	 0.722995 	 & 	 0.617980 	 & 	 0.541753 	 & 	 0.775500 	 \\
 	 & 	 5.553980 	 & 	 4.326916 	 & 	 0.101820 	 & 	 0.149737 	 & 	 0.369190 	 & 	 0.205370 	 & 	 4.197220 	 &  0.175184 	 & 	 0.081658 	 & 	 0.193212 	 & 	 0.331766 	 & 	 0.280110 	 & 	 0.703816 	 & 	 0.368377 	 \\
\multirow{ 2}{*}{$D^0 \rightarrow [\phi(1020)(\rho-\omega)^0]_{L=1}$} 	  	 & 	 5.173380 	 & 	 0.759143 	 & 	 0.168722 	 & 	 0.075294 	 &  0.195510 	 & 	 0.020072 	 & 	 0.072136 	 & 	 0.279133 	 & 	 0.242647 	 & 	 0.200112 	 & 	 0.279190 	 & 	 0.311538 	 & 	 0.269976 	 & 	 0.271299 	 \\
 	 & 	 5.468470 	 & 	 0.610764 	 & 	 0.060390 	 & 	 0.072415 	 & 	 0.243316 	 & 	 0.000920 	 & 	 0.116731 	 &  0.099731 	 & 	 0.032493 	 & 	 0.170320 	 & 	 0.240194 	 & 	 0.263124 	 & 	 0.251792 	 & 	 0.246929 	 \\
\multirow{ 2}{*}{$D^0 \rightarrow [K^*(1680)^0\Kstarb(892)^0]_{L=2}$} 	  	 & 	 7.063650 	 & 	 1.871951 	 & 	 0.284420 	 & 	 0.331382 	 & 	 1.062990 	 & 	 0.075600 	 & 	 1.212430 	 & 	 0.175171 	 & 	 0.122311 	 & 	 0.320587 	 & 	 0.361601 	 & 	 0.392309 	 & 	 0.360778 	 &  0.380264 	 \\
 	 & 	 8.123840 	 & 	 1.312437 	 & 	 0.109500 	 & 	 0.479333 	 & 	 0.097900 	 & 	 0.144700 	 & 	 0.417500 	 &  0.324819 	 & 	 0.179722 	 & 	 0.260820 	 & 	 0.435344 	 & 	 0.709209 	 & 	 0.391011 	 & 	 0.474009 	 \\
\multirow{ 2}{*}{$D^0 \rightarrow [K^+K^-]_{L=0}(\rho-\omega)^0$} 	  	 & 	 5.999940 	 & 	 1.866346 	 & 	 0.005432 	 & 	 0.158124 	 & 	 1.023504 	 & 	 0.033121 	 & 	 0.540066 	 & 	 0.293479 	 & 	 0.066413 	 & 	 0.479565 	 & 	 0.500939 	 & 	 0.825839 	 & 	 0.465382 	 &  0.804880 	 \\
 	 & 	 6.249570 	 & 	 1.124705 	 & 	 0.185250 	 & 	 0.167854 	 & 	 0.180580 	 & 	 0.025350 	 & 	 0.729620 	 &  0.409648 	 & 	 0.252538 	 & 	 0.200945 	 & 	 0.273119 	 & 	 0.280635 	 & 	 0.317137 	 & 	 0.333134 	 \\
\multirow{ 2}{*}{$D^0 \rightarrow [\phi(1020)f_2(1270)^0]_{L=1}$} 	  	 & 	 6.709620 	 & 	 1.686184 	 & 	 0.065929 	 & 	 0.177891 	 & 	 0.275539 	 & 	 0.159122 	 & 	 1.380983 	 & 	 0.108916 	 & 	 0.162702 	 & 	 0.232080 	 & 	 0.405417 	 & 	 0.326074 	 & 	 0.489988 	 & 	 0.442601 	 \\
 	 & 	 6.038370 	 & 	 1.687668 	 & 	 0.161308 	 & 	 0.119009 	 & 	 0.253140 	 & 	 0.034469 	 & 	 1.354552 	 &  0.128381 	 & 	 0.080633 	 & 	 0.240125 	 & 	 0.534701 	 & 	 0.294036 	 & 	 0.367788 	 & 	 0.565409 	 \\
\multirow{ 2}{*}{$D^0 \rightarrow [K^*(892)^0\Kbar{}^*_2(1430)^0]_{L=1}$} 	  	 & 	 5.194470 	 & 	 1.037510 	 & 	 0.235560 	 & 	 0.086162 	 & 	 0.400670 	 & 	 0.015710 	 & 	 0.585700 	 & 	 0.298352 	 & 	 0.114594 	 & 	 0.170917 	 & 	 0.380521 	 & 	 0.313873 	 & 	 0.229385 	 &  0.287054 	 \\
 	 & 	 6.351030 	 & 	 1.363858 	 & 	 0.036430 	 & 	 0.129575 	 & 	 0.306930 	 & 	 0.048880 	 & 	 0.564840 	 &  0.473326 	 & 	 0.164380 	 & 	 0.334675 	 & 	 0.433667 	 & 	 0.356314 	 & 	 0.723216 	 & 	 0.474572 	 \\
\bottomrule
\end{tabular}
}

\end{table}

% This should be taken out in the final paper
%\input{ancillary-app}
%\input{supplementary-app}

\clearpage
\addcontentsline{toc}{section}{References}
\setboolean{inbibliography}{true}
\bibliographystyle{LHCb}
\bibliography{main,standard,LHCb-PAPER,LHCb-CONF,LHCb-DP,LHCb-TDR}

\newpage
% LHCb Collaboration author list
% Data extracted on January 29th, 2019 at 10:05am for reference date 25-Sep-2018
\centerline
{\large\bf LHCb Collaboration}
\begin
{flushleft}
\small
R.~Aaij$^{29}$,
C.~Abell{\'a}n~Beteta$^{46}$,
B.~Adeva$^{43}$,
M.~Adinolfi$^{50}$,
C.A.~Aidala$^{77}$,
Z.~Ajaltouni$^{7}$,
S.~Akar$^{61}$,
P.~Albicocco$^{20}$,
J.~Albrecht$^{12}$,
F.~Alessio$^{44}$
% ...

{\footnotesize \it

% ...
$ ^{7}$Universit{\'e} Clermont Auvergne, CNRS/IN2P3, LPC, Clermont-Ferrand, France\\
% ...
$ ^{12}$Fakult{\"a}t Physik, Technische Universit{\"a}t Dortmund, Dortmund, Germany\\
% ...
$ ^{20}$INFN Laboratori Nazionali di Frascati, Frascati, Italy\\
% ...
$ ^{29}$Nikhef National Institute for Subatomic Physics, Amsterdam, Netherlands\\
% ...
$ ^{43}$Instituto Galego de F{\'\i}sica de Altas Enerx{\'\i}as (IGFAE), Universidade de Santiago de Compostela, Santiago de Compostela, Spain\\
$ ^{44}$European Organization for Nuclear Research (CERN), Geneva, Switzerland\\
% ...
$ ^{46}$Physik-Institut, Universit{\"a}t Z{\"u}rich, Z{\"u}rich, Switzerland\\
% ...
$ ^{50}$H.H. Wills Physics Laboratory, University of Bristol, Bristol, United Kingdom\\
% ...
$ ^{61}$University of Cincinnati, Cincinnati, OH, United States\\
% ...
$ ^{63}$Syracuse University, Syracuse, NY, United States\\
% ...
$ ^{77}$University of Michigan, Ann Arbor, United States, associated to $^{63}$\\
% ...
}
\end{flushleft}

\end{document}